\documentclass[twocolumn]{aastex702}
\usepackage{amsmath}
\usepackage{array}
\usepackage{longtable}
\usepackage{graphicx}
\usepackage{placeins}
\usepackage{etoolbox}
\newcommand{\dodoi}[1]{doi:~\href{https://doi.org/#1}{\nolinkurl{#1}}}
\hypersetup{pdftitle={Testing star formation models through compression and expansion with a local deforming patch},pdfauthor={Ignacio Mosqueira},colorlinks=true,citecolor=blue,linkcolor=blue,urlcolor=blue}
\apptocmd{\thebibliography}{\baselineskip=12pt\relax\parskip=0pt\relax\bibskip=0pt\relax}{}{}
\shorttitle{Star formation through compression and expansion}
\shortauthors{Mosqueira}
\begin{document}

\title{Testing star formation models through compression and expansion with a local deforming patch}

\author{Ignacio Mosqueira}

\affiliation{San Jos\'e State University, One Washington Square, San Jos\'e, CA 95192, USA}
\email[show]{Ignacio.Mosqueira@sjsu.edu}

\raggedbottom
\renewcommand{\dbltopfraction}{0.95}
\renewcommand{\dblfloatpagefraction}{0.85}
\renewcommand{\textfraction}{0.05}
\setcounter{dbltopnumber}{2}
\setlength{\dblfloatsep}{4pt plus 1pt minus 1pt}
\setlength{\dbltextfloatsep}{0pt}
\makeatletter
\renewcommand{\frontmatter@postabstractspace}{0pt}
\setlength{\@dblfptop}{0pt}
\setlength{\@dblfpsep}{18pt}
\setlength{\@dblfpbot}{0pt plus 1fil}
\makeatother

\makeatletter
\newenvironment{widetableblock}{\onecolumngrid\par\addvspace{8pt}\noindent
  \begin{minipage}{\textwidth}\def\@captype{table}}
  {\end{minipage}\par\addvspace{8pt}\twocolumngrid}
\newenvironment{widefigureblock}{\onecolumngrid\par\addvspace{8pt}\noindent
  \begin{minipage}{\textwidth}\def\@captype{figure}}
  {\end{minipage}\par\addvspace{8pt}\twocolumngrid}
\newenvironment{columntableblock}{\par\addvspace{7pt}\noindent
  \begin{minipage}{\linewidth}\def\@captype{table}}
  {\end{minipage}\par\addvspace{7pt}}
\newenvironment{appendixfigure}{\par\addvspace{12pt}\noindent
  \begin{minipage}{\linewidth}\def\@captype{figure}}
  {\end{minipage}\par\addvspace{10pt}}
\makeatother
\begin{abstract}

We use the deforming-box method of \citet{Mosqueira1996} to follow a stratified patch of a galactic disk through compression--expansion cycles, with ideal magnetohydrodynamics (MHD), cooling, self-gravity, star-cluster formation, far-ultraviolet heating and supernova feedback. In a validation test, a single patch reproduces a domain of 16 periodic copies at the same resolution and timestep, and runs 15.6 times faster. Starting from a common state, we compress the patch to half its radial width and return it to its original width over 20, 50 or 100 Myr. We compare all three cycling runs to one uncompressed control. Over the complete passages, the cycling runs form 37\% less, 202\% more and 29\% more stellar mass than the control over the same intervals. We evaluate the Kennicutt--Schmidt (KS) relation and the vertically integrated pressure-regulated, feedback-modulated (PRFM) prescription on each simulated history and compare each prediction with the stellar mass formed in the same run. The prescriptions underestimate the stellar mass formed in all four runs and how it varies among the cycling runs. Developing a physical compression--expansion scaling is left to future work.

\end{abstract}


\section{Introduction}\label{sec:1}

Star-formation models must describe gas in environments shaped by galactic spiral arms, and testing them requires simulations that resolve the small scales on which stars form.
The deforming patch technique was first developed for local simulations of perturbed dense planetary rings \citep{Mosqueira1996}. \citet{OgilvieLatter2013} and \citet{OgilvieBarker2014} developed closely related local descriptions of global deformations for warped and eccentric disks, and \citet{WienkersOgilvie2018} implemented a local numerical model to study nonlinear instability and turbulence in vertically structured eccentric disks. We revisit this numerical technique as an inexpensive test environment for star-formation simulations.

Galaxies impose different compression histories through spiral arms, bars and interactions. \citet{Semenov2017} follow gas through compression and expansion as it enters and leaves spiral arms. The spiral-arm application is also motivated by \citet{KimKimOstriker2020}, whose arm/interarm environment provides strongly varying conditions for local star formation.
The Kennicutt--Schmidt (KS) relation links gas surface density and star-formation rate per unit area \citep{Kennicutt1998}. \citet{KimKimOstriker2020} examine this relation in their spiral-arm simulations, finding a roughly linear trend with substantial scatter across spatial bins sampled over time. The pressure-regulated, feedback-modulated (PRFM) model seeks to explain star formation through the pressure required to support the gas and the feedback that maintains that support \citep{OstrikerKim2022}. 

Estimates of the time gas takes to cross a spiral arm in M51 inform our 20 and 50 Myr passage timescales \citep{Koda2009}, and its arm--interarm contrast in gas surface density, about two to three \citep{Hitschfeld2009}, motivates halving the patch width over one cycle. We also test a slower, 100 Myr passage \citep{KimKimOstriker2020}. We compare each prescription's prediction with the stellar mass formed in the same run over the same interval, and each cycling run with the uncompressed control.

Our approach involves following a small region as the flow carries and deforms it, rather than modeling a fixed region through which the flow passes. The supplied global flow carries the galactic mode pattern, so the local code need not simulate the whole arm-to-arm interval to follow matter through compression and expansion. Once the local approximation and its image conditions are specified, tracking the deforming domain involves a change of coordinates plus the required globally consistent geometric source terms. Our independent solver combines the ideal magnetohydrodynamics (MHD) and other local physics of standard TIGRESS (Three-phase Interstellar medium in Galaxies Resolving Evolution with Star formation and Supernova feedback; \citealt{KimOstriker2017}) with the deformation prescribed by a galactic mode. The imposed flow sets the contraction and expansion of the patch, within which the gas, star clusters and their feedback evolve (Section~\ref{sec:physics}).
This is Part I of a two-paper study. Part II \citep{MosqueiraCloudAssembly} examines the evolving-cloud simulations and the assembly and binding histories of the control's star-forming regions, which provide the basis for further model development.

Section~\ref{sec:2} defines the deforming patch, and Section~\ref{sec:validation} verifies it, including a test of whether one patch reproduces an independently evolved larger domain containing 16 periodic copies and the resulting computational saving (Section~\ref{sec:validation_cost}). Section~\ref{sec:9} compares the KS and PRFM predictions with each run and the cycling runs with the control, follows the control through 135 Myr and tests feedback during expansion with a matched restart in the 20 Myr passage. Section~\ref{sec:magnetic_response} examines the magnetic response through the 20 Myr passage, and Section~\ref{sec:field_strength} compares fixed and cycling boxes at several field strengths; Section~\ref{sec:field_compression} shows that the cycling response also depends on how the sink length is set. Section~\ref{sec:10} discusses what the method enables and its limits, and Section~\ref{sec:11} concludes. Table~\ref{tab:roadmap} lists the tests with and without imposed compression; the appendices hold the detailed verification, earlier experiments with imposed compression and their fixed-size controls, results for uncompressed boxes and a precursor solver-matching control.

\section{The prescribed-flow local calculation}\label{sec:2}

\subsection{Coordinates, physical resolution and MHD evolution}\label{sec:2.1}

The three-dimensional box deforms along $x$. Translation can be removed by following the box center; the present compression tests set that translation to zero. Let $(\xi,\eta,\zeta)$ be coordinates fixed to the box, $t$ physical time, $a(t)$ its dimensionless stretch along $x$, and $\mathbf{v}$ the physical gas velocity. A dot denotes a time derivative and a hat a Cartesian unit vector. The physical coordinates $(x,y,z)$, background expansion rate $H$ and residual velocity $\mathbf{w}$ are

\begin{equation}
\begin{gathered}
x=a(t)\xi,\quad y=\eta,\quad z=\zeta,\\
H=\dot a/a,\qquad \mathbf{w}=\mathbf{v}-Hx\,\hat{\mathbf{x}}.
\end{gathered}
\end{equation}

The scale factor $a(t)$ specifies the box size: $a<1$ denotes a compressed state and $a=1$ the reference width. Contraction has $H<0$ and expansion $H>0$; the box remains compressed during expansion until it returns to $a=1$. Local collapse, cloud convergence and background shear can remain active at either size. The imposed background flow is spatially linear: its velocity gradient is uniform across the patch and can vary with time, while the local MHD dynamics remain nonlinear. With reference mesh spacing $\Delta$, spatial derivatives use the physical spacings $(a\Delta,\Delta,\Delta)$.

The following equations use coordinates fixed to the box. Here $\rho$ is mass density, $P$ thermal pressure, $e$ thermal energy density, $\gamma$ the adiabatic index, $\Phi$ the gravitational potential per unit mass, and $\mathbf{B}$ the magnetic field. We use rationalized magnetic units, so $B^2/2$ is magnetic energy density, with $B^2=\mathbf{B}\cdot\mathbf{B}$. The gradient $\nabla=\nabla_x$ acts on physical coordinates, while $\partial_t|_{\xi}$ holds the box coordinates fixed. With the residual material derivative $D'=\partial_t|_{\xi}+\mathbf{w}\cdot\nabla_x$ and $\nabla\cdot\mathbf{B}=0$, the homogeneous ideal-MHD equations are

\begin{equation}
\begin{gathered}
D^\prime\rho=-\rho(\nabla\cdot\mathbf{w}+H),\\
P=(\gamma-1)e,\quad\gamma=5/3,
\end{gathered}
\end{equation}

\begin{equation}
\begin{aligned}
D^\prime\mathbf{w}={}&-\rho^{-1}\nabla(P+B^2/2)\\
&+\rho^{-1}(\mathbf{B}\cdot\nabla)\mathbf{B}-\nabla\Phi\\
&-Hw_x\hat{\mathbf{x}},
\end{aligned}
\end{equation}

\begin{equation}
\begin{aligned}
D^\prime\mathbf{B}={}&(\mathbf{B}\cdot\nabla)\mathbf{w}-\mathbf{B}\nabla\cdot\mathbf{w}\\
&+H(B_x\hat{\mathbf{x}}-\mathbf{B}).
\end{aligned}
\end{equation}

Electrons, ions and neutrals are treated as one perfectly coupled conducting fluid. With Ohmic resistivity, ambipolar diffusion and the Hall effect omitted, magnetic flux is frozen into the gas in the continuum equations. TIGRESS and its spiral-arm calculations also adopt ideal MHD \citep{KimOstriker2017,KimKimOstriker2020}. Constrained transport preserves the discrete divergence constraint. Finite-resolution transport and remapping introduce numerical diffusion that depends on the scheme, resolution and flow. The background-deformation source step conserves the flux through each moving cell face to round-off.

The background flow requires the nonrotating acceleration $\mathbf f_{\rm imposed}=(\ddot a/a)x\hat{\mathbf x}=(\dot H+H^2)x\hat{\mathbf x}$. The code subtracts it together with the prescribed background; replacing it with an unrelated spiral force while keeping the same prescribed history would make the forcing inconsistent. Gravity, radiative heating and cooling, mass conversion into particles, and feedback are added explicitly. In these calculations both gas and particles receive the imposed linear forcing; stars drifting through a separately forced gas spiral are not yet included.

Transport uses the MHD momentum-stress tensor $\mathbf{T}$, total residual energy density $E$, and its flux $\mathbf{F}_E$. With $\mathbf{I}$ the identity tensor, juxtaposed vectors denoting outer products, and $w^2=\mathbf{w}\cdot\mathbf{w}$, these are

\begin{equation}
\begin{gathered}
\mathbf{T}=\rho\mathbf{w}\mathbf{w}+(P+B^2/2)\mathbf{I}-\mathbf{B}\mathbf{B},\\
E=e+\rho w^2/2+B^2/2,
\end{gathered}
\end{equation}

\begin{equation}
\mathbf{F}_E=(E+P+B^2/2)\mathbf{w}-(\mathbf{w}\cdot\mathbf{B})\mathbf{B}.
\end{equation}

Subscripts 1 and 2 denote the states before and after a source step. For a uniform linear-deformation source step from $a_1$ to $a_2$, with $r=a_2/a_1$, the code applies the exact maps

\begin{equation}
\begin{gathered}
\rho_2=\rho_1/r,\quad w_{x,2}=w_{x,1}/r,\\
w_{y,z,2}=w_{y,z,1},\quad P_2=P_1/r^\gamma,
\end{gathered}
\end{equation}

\begin{equation}
B_{x,2}=B_{x,1},\qquad B_{y,z,2}=B_{y,z,1}/r.
\end{equation}

The magnetic map is flux freezing under the uniform deformation: the flux through each moving cell face is conserved, so the components normal to the $y$ and $z$ faces, whose areas scale with $a$, change as $1/r$, while $B_x$, normal to faces of fixed area, is unchanged. These maps are split symmetrically around transport on the midpoint geometry. Physical volume and horizontal area scale as $a$, so the residual gas-plus-magnetic energy ledger includes the compression work rate $\dot W$, integrated over physical volume $dV$:

\begin{equation}
\dot W=-H\int\left[\rho w_x^2+P+\frac{B_y^2+B_z^2-B_x^2}{2}\right]dV.
\end{equation}

The magnetic term is the normal Maxwell stress in the direction of compression. This energy is accounted for as work done by the imposed flow, not as numerical heating. Because the cell count is fixed, compression increases spatial resolution without shortening the physical evolution time.

The magnetic field evolves with the local gas: magnetic pressure and tension act in the momentum equation, and magnetic energy enters the total-energy equation. The prescribed flow sets the large-scale deformation, while collapse, shear, turbulence and feedback can advect and bend the field within the patch, so its strength and geometry are part of the simulated response.

\begin{figure*}[!tp]
\centering
\includegraphics[width=\textwidth,height=0.63\textheight,keepaspectratio]{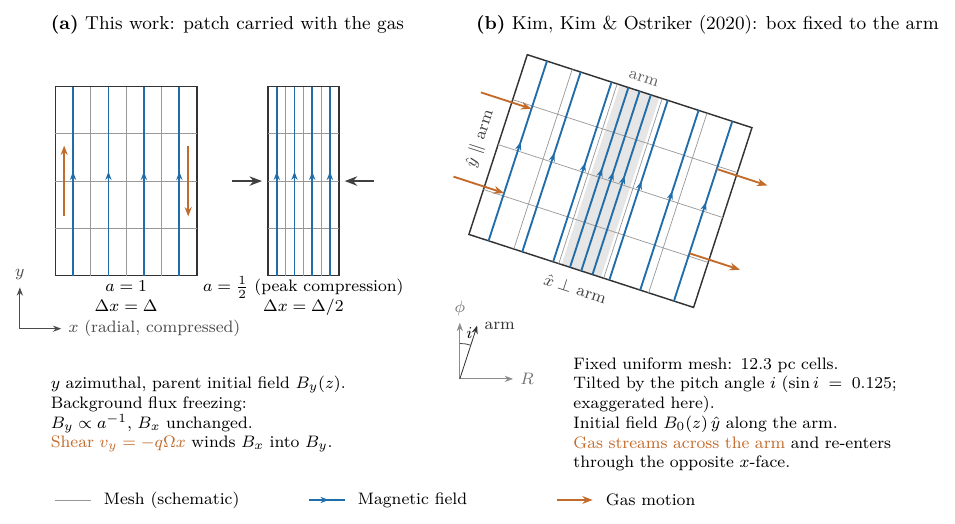}
\caption{\textbf{Field and compression geometry.} Schematic top views, not to scale: blue arrowed lines denote the magnetic field, light gray lines the computational mesh (not the actual cell counts) and orange arrows gas motion; $z$ is vertical, with diode outflow boundaries in both setups. (a) Our patch uses radial $x$ and azimuthal $y$ reference axes. Its parent calculation is initialized with $B_y(z)=[2\bar P(z)/\beta]^{1/2}$, where $\bar P$ is the horizontally averaged thermal pressure and $\beta=2\bar P/B^2=10$; all three cycling runs and the control inherit the evolved field at 35 Myr. The drawing shows background compression alone (left, $a=1$; right, $a=1/2$), omitting the accumulated mesh shear. As the radial mesh spacing shrinks, conserved moving-face flux packs the same field lines closer: $B_y\propto a^{-1}$ and $B_x$ is unchanged. Shear, $v_y=-q\Omega x$, winds radial into azimuthal field, with shear parameter $q$ and angular frequency $\Omega$. (b) The spiral-arm setup of \citet{KimKimOstriker2020}: a box fixed to the arm pattern and tilted by the pitch angle $i$ ($\sin i=0.125$, exaggerated here), with $\hat x$ perpendicular and $\hat y$ parallel to the arm. The uniform 12.3 pc mesh stays fixed; the closer field lines schematically indicate field amplification. The initial field $B_0(z)\hat y$ lies along the arm; arrows show the cross-arm streaming component, with gas re-entering through the opposite $x$-face. Both setups compress perpendicular to the initial field; (a) follows the gas, whereas (b) records the arm region through which it flows.\label{fig:field_geometry}}
\end{figure*}

Figure~\ref{fig:field_geometry} compares our background field geometry with the magnetized spiral-arm calculations of \citet{KimKimOstriker2020}. Our parent calculation and their magnetized models are initialized with a horizontal field set by the pressure profile and thermal-to-magnetic pressure ratio $\beta=10$, and both compress perpendicular to that field. Their box is fixed to the arm pattern and tilted by the pitch angle, so gas streams across the arm through it; \citet[][Section II.3]{KimKimOstriker2020} use a fixed, uniform mesh with 12.3 pc cells throughout the arm and interarm regions, within which compression and magnetic amplification occur. Our patch uses radial--azimuthal reference axes and is carried with the gas, and its mesh contracts with the imposed flow: $\Delta x=a(t)\Delta$, so twofold compression halves the physical cell width along $x$. A fixed arm-frame formulation could also use a nonuniform or adaptive mesh; the distinction concerns the implementations compared in the figure.

\subsection{Image conditions and open vertical boundaries}\label{sec:2.2}

The box follows the prescribed large-scale flow, while gas and stars remain free to move relative to it. Random or turbulent motions can carry material across a side in either horizontal direction; that material re-enters through the corresponding opposite side. In a shearing, deforming box, re-entry includes the required position and background-velocity shifts, preserving motion relative to the local background flow. The top and bottom are open: material escaping vertically does not re-enter.

Without shear, residual gas variables and magnetic fields are periodic in the box coordinates, while the physical velocities of periodic images differ by the imposed flow. These image conditions represent residual disturbances that repeat over the patch, with the supplied background carrying the global flow; Appendix~\ref{sec:3.2} quantifies the disturbance content that this periodicity selects. Across the $x$ boundary, let $L_x$ be the current physical box length and $\Delta v_x$ the velocity jump between images. The physical total-energy density is $E_{\rm phys}=e+\rho v^2/2+B^2/2$. The image conditions are

\begin{equation}
\begin{gathered}
\Delta v_x=HL_x,\\
v_x(x+L_x)=v_x(x)+\Delta v_x,
\end{gathered}
\end{equation}

\begin{equation}
\begin{aligned}
&E_{\rm phys}(x+L_x)-E_{\rm phys}(x)\\
&\qquad=\rho v_x\Delta v_x+\frac{1}{2}\rho(\Delta v_x)^2.
\end{aligned}
\end{equation}

The independent physical-image check constructs the donor physical velocity, applies this jump, and subtracts the background at the ghost position before rebuilding the conserved state. Magnetic fields follow the area scaling under the deformation above; the velocity jump between images does not multiply the magnetic flux. At the open $z$ surfaces, a diode prohibits inflow both in the gas Riemann states and in the vertical velocity used by the edge electromotive force. The second condition is essential for a consistent magnetic boundary.

All three principal cycling runs and their shared uncompressed control include rotation and shear. The map is $x=a\xi$, $y=\eta+s\xi+y_0$, $z=\zeta$, with dimensionless shear displacement $s$, lattice-origin offset $y_0$ and $\dot s=-q\Omega a$ between lattice remaps. Here $\Omega$ is the fixed angular frequency and $q=-d\ln\Omega/d\ln R$ the shear parameter at galactocentric radius $R$. The stored velocity subtracts the full background $\mathbf v_0=(Hx,-q\Omega x,0)$. Radial images are displaced by $(aL_\xi,sL_\xi,0)$, where $L_\xi$ is the reference radial length, and differ in velocity by $(HL_x,-q\Omega L_x,0)$, with $L_x=aL_\xi$. Thus periodicity on this moving lattice is shearing periodicity in physical space. Appendix~\ref{app:principal_spec} gives the combined equations, magnetic mapping, remap, particle/source frames, gravity and component checks.

Shear also winds radial field into azimuthal field through the $-q\Omega B_x\hat{\mathbf y}$ term of the sheared induction equation (Appendix~\ref{app:principal_map}); the stored face fluxes and the physical field are related by the evolving mesh geometry (Appendix~\ref{app:principal_map}). Constrained transport updates neighboring faces with shared edge electromotive forces, so their contributions cancel in the discrete magnetic divergence, and the periodic images and shear remaps use the same face-flux representation. Expansion reverses the geometric scaling but does not reset field changes produced by the intervening local evolution.

\begin{figure*}[!tp]
\centering
\includegraphics[width=\textwidth,height=0.63\textheight,keepaspectratio]{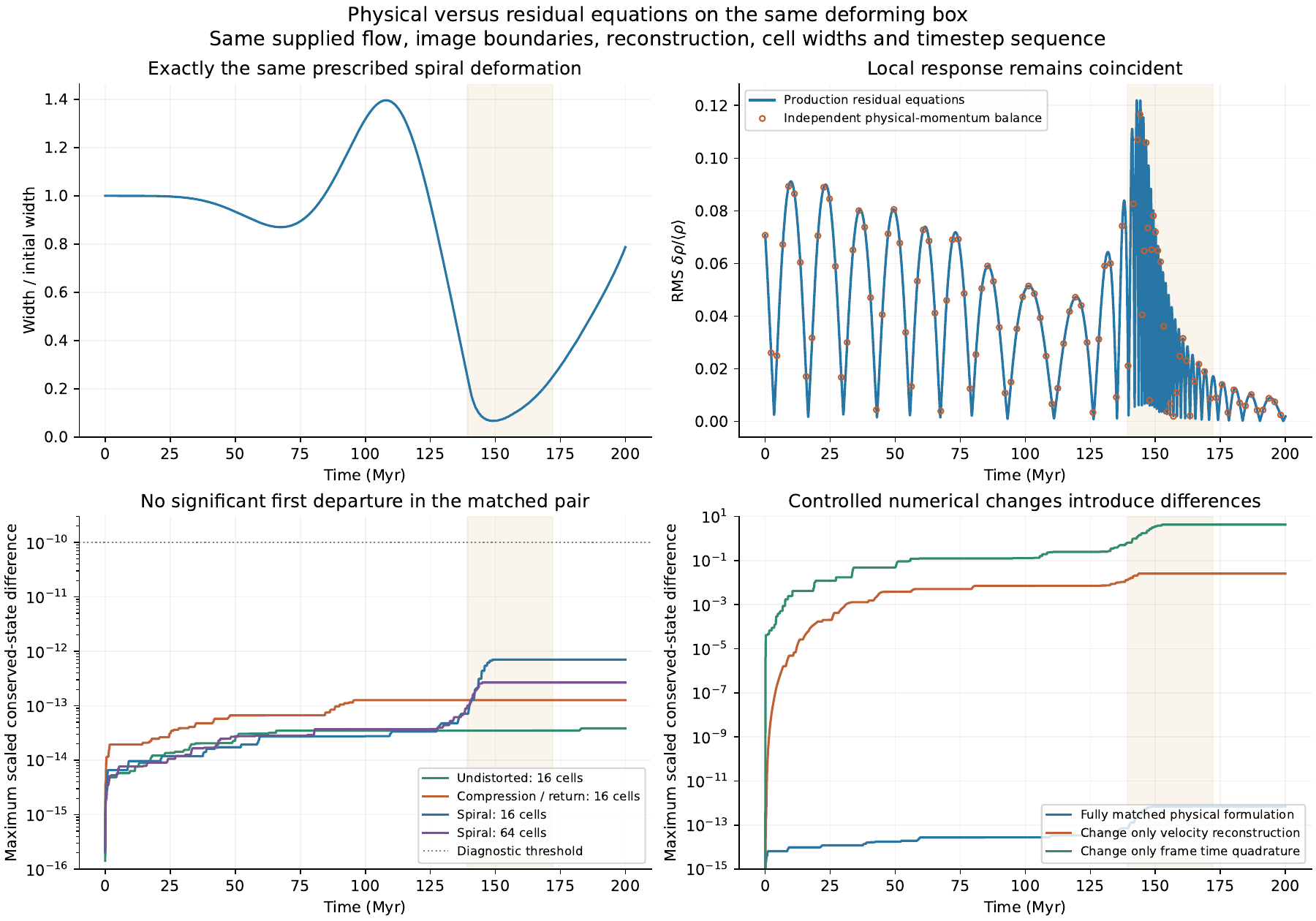}
\caption{\textbf{Independent physical and residual evolution.} The two formulations evolve independently with the same prescribed flow, image conditions, reconstruction and timesteps. The suite includes an undistorted control, eightfold compression and return, and a supplied spiral history. Deliberate numerical ablations test sensitivity to reconstruction and to time quadrature of the background-momentum terms. Appendix~\ref{sec:3.1} defines the scaled state norm and gives the component results.\label{fig:2}}
\end{figure*}

\subsection{Finite-region sources in a common physical frame}\label{sec:2.3}

Periodic transport alone does not complete the transformation, because collapse criteria, accretion and a supernova kernel act across a finite region whose background velocities differ. The physical $x$-convergence is $\partial_x w_x+H$. Gas with residual velocity $\mathbf{w}_g$ at physical separation $\delta x$ from a particle is transferred in the particle's background frame with velocity $\mathbf{w}_g+H\delta x\,\hat{\mathbf{x}}$, using the nearest periodic image. In the three principal cycling runs and their shared uncompressed control, a sink removes gas mass, momentum and energy from its 27-cell control volume. The magnetic flux that threaded the accreted gas stays on the mesh. Removing mass at fixed flux lowers the mass-to-flux ratio of the gas remaining in the accretion region. Transport of flux below this scale is unresolved. \citet{KimKimOstriker2020} likewise do not reset magnetic fields in their sink control volumes. For feedback, every target cell is boosted to a common physical frame centered on the progenitor, the source operation is applied there, and the result is transformed back. The binding test in the stratified runs of Appendix~\ref{sec:5} and the binding diagnostics in the reduced cloud model of Part~II use physical velocities, including the background gradients. The principal runs use the sink formation criteria in Table~\ref{tab:principal_sources} without a separate binding test.

The frame change also enters the energy bookkeeping. For this source update, $\Delta E_{\rm res}$ and $\Delta E_{\rm phys}$ denote volume-integrated energy changes, whereas $E$ and $E_{\rm phys}$ above are densities. Let $i$ index target cells of physical volume $\Delta V_i$, $m^{\rm phys}_{x,i}$ be their physical $x$-momentum density, and $v_{0,i}$ their background-flow velocity offset in the common frame. A $\Delta$ on a state variable denotes its change during the source operation. Then

\begin{equation}
\begin{gathered}
\Delta E_{\rm res}-\Delta E_{\rm phys}=\\
\sum_i\left[-v_{0,i}\Delta m^{\rm phys}_{x,i}+\frac{1}{2}v_{0,i}^2\Delta\rho_i\right]\Delta V_i.
\end{gathered}
\end{equation}

The energy budget records this coordinate-conversion contribution separately; it is not additional feedback energy. The local source transformation leaves magnetic fields unchanged.

\begin{table*}[!tp]
\caption{Numerical validation and model tests. Grid spacings are initial cell widths.\label{tab:roadmap}\label{tab:1}\label{tab:2}}
\centering\footnotesize\setlength{\tabcolsep}{4pt}\renewcommand{\arraystretch}{1.2}
\begin{tabular}{>{\raggedright\arraybackslash}p{\dimexpr 0.48\textwidth-2\tabcolsep\relax}>{\raggedright\arraybackslash}p{\dimexpr 0.52\textwidth-2\tabcolsep\relax}}
\hline
\textbf{Test} & \textbf{Purpose and location} \\ \hline
\multicolumn{2}{l}{\textbf{Without imposed box compression}} \\
Undistorted physical/residual test; MHD waves and shocks & Transport and representation controls (Appendices~\ref{sec:3.1} and~\ref{sec:3.3}). \\
Fixed-box hydrodynamic and zero-field tests & Test sensitivity to feedback, grid spacing and solver choice (Appendices~\ref{sec:4}--\ref{sec:6}). \\
Shared uncompressed control, 35--135 Myr & Stellar mass formed and predicted over the same intervals as the cycling runs (Section~\ref{sec:9.1}). \\ \hline
\multicolumn{2}{l}{\textbf{With prescribed compression or expansion}} \\
Linear-flow and supplied-spiral equation tests; flux freezing & Coordinate, boundary and work verification (Section~\ref{sec:validation_geometry}; Figure~\ref{fig:2}; Appendix~\ref{sec:3}). \\
Smooth compression--return and fixed-event refinement & Numerical convergence through deformation (Figure~\ref{fig:16}; Appendix~\ref{sec:7}). \\
Local box versus a domain containing 16 periodic copies & Equivalence and cost at matched resolution and timesteps (Section~\ref{sec:validation_cost}; Figure~\ref{fig:20}). \\
30 Myr stratified pilot runs & Test how magnetic stresses and stellar mass change with grid spacing (16, 8 and 4 pc; Section~\ref{sec:validation_stars}; Figures~\ref{fig:11}, \ref{fig:17}; Appendices~\ref{sec:5}, \ref{sec:7}). \\
Principal 20, 50 and 100 Myr passages (initially 8 pc cells) & Stellar mass in the cycling runs and the shared uncompressed control, with KS and PRFM predictions (Tables~\ref{tab:prediction_errors} and~\ref{tab:cycle_change}). \\
Field-strength and sink-length tests & Vary the initial field strength and the sink length; repeat fixed-box cases with different perturbation seeds (Sections~\ref{sec:field_strength} and~\ref{sec:field_physics}). \\
Three successive passages & Test PRFM over repeated cycles and compare grid spacings at fixed physical source lengths (Appendix~\ref{sec:8}). \\
Expansion gradient at fixed gas state & Cold gas passing three-axis convergence with the gradient retained, omitted or reversed (Table~\ref{tab:expansion_gradient}). \\
Matched far-ultraviolet intervention during expansion & Feedback-history response on the same flow (Section~\ref{sec:fuv}). \\ \hline
\end{tabular}
\end{table*}

\subsection{Local physics and comparison with standard TIGRESS}\label{sec:physics}

The three principal cycling runs and their shared uncompressed control use the main local physics of standard TIGRESS \citep{KimOstriker2017,KimKimOstriker2020}. Our runs and TIGRESS both evolve a vertically stratified, differentially rotating patch with ideal MHD, gas and stellar self-gravity, external gravity from old stars and dark matter, and open vertical boundaries. Both include radiative cooling at solar metallicity, far-ultraviolet (FUV) heating from young clusters, and supernovae from clusters and runaway stars. We use the tabulated TIGRESS stellar luminosities and supernova rates as functions of age (Appendix~\ref{app:principal_sources}). In both calculations a cluster can form where the gas exceeds the Larson--Penston density threshold, lies at a local gravitational potential minimum and converges along all three axes. Clusters accrete gas and supply feedback independently of PRFM.

Two local prescriptions differ from those in the standard TIGRESS spiral-arm calculation of \citet[][Sections II.2--II.3]{KimKimOstriker2020}. \citet{KimKimOstriker2020} merge growing cluster particles when their $3\times3\times3$-cell control volumes overlap. With their 12.3 pc cells, each volume is about 37 pc on a side. Our principal runs start with 8 pc cells. We merge two growing cluster particles when their physical center-to-center separation is less than 24 pc. Both must be younger than 40 Myr and neither may have produced its first supernova. The merged particle carries their combined stellar mass. Gas accretion transfers mass and momentum through a region of 27 mapped cells (Table~\ref{tab:principal_sources}). For each supernova, both calculations compare the mass in the injection region with the expected shell-formation mass, $M_{\rm sf}$, to select the feedback stage. At an ambient hydrogen density of $n_{\rm H}=1\,{\rm cm}^{-3}$, we use $M_{\rm sf}=1540\,M_\odot$ and they use $1679\,M_\odot$. Our shell-formation mass scales as $n_{\rm H}^{-0.33}$ and theirs as $n_{\rm H}^{-0.26}$. Both treatments inject energy, terminal momentum or ejecta according to the stage. Section~\ref{sec:pair_setup} specifies the source prescriptions shared by the three principal cycling runs and their shared control.

Even without imposed compression, comparing individual star-formation episodes with TIGRESS is difficult. The TIGRESS solar-neighborhood simulation, model R8, follows a patch centered 8 kpc from the galactic center. R8 and our uncompressed control both start with a gas column of $12\,M_\odot\,{\rm pc}^{-2}$, a vertical turbulent dispersion of $10\,{\rm km\,s}^{-1}$ and an azimuthal field with $\beta=10$ \citep{KimSMAUG2020,OstrikerKim2022}. Our patch is 256 pc across with 8 pc cells, whereas R8 is 1024 pc across with 4 pc cells. Each calculation follows its own cluster births and supernovae. Section~\ref{sec:10} compares their midplane stress.

Transport uses minmod piecewise-linear reconstruction and the Harten--Lax--van Leer discontinuities (HLLD) approximate Riemann solver \citep{MiyoshiKusano2005}, which computes fluxes across cell faces while resolving contact and Alfv\'{e}n waves. Time integration is unsplit second-order Runge--Kutta, with face-staggered upwind constrained transport following \citet{MignoneDelZanna2021}. Invalid intermediate Riemann states fall back to the more diffusive, two-wave Harten--Lax--van Leer--Einfeldt (HLLE) solver \citep{Einfeldt1988}; failed transport steps are retried with first-order HLLE, with recorded timestep subdivision where required. There is no Alfv\'{e}n-speed cap. The stellar application in Section~\ref{sec:9} also enables a selective magnetic-energy consistency correction, which is recorded and described in Section~\ref{sec:9.3}.

\section{Numerical validation and computational cost}\label{sec:validation}

Three tests support the application: agreement of the physical and residual representations, convergence through deformation, and cost at matched physical resolution and timesteps. They concern the numerical representation, not convergence of a self-generated stellar population, which Section~\ref{sec:validation_stars} addresses. Table~\ref{tab:roadmap} lists each validation and model test and whether it imposes compression (changing $a(t)$, physical cell widths and image and source terms) or keeps $a=1$.

\label{sec:1.1}

\subsection{Transformation, boundaries and convergence}\label{sec:validation_geometry}

Independent physical and residual formulations are run with the same prescribed flow, image conditions, reconstruction and timestep sequence. Across 14 one-dimensional runs, including an undistorted control, eightfold compression and return, and a supplied spiral history, the maximum scaled state difference is $1.02\times10^{-12}$. Figure~\ref{fig:2} shows the matched representations and deliberate reconstruction/time-quadrature changes, which separate the solutions at identifiable stages. In the MHD uniform compression--return test, the deformation scalings are recovered to $1.87\times10^{-15}$. Source-frame tests cover momentum-conserving, Sedov--Taylor and ejecta-stage feedback at nonzero compression or expansion rate, including a source across the periodic seam. Appendix~\ref{sec:3} collects the representation requirements, component checks, refinement and cost evidence, including a deliberately incorrect boundary operator that tests sensitivity to the image velocity jump.

\begin{figure*}[!tp]
\centering
\includegraphics[width=\textwidth,height=0.36\textheight,keepaspectratio]{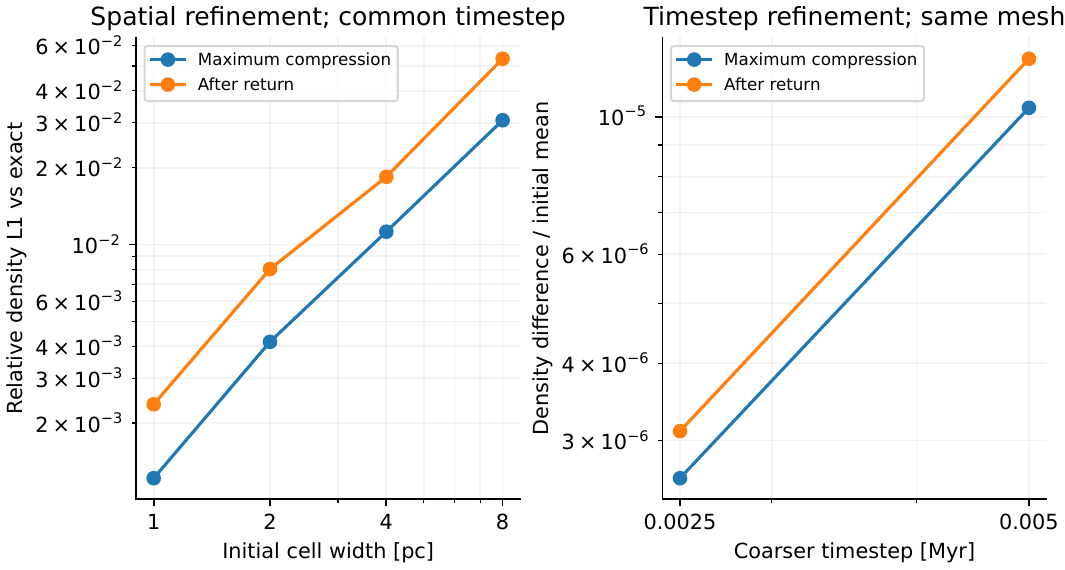}
\caption{\textbf{Numerical convergence through prescribed compression and return.} A smooth MHD pattern undergoes fourfold compression in a 64 pc periodic cube; gravity, cooling and stars are off. Left: four meshes with the same 1600 timesteps, compared with analytic cell averages. Right: a fixed $16^3$ mesh with 400, 800 and 1600 steps. Spatial errors decrease through compression and return, and the temporal order is approximately two. Definitions and full values are in Appendix~\ref{sec:7.1}.\label{fig:16}}
\vspace{10pt}
\centering
\includegraphics[width=\textwidth,height=0.24\textheight,keepaspectratio]{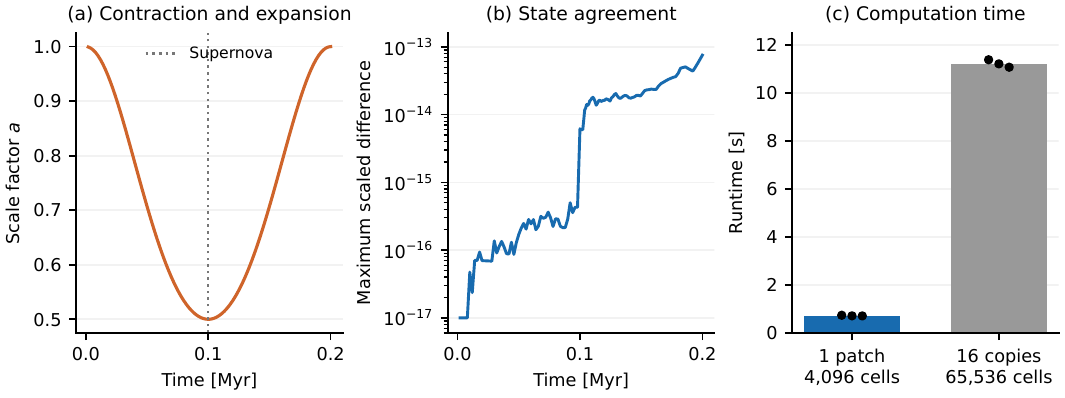}
\caption{\textbf{Equivalence and computational cost through contraction and expansion.} Independently evolved 4096-cell and 65536-cell domains contain one and 16 copies of the same local problem, with the same scheme, 121 timesteps, physical mesh widths, cooling, gravity and fixed supernova history. Panels show (a) the box width through the cycle, with the prescribed supernova marked, (b) the maximum scaled state difference, and (c) the computation time. Bars show the median time for advancing the solution and evaluating the timestep; points show all three serial repetitions. Initialization and output are excluded equally. Each copy has one prescribed supernova; new star formation is disabled. The maximum scaled state difference is $7.53\times10^{-14}$ and the median cost ratio is 15.6 (range 15.5--15.8).\label{fig:20}}
\end{figure*}

A smooth advected MHD pattern has an analytic solution through fourfold compression and return (Figure~\ref{fig:16}). With the same 1600 timesteps, refining from 8 to 64 cells per axis reduces the normalized density L1 error (summed absolute cell-average error divided by summed exact density) from 3.07\% to 0.122\% at maximum compression and from 5.34\% to 0.237\% after return. The finest spatial orders are 1.77 and 1.76; timestep refinement on a fixed mesh gives orders 1.99 and 2.00. A separate compressed source problem retains exactly one prescribed supernova while the timestep is refined; adjacent density and field L1 differences decrease to 0.0109\% and 0.0089\%. Because this is a shocked source test, we do not claim second-order convergence for it. Appendix~\ref{sec:7} gives the analytic definitions, norms and refinement tables.

\subsection{Equivalence and computational cost through contraction and expansion}\label{sec:validation_cost}

This test asks whether one deforming patch reproduces the local evolution inside a larger domain throughout compression and expansion. We compare one $16^3$ patch with a $64\times64\times16$ domain initialized with 16 horizontal copies of the same periodic problem. Each domain advances its own gas and magnetic fields; the larger calculation is not filled from the evolving small-box solution. Both are evolved independently for 0.2 Myr through twofold compression and return, with identical physical cell widths, numerical scheme and 121 timesteps; initial 8 pc cells become $4\times8\times8$ pc at peak compression. Cooling, self-gravity, particle gravity and drift, and mapped feedback are active. Each copy has one prescribed supernova, so the source history is the same in both domains; new star formation is disabled in this benchmark.

The maximum scaled gas and field difference over the entire cycle is $7.53\times10^{-14}$ (Figure~\ref{fig:20}b), so the small patch reproduces the evolving local state of the larger periodic domain through compression, feedback and return.

With that equivalence established, three serial timing pairs give a median runtime ratio of 15.6, with range 15.5--15.8 (Figure~\ref{fig:20}). Both timings include the advance and the evaluation of the timestep stability limit and exclude initialization and output, so the benchmark measures the saving from eliminating repeated volume at equal compressed resolution and physical duration. It is separate from the stellar application, whose adaptive timesteps and feedback histories differ. Appendix~\ref{sec:7.5} gives the setup, finite-source qualifications and individual timings.

\subsection{Stratified contraction and expansion at three resolutions}\label{sec:validation_stars}

\begin{figure*}[!tp]
\centering
\includegraphics[width=\textwidth,height=0.63\textheight,keepaspectratio]{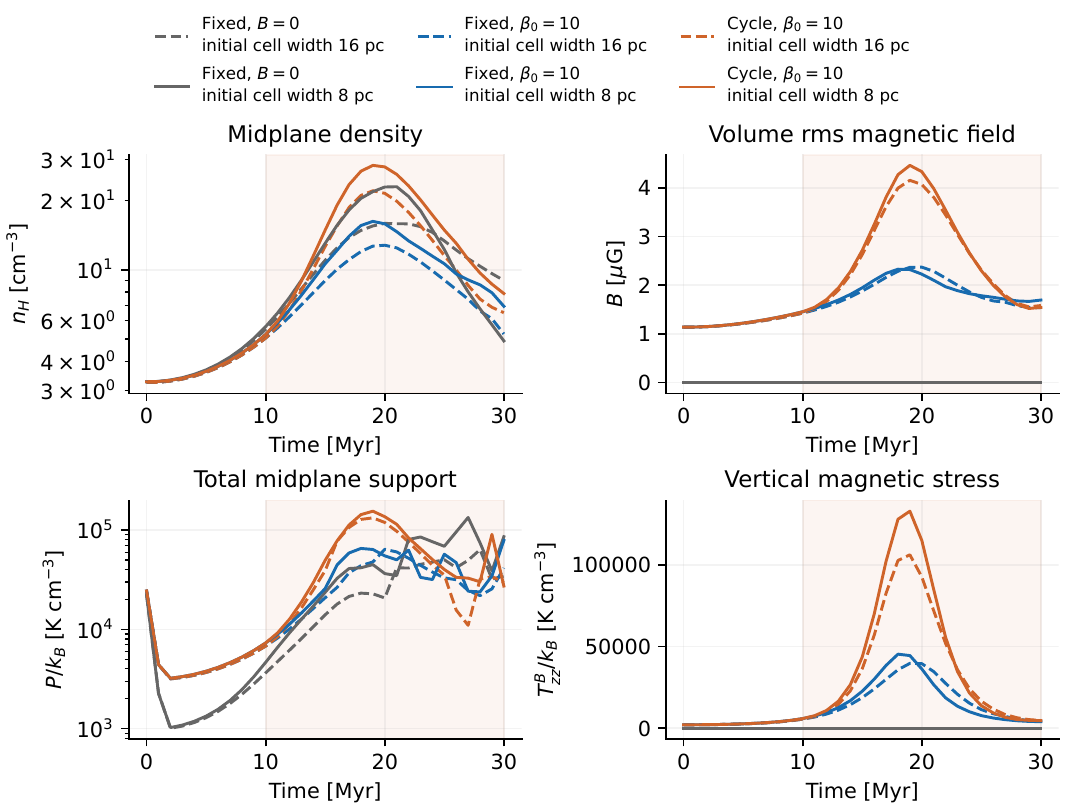}
\caption{\textbf{Gas and magnetic stresses during contraction and expansion.} Grid-spacing test at an initial gas column of $20\,M_\odot\,{\rm pc}^{-2}$, with 16 pc (dashed) and 8 pc (solid) initial cells. The shaded interval is the twofold $x$ compression and return. Blue and orange denote the fixed and cycling boxes at initial plasma $\beta_0=10$. Gray denotes the fixed box with zero field. The cycling run transforms sources to a common physical frame. Magnetic stress is $(B_x^2+B_y^2-B_z^2)/2$. Field strength is a whole-volume rms; density and stresses are native midplane averages spanning two vertical cells (32 or 16 pc), so this figure does not compare equal physical slabs across resolutions; Appendix~\ref{sec:7.3} and Figure~\ref{fig:18} do. Only the $x$ cell width contracts. These runs span 0--30 Myr. The 20 Myr passage in Section~\ref{sec:pair_setup} spans 35--55 Myr.\label{fig:11}}
\end{figure*}

These 30 Myr pilot runs test how changing the initial cell width affects the magnetic stresses through compression and return, and the total stellar mass formed. Figure~\ref{fig:11} follows density, volume root-mean-square (rms) field strength and vertical support in cycling and fixed boxes with initial cell widths of 16 and 8 pc.

\begin{figure*}[!tp]
\centering
\includegraphics[width=\textwidth,height=0.63\textheight,keepaspectratio]{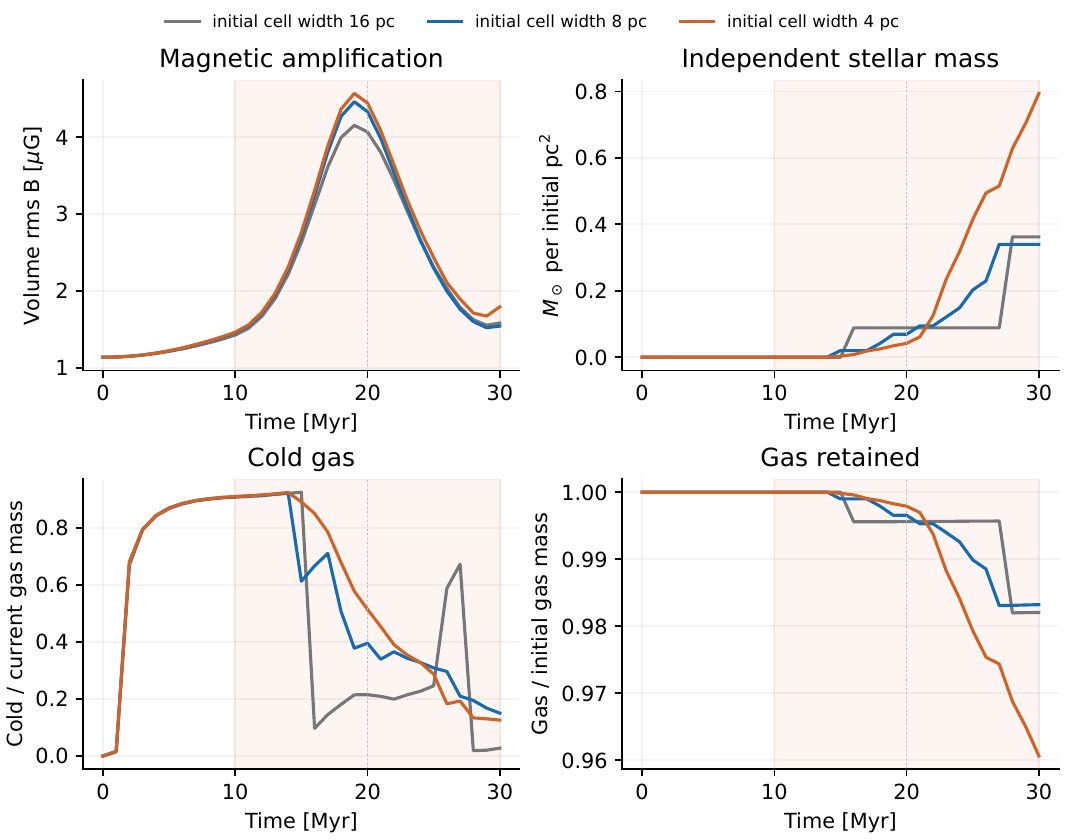}
\caption{\textbf{Gas and stellar histories through contraction and expansion at three resolutions.} The 30 Myr refinement test uses initial cell widths of 16, 8 and 4 pc, with $x$ widths halved at 20 Myr. Initial domain, column, field parameter and perturbation seed are shared; adaptive timesteps and stellar events evolve independently. At maximum compression (20 Myr) the field amplification changes less with each refinement. After return (30 Myr) the final stellar mass and cold-gas fraction remain sensitive to resolution. Appendix~\ref{sec:7.3} gives common-volume measurements and source-scale qualifications.\label{fig:17}}
\end{figure*}

We repeated the 30 Myr stratified cycle with initial cell widths of 16, 8 and 4 pc in the same physical domain (Figure~\ref{fig:17}). At peak compression, successive changes in the volume-rms field strength decrease from 6.49\% to 2.60\%. Final stellar mass, in contrast, changes by $-6.2\%$ and $+134\%$: the full stellar response is not converged. The unresolved source scales shrink with the mesh; Appendices~\ref{sec:7.4} and~\ref{sec:7.3} give the timestep and source-scale controls and equal-volume refinement diagnostics. The repeated-passage test likewise changes the sign of the cycle-minus-control stellar response under refinement (Appendix~\ref{sec:8.3}). These sensitivities limit the astrophysical conclusions; they do not affect the separately verified coordinate transformation or the matched cost result.

\section{Testing star formation through compression and expansion}\label{sec:9}

For the 20, 50 and 100 Myr passages and the shared uncompressed control, we compare the stellar mass formed in each MHD calculation with the KS and PRFM predictions for that calculation. The MHD calculation evolves gas, fields and star particles whose ages determine their feedback. At each saved time KS uses the gas column, and conditional PRFM infers an equilibrium rate from the gas and stellar columns and stellar vertical motions of the same simulation. A separate restart during the 20 Myr passage tests continued FUV feedback from the early cluster. Reported measurements generally have three significant figures, and event times are given to two decimal places in Myr. Totals, differences and ratios are calculated from unrounded records.

\subsection{Shared state, physics and prescribed passage}\label{sec:pair_setup}

The reference 20 Myr comparison spans absolute times 35--55 Myr. One uncompressed control is followed from 35 to 135 Myr; all three passage comparisons reuse this history over matching intervals. All three cycling runs and the control use the same $32\times32\times896$ mesh, spanning $256\times256\times7168$ pc, with cells initially 8 pc on a side. At maximum compression the radial cell width is 4 pc and the other two widths remain 8 pc. The control keeps 8 pc cells throughout. In all four runs eligible clusters merge when their physical center-to-center separation is less than 24 pc (Table~\ref{tab:principal_sources}).

The preparation uses an initial gas column of $12\,M_\odot\,{\rm pc}^{-2}$, initial plasma $\beta_0=2P/B^2=10$ (thermal-to-magnetic pressure ratio), and an external potential modeled on the TIGRESS solar-neighborhood simulation (R8), with old-stellar column $42\,M_\odot\,{\rm pc}^{-2}$, scale height 245 pc and halo density $\rho_{\rm dm}=0.0064\,M_\odot\,{\rm pc}^{-3}$. The orbital period is 219 Myr and the shear parameter (Section~\ref{sec:2.2}) is $q=1$. The shared state retains gas, magnetic fields, clusters, stellar ages, scheduled feedback and random-generator lineage. Statistical stationarity was not tested at this checkpoint.

The three cycling runs and the uncompressed control use the same executable, source tables, numerical scheme and adaptive timestep rule. The source treatment uses pinned cooling and STARBURST99 FUV/supernova tables, age-dependent attenuated FUV heating, cluster formation and accretion, and runaway progenitors. By contrast, in the 30 Myr field-strength tests, heating scales with the surface density of young stars and supernova delays are sampled uniformly from 3 to 40 Myr (Appendices~\ref{sec:4.1} and~\ref{sec:5.1}). The original seed population remains active in feedback but is excluded from new stellar mass. No measured or predicted PRFM pressure, rate or efficiency drives the MHD calculation. Table~\ref{tab:principal_sources} specifies the spatial source choices; Appendix~\ref{app:principal_sources} identifies the cooling and stellar tables, their normalization, FUV attenuation and runaway sampling. Source recipes and the common checkpoint are retained in the provenance records (Appendix~\ref{app:provenance}).

The control has $a=1$. For a pulse of duration $T$ starting at $t_0=35$ Myr, each cycling run uses
\begin{equation}
\begin{gathered}
a(t)=\exp\!\left\{-\frac{\ln 2}{2}\left[1-\cos\!\left(2\pi\frac{t-t_0}{T}\right)\right]\right\},\\
t_0\leq t\leq t_0+T.
\end{gathered}
\end{equation}
In M51, the total gas surface density in the spiral arms is about two to three times that in the interarm regions \citep{Hitschfeld2009}. We adopt a twofold radial contraction as a representative amplitude: at fixed gas mass, halving the patch width doubles its mean gas surface density. For the reference pulse, $T=20$ Myr. Maximum (twofold) compression is at 45 Myr, and the box returns to its initial radial width at 55 Myr while the lattice continues to shear. The gas and stellar histories are not reset. Accepted timesteps and feedback events may differ between the cycling runs and the control as the local response develops. The full-passage stellar masses are compared at that return.

In M51, \citet{Koda2009} estimate that gas crosses the width of a spiral arm in about 20--50 Myr. At a galactocentric radius of 3 kpc, \citet{Kuno1995} estimate a crossing time of 48 Myr from a molecular arm width of about 1 kpc and a gas speed across the arm of $20\,{\rm km\,s^{-1}}$ in the frame of the spiral pattern. In the strong-arm magnetized simulation of \citet{KimKimOstriker2020}, gas moving at the background speed of $15\,{\rm km\,s^{-1}}$ would cross their arm region, about 0.9 kpc wide, in about 60 Myr, and their mean velocity across the arm decreases as the gas enters the dense ridge. The travel time from one arm crest to the next also includes the interarm region; for a steady $m$-armed pattern it is $2\pi/(m|\Omega-\Omega_{\rm p}|)$, where $\Omega$ and $\Omega_{\rm p}$ are the gas orbital and pattern angular speeds. The parameters of \citet{KimKimOstriker2020} give about 205 Myr, and \citet{Koda2021} obtains arm-to-arm times of 153 and 243 Myr at radii of 2.43 and 2.70 kpc by applying a rotating pattern to M51. The 20 and 50 Myr pulses span the estimated arm-crossing range; the 20 Myr duration was carried forward from the faster of the 20 and 40 Myr forcing experiments in Appendix~\ref{sec:8.1}, and the 100 Myr pulse tests a slower contraction and return at the same amplitude. Each cycling calculation ends when the patch regains its original width.

\begin{table*}[!tp]
\caption{Source prescriptions shared by the three cycling runs and their shared uncompressed control. The initial cell width is $\Delta=8$ pc. The radial cell width follows $a\Delta$; the 24 pc center-to-center merger threshold stays fixed. The physical box lengths are $L_x$ (radial) and $L_y=256$ pc (azimuthal).\label{tab:principal_sources}}
\centering\footnotesize\setlength{\tabcolsep}{4pt}\renewcommand{\arraystretch}{1.15}
\begin{tabular}{>{\raggedright\arraybackslash}p{\dimexpr 0.28\textwidth-2\tabcolsep\relax}>{\raggedright\arraybackslash}p{\dimexpr 0.72\textwidth-2\tabcolsep\relax}}
\hline
\textbf{Operation} & \textbf{Implemented choice} \\ \hline
Collapse threshold & $\rho_{\rm LP}=8.86(P/\rho)/(\pi G\ell_{\rm LP}^2)$, with fixed $\ell_{\rm LP}=\Delta=8$ pc. Radial compression does not change this length. \\
Formation tests & Potential minimum in 27 mapped cells and physical convergence along all three axes. No additional energy-binding gate is applied to principal-run sink creation; the isolated cloud-binding test belongs to the reduced cloud model of Part~II. \\
Accretion region & A $3\times3\times3$ mapped-cell region, with conserved flux, swept-volume and reset accounting. Its radial extent contracts and its shape shears with the grid. \\
New-sink exclusion and cluster merging & A cluster younger than 40 Myr excludes new sinks at center-to-center separations less than $3\Delta=24$ pc. Two such clusters merge when their physical center-to-center separation is less than 24 pc and neither has had its first supernova. Both tests use the nearest sheared image and exclude runaways. \\
Particle gravity & Triangular-shaped-cloud deposition and force interpolation over 27 mapped cells. Gravitational smoothing follows the mesh; there is no separately imposed point-particle softening radius. \\
Supernova injection & Physical spheres sampled at mapped cell centers; minimum injection radius $3\Delta=24$ pc, radius increments $\Delta/2=4$ pc, maximum $\min(128\,{\rm pc},L_x/2,L_y/2)$. The cap falls from 128 to 64 pc at peak compression. \\
\hline
\end{tabular}
\end{table*}

\subsection{Model inputs and comparison measures}\label{sec:pair_measurements}

The vertically integrated PRFM calculation uses the simulated gas and stellar columns and stellar vertical velocity moments to infer equilibrium total pressure $P_{\rm eq}$, thickness and effective dispersion. This equilibrium pressure represents the combined thermal, turbulent and magnetic support of the column. Its pressure-dependent feedback yield is $\Upsilon_{\rm tot}=\Upsilon_0(P_{\rm eq}/P_0)^{-0.21}$, with $\Upsilon_0=1030\,{\rm km\,s^{-1}}$ and $P_0/k_B=10^4\,{\rm K\,cm^{-3}}$, where $k_B$ is Boltzmann's constant; the predicted star-formation rate per unit area is $P_{\rm eq}/\Upsilon_{\rm tot}$. These are the \citet{OstrikerKim2022} TIGRESS fits adopted by \citet{Jeffreson2026}, who present fits to TIGRESS-NCR (non-equilibrium cooling and radiation) separately. The TIGRESS calibration runs and all our principal runs include age-dependent FUV heating, supernovae and runaway progenitors (Section~\ref{sec:physics}).

PRFM is evaluated conditionally: at each saved time it receives the run's gas and stellar columns and stellar vertical motions from the MHD calculation, excluding the measured pressure and star-formation rate, and returns an equilibrium rate. Those inputs already reflect past cooling and feedback, but PRFM carries no separate cloud-assembly state or stellar-age feedback between samples; it does not forecast the supplied history, and its output never drives the MHD calculation.

For the primary PRFM prediction we group the saved gas into horizontal layers 16 pc thick (two vertical grid cells) and select layers with mean $n_{\rm H}>0.13\,{\rm cm^{-3}}$. Their gas mass $M_{\rm elig}$ is divided by the predicted column depletion time $t_{\rm dep}$: $\dot M_\star=M_{\rm elig}/t_{\rm dep}=f_{\rm elig}A P_{\rm eq}/\Upsilon_{\rm tot}$, where $A$ is the physical horizontal patch area and $f_{\rm elig}$ is the eligible fraction of the full gas mass. The atmosphere therefore does not dilute the density used for eligibility. The fixed old-stellar dispersion is $19.1\,{\rm km\,s^{-1}}$, obtained from a Jeans integral over the 35 Myr starting state in the simulation potential; Appendix~\ref{app:prfm_spec} specifies this integral, its combination with live-star velocity moments, and the pressure--thickness--dispersion solve. These local filters differ from the 500 pc kernels of \citet{Jeffreson2026}, so this is not an exact reproduction of their Arepo implementation. Appendix~\ref{sec:6} gives formula checks and a distinct measured-pressure diagnostic.

Two sensitivity tests check these choices over the 20 Myr passage (35--55 Myr). The primary PRFM prediction underestimates the simulated stellar mass by 35\% in the cycling run and 69\% in the control. The full column ($f_{\rm elig}=1$; $7{,}780$ and $6{,}020\,M_\odot$) gives 31\% and 66\%, and doubling the halo density supplied to PRFM, which matches the simulation's near-midplane halo restoring coefficient (Appendix~\ref{app:prfm_spec}), gives 31\% and 67\%.

New stellar mass sums births and accretion after 35 Myr. PRFM stellar mass integrates $f_{\rm elig}(t)A(t)P_{\rm eq}(t)/\Upsilon_{\rm tot}(t)$ over the same interval, where $A(t)$ is physical horizontal area, and timing uses exact event times and common one-Myr bins. The \citet{OstrikerKim2022} calibration of the pressure--rate relation and feedback yield uses a trailing 40 Myr star-formation rate average, whereas we integrate the predictions and sum new stellar mass over each complete passage.

For the KS comparison we use the disk-averaged law of \citet{Kennicutt1998} as an instantaneous whole-patch prescription. The hydrogen column sets the surface rate through
\begin{equation}
\frac{\Sigma_{\rm SFR,KS}}{M_\odot\,{\rm yr}^{-1}\,{\rm kpc}^{-2}}=2.5\times10^{-4}
\left(\frac{\Sigma_{\rm H}}{M_\odot\,{\rm pc}^{-2}}\right)^{1.4}.
\end{equation}
The observational fit uses atomic and molecular hydrogen. We use the whole-patch hydrogen column $\Sigma_{\rm H}=M_{\rm gas}/(1.4271\,A)$ as its proxy, with the simulated gas mass $M_{\rm gas}$ summed over all phases and $A$ the physical patch area. The factor 1.4271 is the gas mass per hydrogen nucleus in units of $m_{\rm H}$ (Appendix~\ref{app:principal_sources}). The tabulated KS masses use Kennicutt's published coefficient, which assumes a Salpeter initial mass function ($0.1$--$100\,M_\odot$).\footnote{The simulation uses a Kroupa initial mass function. Converting the KS predictions to that convention multiplies them by about 0.67 \citep{MadauDickinson2014}, which makes every KS shortfall in Table~\ref{tab:prediction_errors} larger: 56.8\%, 78.0\% and 67.4\% in the 20, 50 and 100 Myr cycling runs and 62.5\% in the control. The conversion therefore does not reduce the discrepancy.} The KS surface rate times $A$ gives the patch rate. Both KS and conditional PRFM rates are integrated by the trapezoidal rule on the half-Myr output grid, with all coefficients fixed across runs and passage durations.

\subsection{Stellar mass through passages of three durations}\label{sec:period_response}

Each passage starts from the same 35 Myr state, halves the radial width and returns it to its original value. The peak strain rates $|H|_{\rm max}=\pi\ln 2/T$ are 0.109, 0.0436 and $0.0218\,{\rm Myr}^{-1}$ for durations $T=20$, 50 and 100 Myr. Table~\ref{tab:period_response} in Appendix~\ref{app:principal_spec} gives the stellar masses and birth counts for each passage and the corresponding interval of the shared control.

Table~\ref{tab:prediction_errors} gives the error in each prediction as a percentage of the stellar mass simulated in the same run. A negative value means that the prescription underestimates the mass. Each row gives the interval over which the stellar mass is summed.

\begin{columntableblock}
\caption{Error in predicted stellar mass relative to the simulated mass in the same run over the same interval.\label{tab:prediction_errors}}
\centering\footnotesize
\begin{tabular}{@{}llrr@{}}
\hline
Run & Interval (Myr) & KS & PRFM \\
\hline
20 Myr cycle & 35--55 & $-35.5\%$ & $-34.7\%$ \\
50 Myr cycle & 35--85 & $-67.2\%$ & $-67.5\%$ \\
100 Myr cycle & 35--135 & $-51.4\%$ & $-52.7\%$ \\
Control & 35--135 & $-44.0\%$ & $-53.4\%$ \\
\hline
\end{tabular}
\end{columntableblock}

Table~\ref{tab:cycle_change} compares only the simulations. For each passage we subtract the stellar mass formed in the shared uncompressed control from the mass formed in the cycling run and divide by the control mass over the same interval. The largest increase occurs in the 50 Myr passage.

\begin{columntableblock}
\caption{Change in simulated stellar mass relative to the shared uncompressed control over the matching interval.\label{tab:cycle_change}}
\centering\footnotesize
\begin{tabular}{@{}lr@{}}
\hline
Cycling passage & Change \\
\hline
20 Myr & $-37\%$ \\
50 Myr & $+202\%$ \\
100 Myr & $+29\%$ \\
\hline
\end{tabular}
\end{columntableblock}

\subsection{Cluster timing in the 20 Myr passage}\label{sec:9.2}

\begin{figure*}[!tp]
\centering
\includegraphics[width=\textwidth,height=0.63\textheight,keepaspectratio]{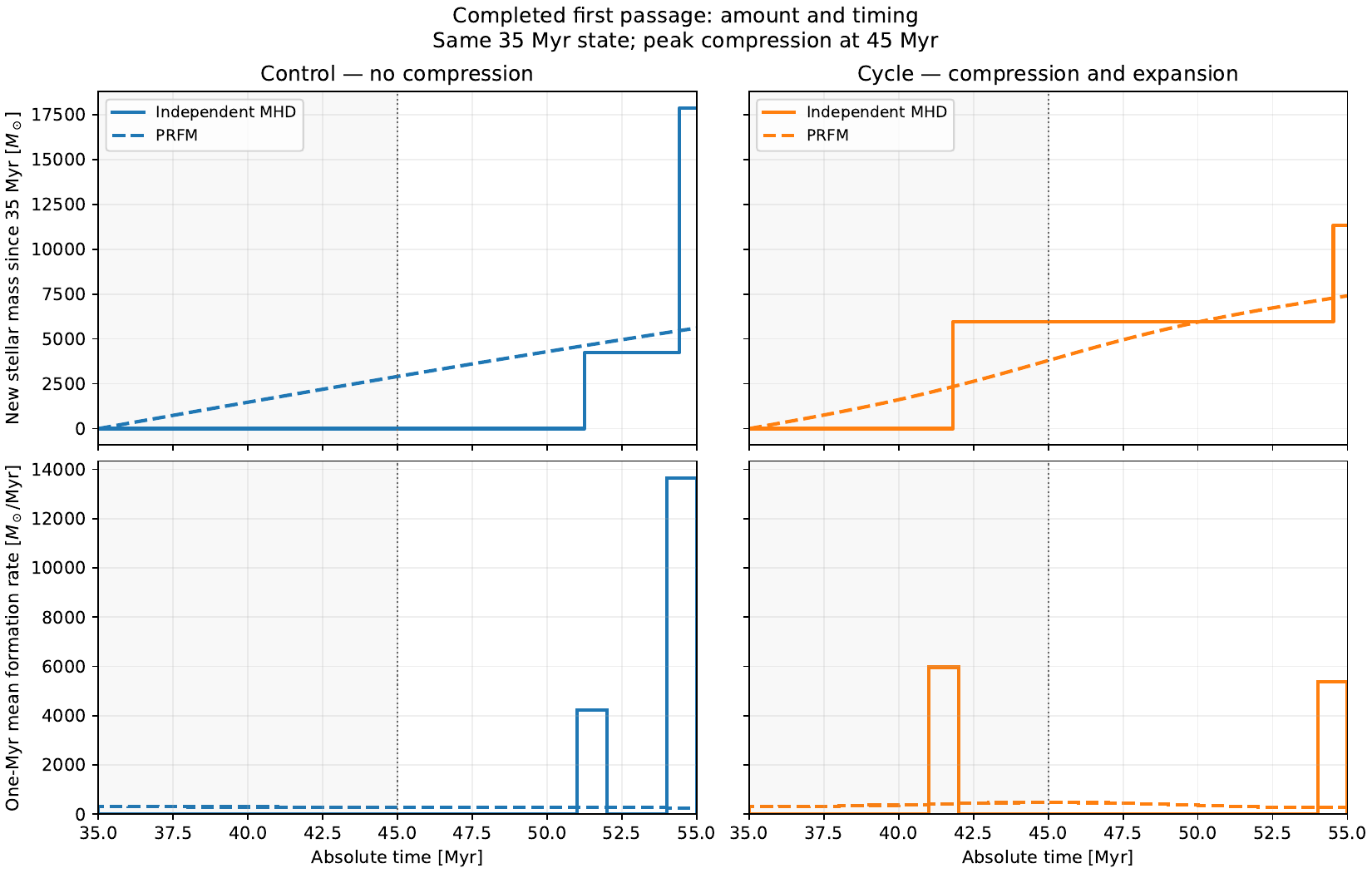}
\caption{\textbf{Stellar mass and cluster births in the 20 Myr comparison.} The 20 Myr cycling run and the shared uncompressed control start from the same 35 Myr state; the figure shows 35--55 Myr, with maximum twofold compression at 45 Myr. The control continues to 135 Myr for the longer passage comparisons. The upper panels show cumulative stellar mass and the lower panels show one-Myr mean rates. The simulated totals include cluster birth masses and subsequent accretion. The cycling run forms stars earlier but less mass overall than the control. Over this interval PRFM underestimates the stellar mass formed by 34.7\% in the cycling run and 68.6\% in the control. The comparison follows one realization. Section~\ref{sec:9.3} gives the numerical checks.\label{fig:26}}
\end{figure*}

Both runs form two clusters over 35--55 Myr (Figure~\ref{fig:26}). The cycling patch forms its first cluster at 41.82 Myr and its second at 54.52 Myr. The control forms its first cluster at 51.25 Myr and overtakes the cycling patch with its second birth at 54.43 Myr. Conditional PRFM underestimates the stellar mass formed in the cycling patch by 34.7\% (Table~\ref{tab:prediction_errors}), with its highest rate in the 44--45 Myr bin. The simulated rate peaks in the 41--42 Myr bin.

\subsection{A matched feedback intervention during expansion}\label{sec:fuv}

\begin{figure*}[!tp]
\centering
\includegraphics[width=\textwidth,height=0.63\textheight,keepaspectratio]{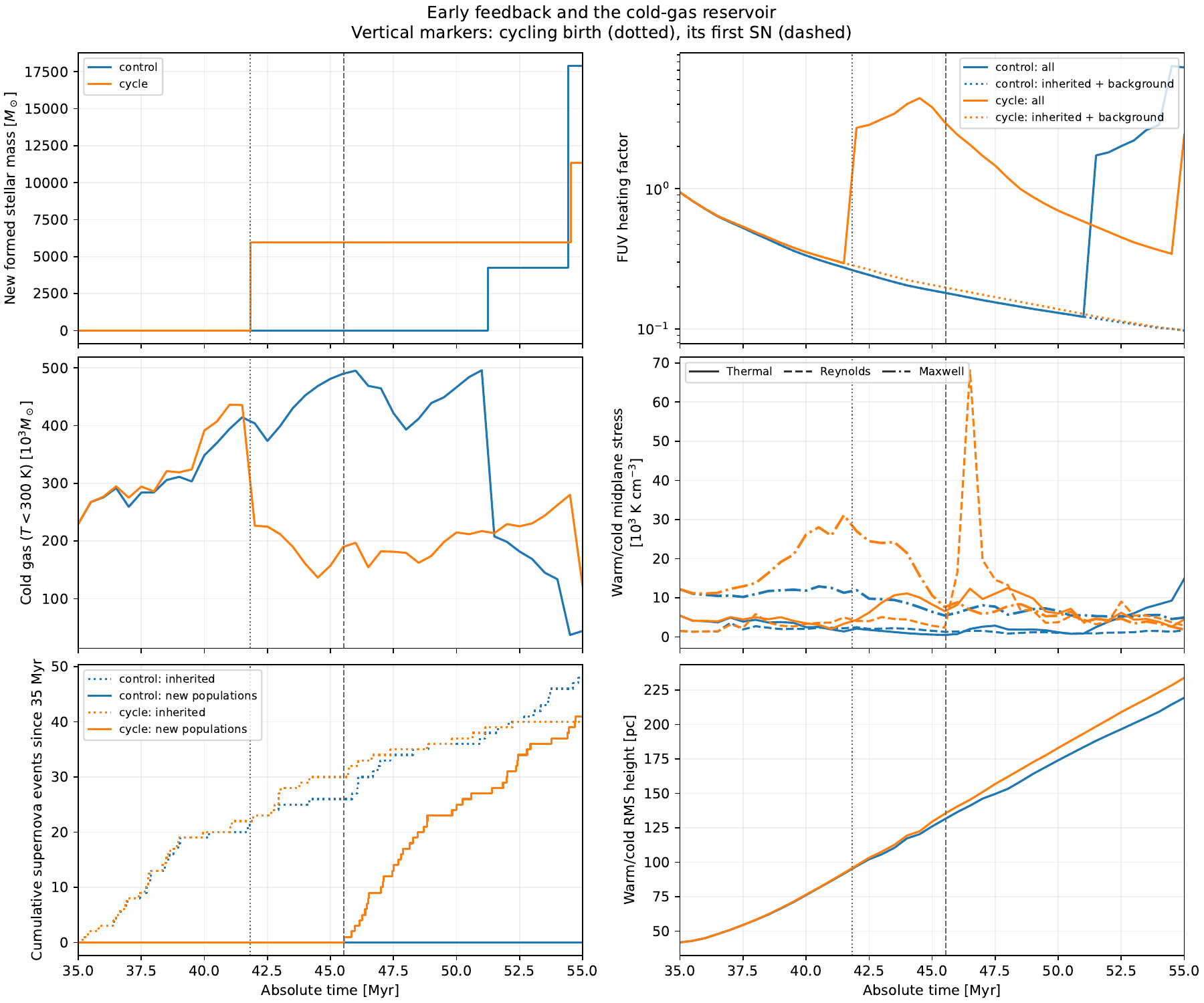}
\caption{\textbf{Feedback history through compression and expansion.} Prescribed compression--expansion cycle (orange) compared with the uncompressed control (blue), 35--55 Myr. Left, top to bottom: newly formed stellar mass, cold gas (below 300 K) and cumulative supernova events from inherited and new populations. Right: FUV heating factor; warm/cold midplane thermal (solid), vertical Reynolds (dashed) and Maxwell (dash-dotted) stresses; and warm/cold rms height. The three stresses are conditional volume means in gas below $2\times10^4$ K within $|z|<8$ pc. The Maxwell term is $(B_x^2+B_y^2-B_z^2)/2$, evaluated before averaging; all three are divided by $k_B$. Vertical dotted and dashed lines mark the cycling birth and its first supernova.\label{fig:39}}
\end{figure*}

Figure~\ref{fig:39} follows feedback and gas state through the passage. After the early cycling birth, the FUV heating factor rises sharply and the cold-gas reservoir (below 300 K) falls, remaining below the control's until the control's first birth. The new cluster's age-dependent luminosity motivates a restart with that source either retained or removed. Whether temperature-selected cold gas assembles into a collapsing region also depends on its density, velocities, magnetic field and gravity.

To test how retained feedback history shapes the response during expansion, we restart the identical 45.5 Myr cycling state twice and, in one branch, remove only the early cluster's subsequent FUV luminosity. Cluster mass, gravity and supernovae remain active. Inherited supernova events, runaway launches and kicks are matched between branches, while gas attenuation, coupling and stellar trajectories can respond. The reference branch reproduces the original final physical and random-generator state.

\begin{figure*}[!tp]
\centering
\includegraphics[width=\textwidth,height=0.63\textheight,keepaspectratio]{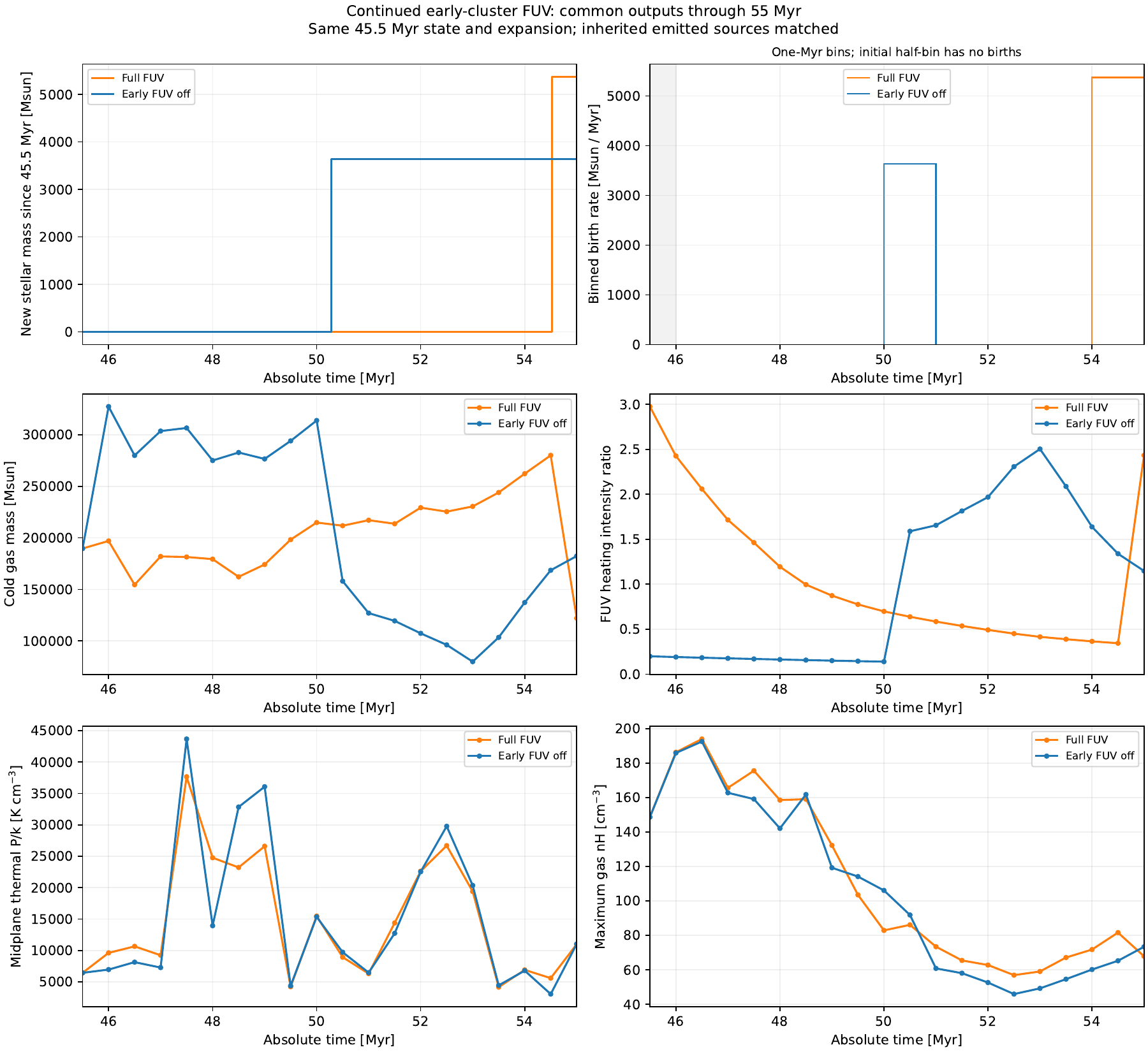}
\caption{\textbf{Matched FUV intervention during expansion.} Both branches restart from the same 45.5 Myr state and expand to 55 Myr. One branch switches off subsequent FUV from the early cluster (cluster 66). Its mass, gravity and supernovae are retained, and inherited supernova events and runaway kicks are matched. Retaining that FUV delays the next birth by 4.24 Myr and increases the stellar mass formed after restart by $1{,}733\,M_\odot$ ($47.6\%$ relative to FUV-off). Both series use a common 45.5--46 Myr initial bin and one-Myr bins thereafter.\label{fig:41}}
\end{figure*}

Between the restart at 45.5 Myr and the return at 55 Myr, each branch forms one cluster. With continued FUV the cluster forms at 54.52 Myr with $5{,}371\,M_\odot$. With the early cluster's subsequent FUV switched off it forms at 50.28 Myr with $3{,}638\,M_\odot$ (Figure~\ref{fig:41}). Retaining that FUV delays the birth by 4.24 Myr and increases the stellar mass formed by $1{,}733\,M_\odot$ ($47.6\%$ relative to FUV-off). At 55 Myr the full-FUV branch contains $1.22\times10^5\,M_\odot$ of cold gas, compared with $1.82\times10^5\,M_\odot$ in the FUV-off branch. Appendix~\ref{app:cycle_diagnostics} gives the feedback and cold-gas histories.

\subsection{Stellar mass and timing in the uncompressed control}\label{sec:9.1}

The uncompressed control forms five clusters between 35 and 135 Myr, at 51.25, 54.43, 97.10, 100.26 and 126.27 Myr. It adds no stellar mass during the 42 Myr from 55 to 97 Myr. The control has therefore formed the same $17{,}888\,M_\odot$ at the ends of the 20 and 50 Myr comparison intervals (55 and 85 Myr, respectively). Over the full 100 Myr, KS underestimates the $54{,}194\,M_\odot$ formed in the control by 44.0\% and PRFM by 53.4\% (Table~\ref{tab:prediction_errors}).

\subsection{Numerical checks of the principal runs}\label{sec:9.3}

The selective constrained-transport energy fallback reconciles the finite-volume and constrained-transport magnetic energies only when their mismatch would otherwise produce negative thermal energy. It changes the gas energy without directly resetting the face magnetic fluxes. It is an explicitly accumulated, nonconservative numerical adjustment, not a physical feedback source. All three principal cycling runs and the control enable this option. Appendix~\ref{sec:8.4} retains the underlying transport-repair tests, and Appendix~\ref{app:cycle_diagnostics} the records of the principal runs.

Over 35--55 Myr, the 20 Myr cycling run and the control meet the analysis targets of $10^{-8}$ for the relative mass ledger, $10^{-10}$ for the normalized magnetic divergence and $10^{-6}$ for the floor-added mass fraction. The cycle's energy-ledger residual reaches $1.01\times10^{-8}$, slightly above the $10^{-8}$ analysis target but below the $10^{-6}$ integration guard; the cause of this miss is unresolved. The maximum saved ratio of cumulative constrained-transport correction to supernova energy is $2.26\times10^{-7}$ in the control and $1.16\times10^{-7}$ in the cycle; these global samples at saved outputs do not bound cellwise or within-output effects. The 50 and 100 Myr cycling runs and the shared control meet the mass, magnetic-divergence and floor-mass targets. The maximum relative energy-ledger residuals for each cycling run and the corresponding interval of the control are $8.19\times10^{-8}$ and $5.68\times10^{-8}$ for the 50 Myr passage and $4.49\times10^{-7}$ and $4.19\times10^{-7}$ for the 100 Myr passage, above the $10^{-8}$ analysis target and below the $10^{-6}$ integration guard.

\section{Magnetic response through compression and expansion}\label{sec:magnetic_response}

The 30 Myr stratified run (Figure~\ref{fig:11}) shows how compression amplifies the field and changes the vertical magnetic stress. Figure~\ref{fig:39} shows that stress alongside the thermal and Reynolds stresses for the 20 Myr cycling run and the control over 35--55 Myr; each output gives a volume average over warm/cold gas (below $2\times10^4$ K) within $|z|<8$ pc. The Maxwell term, $(B_x^2+B_y^2-B_z^2)/2$, combines horizontal magnetic pressure with opposing vertical tension and is computed from the local field before averaging. The means reflect both field evolution and changes in the gas occupying the selected band and temperature range.

At the saved output at 41.5 Myr, just before the first cycling birth, the cycling Maxwell stress divided by $k_B$ reaches $3.11\times10^4\,{\rm K\,cm^{-3}}$, 2.8 times the control's $1.13\times10^4\,{\rm K\,cm^{-3}}$. It contributes 78.8\% of the cycling sum of thermal, Reynolds and Maxwell stresses, compared with 75.7\% in the control and 63.6\% in the shared initial state: magnetic stress already dominates without compression. This sampled cycling maximum precedes maximum compression at 45 Myr. After the first birth, the cycling stress falls below the control's by 50 Myr. At 55 Myr the cycle and control have declined to $1.87\times10^3$ and $4.95\times10^3\,{\rm K\,cm^{-3}}$, respectively, both below their shared initial value of $1.22\times10^4\,{\rm K\,cm^{-3}}$. The cycle ends at 38\% of the control's stress. Relative to the control, the cycling run therefore shows an early enhancement of magnetic stress followed by a larger decline.

The magnetic contribution to the measured midplane stress is substantial in both runs. The vertical magnetic force also depends on spatial gradients and boundary stresses, and assessing the binding of an individual cloud requires its own mass, motions and stresses. Section~\ref{sec:field_strength} varies the initial field in separate 30 Myr tests that include a zero-field pair.

\section{Field strength and the compression--expansion response}\label{sec:field_strength}

All three principal cycling runs and their shared control start with the same field. To measure how the star-formation response depends on field strength, we vary the initial field in paired fixed and cycling boxes.

\subsection{Paired field-strength runs}\label{sec:field_runs}

\begin{figure*}[!tp]
\centering
\includegraphics[width=\textwidth,height=0.63\textheight,keepaspectratio]{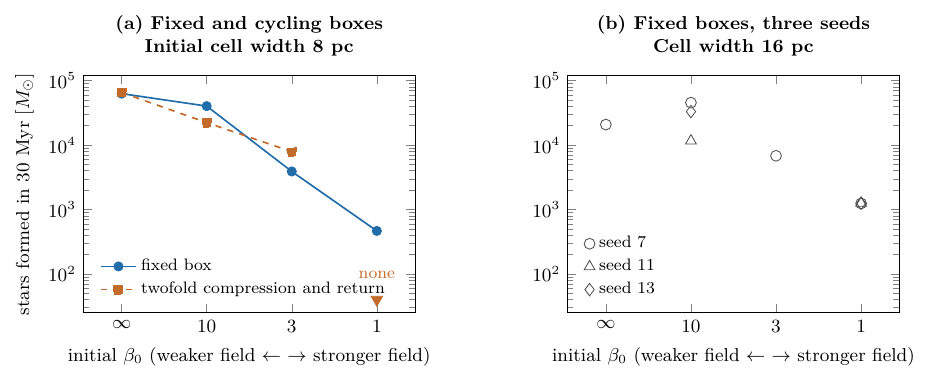}
\caption{\textbf{Stellar mass formed in 30 Myr against initial field strength.} (a) Field-strength tests with initially 8 pc cells ($\Sigma_0=20\,M_\odot\,{\rm pc}^{-2}$, seed 7) in a fixed box (blue) and through twofold compression and return between 10 and 30 Myr (orange), with a sink length that shrinks with the mesh; holding it fixed at 8 pc reverses the $\beta_0=10$ comparison (Section~\ref{sec:field_compression}). The downward triangle marks a run in which no stars form. (b) Fixed-box tests with 16 pc cells and three perturbation seeds. At $\beta_0=1$ each fixed-box run forms a single sink and the cycling run forms none. Each sink receives its birth mass from the excess density in one cell, with no subsequent accretion recorded.\label{fig:field_strength}}
\end{figure*}

We repeated the 30 Myr stratified calculation of Section~\ref{sec:validation_stars} (initial gas column $\Sigma_0=20\,M_\odot\,{\rm pc}^{-2}$, 8 pc cells, perturbation seed 7) with initial $\beta_0=\infty$ (no field), 10, 3 and 1, where $B_y(z)=[2\bar P(z)/\beta_0]^{1/2}$ as in Figure~\ref{fig:field_geometry}. Each field strength was evolved in a fixed box and through the same twofold compression and return between 10 and 30 Myr. Fixed boxes with 16 pc cells test whether the reduction in stellar mass between $\beta_0=10$ and 1 persists across three perturbation seeds (7, 11 and 13). Rerunning two cases with the archived executable reproduces their recorded stellar masses exactly. The new field-strength tests use that same executable. Figure~\ref{fig:field_strength} shows the stellar mass formed in 30 Myr.

\subsection{Strong fields suppress star formation in fixed boxes}\label{sec:field_suppression}

In the fixed box the stellar mass formed in 30 Myr falls from $6.27\times10^4\,M_\odot$ without a field to $4.02\times10^4$, $3.89\times10^3$ and $4.65\times10^2\,M_\odot$ at $\beta_0=10$, 3 and 1. At $\beta_0=1$ a single sink forms near the end of the run, with its entire mass supplied by the birth cell and no subsequent accretion. The same holds in fixed boxes with 16 pc cells for all three seeds, each forming one such sink between about 22 and 27 Myr, whereas $\beta_0=10$ forms $(1.16$--$4.55)\times10^4\,M_\odot$ across those seeds. The weaker field has no consistent sign relative to the zero-field case under refinement (Section~\ref{sec:validation_stars} and Appendix~\ref{sec:5}), but strong-field suppression appears at both resolutions and for every seed.

\subsection{Field-strength dependence over a compression--expansion cycle}\label{sec:field_compression}

The complete compression--expansion cycle changes this dependence. At $\beta_0=10$, the fixed-box control formed $40{,}200\,M_\odot$ of stars over 30 Myr. With the sink length held at 8 pc, the cycling box formed $43{,}200\,M_\odot$, more than the control. When the sink length shrank to 4 pc at peak contraction, the cycling box formed $22{,}300\,M_\odot$, less than the same control (Appendix~\ref{sec:7.4}, Table~\ref{tab:13}). At $\beta_0=3$, with the sink length set by the smallest physical cell width, the cycling and fixed boxes formed $7.90\times10^3$ and $3.89\times10^3\,M_\odot$, respectively. The cycling run formed no sink at $\beta_0=1$, whereas the fixed-box run formed one late sink. Without a field the cycling total is $6.53\times10^4\,M_\odot$, $4.1\%$ above the control. All totals use the same elapsed interval in each pair.

The field-strength tests with initially 8 pc cells use one seed and add the field to a hydrostatic profile without re-equilibration; they show how the response changes with field strength but do not measure a threshold. Section~\ref{sec:field_physics} compares two sink-length prescriptions at $\beta_0=10$.

\section{Discussion}\label{sec:10}

In the standard TIGRESS spiral-arm simulations, gas streams through a prescribed stellar spiral potential and develops an arm and interarm flow \citep{KimKimOstriker2020}. Here we prescribe the radial contraction and return experienced by a patch moving with the imposed large-scale flow. Both approaches follow gas as it cools and collapses, the formation of star clusters, and the FUV heating and supernovae supplied by the stellar populations of those clusters. Prescribing the deformation allows us to vary the duration of a passage from a common initial state while retaining the evolving gas, field and stellar populations.

Our 24 pc merger distance equals three initial cell widths and stays fixed throughout each passage, whereas at peak compression the radial cell width falls to 4 pc, halving the radial extent of the accretion region (Table~\ref{tab:principal_sources}). In the standard TIGRESS spiral-arm calculation, growing clusters merge when their accretion regions overlap on the fixed grid (Section~\ref{sec:physics}). For supernovae, the two shell-formation masses can place a remnant on different sides of the transition between energy and terminal-momentum injection at the same ambient density and enclosed mass. All four of our runs use the same feedback prescription.

The difference in midplane stress between our uncompressed control and the TIGRESS solar-neighborhood model R8 (Section~\ref{sec:physics}) depends on the time interval sampled in our control. Their independent feedback histories, patch sizes and resolutions make a direct comparison difficult; we return to it in Part~II \citep{MosqueiraCloudAssembly}, where we follow the histories of individual star-forming regions.

\subsection{Star formation over a complete passage}\label{sec:10.1}

The cycling runs form less stellar mass than the shared uncompressed control over the complete 20 Myr passage and more over the 50 and 100 Myr passages (Table~\ref{tab:cycle_change}). Both the 20 and 50 Myr passages lie within the arm-crossing range estimated for M51, and at fixed gas mass the twofold radial contraction would double the mean gas surface density of the patch, matching the low end of the arm--interarm contrast of two to three measured there (Section~\ref{sec:pair_setup}). All three principal passages use the same twofold contraction.

KS and PRFM underestimate the stellar mass formed in all four principal runs. The two prescriptions also understate the level of variation: the fractional shortfall changes with passage duration, nearly doubling from the 20 Myr to the 50 Myr cycling run (Table~\ref{tab:prediction_errors}).

Neither prescription includes the signed radial velocity gradient of the compression--expansion cycle. The MHD sink test uses this physical velocity gradient and requires convergence along all three axes at a dense local potential minimum, whereas KS uses gas column alone and PRFM uses gas and stellar columns and stellar vertical motions \citep{Jeffreson2026}. Approaching and separating gas with otherwise identical inputs therefore receive the same instantaneous star-formation rate. With density, gravity, magnetic field and residual velocities held fixed, omitting the outward gradient at 46 Myr increases the cold mass classified as converging by $20\%$, from 12.9\% to 15.5\% of the cold gas (Table~\ref{tab:expansion_gradient}; Appendix~\ref{app:cycle_diagnostics}). The test measures this change in converging cold mass; no snapshot contains a qualifying collapse candidate. If sustained convergence feeds a region already close to collapse, it can change when a cluster forms and when its feedback reaches the surrounding gas, and so alter the full-cycle history.

\begin{columntableblock}
\caption{Cold mass converging along all three axes as a percentage of gas colder than 300 K in the sink-search domain. For duration $T$ and snapshot time $t$, the same cells are tested with the measured outward gradient $+H$, with that gradient omitted, and with its sign reversed. None of these snapshots contains a dense local potential minimum that qualifies for sink formation.\label{tab:expansion_gradient}}
\centering\footnotesize\setlength{\tabcolsep}{3pt}
\begin{tabular}{@{}rrrrrr@{}}
\hline
& & & \multicolumn{3}{c}{Converging cold mass (\%)} \\
\cline{4-6}
$T$ (Myr) & $t$ (Myr) & $H$ (Myr$^{-1}$) & $+H$ & $0$ & $-H$ \\
\hline
20 & 45.5 & $+0.0170$ & 10.1 & 10.8 & 12.5 \\
20 & 46.0 & $+0.0336$ & 12.9 & 15.5 & 18.2 \\
20 & 54.5 & $+0.0170$ & 28.2 & 32.7 & 35.7 \\
50 & 84.5 & $+0.00273$ & 0.0 & 0.0 & 0.0 \\
100 & 134.5 & $+0.000684$ & 25.3 & 25.5 & 25.5 \\
\hline
\end{tabular}
\end{columntableblock}

\subsection{Field strength and the sink prescription}\label{sec:field_physics}

In ideal MHD, magnetic flux through a material tube is conserved. If the tube also retains its mass, its mass-to-flux ratio stays constant.
When a sink accretes gas, the magnetic flux threading that gas stays on the mesh. This lowers the mass-to-flux ratio of the gas left in the accretion region in both the principal runs and the field-strength tests (Section~\ref{sec:2.3}).

In the 30 Myr field-strength test, the cycling box formed more stellar mass than its fixed-box control when the sink length was held fixed and less when it shrank during contraction (Section~\ref{sec:field_compression}). Changing the sink length affects how the simulation treats collapse, binding and accretion together; determining when the simulated star-formation rate is robust to that choice is left to future work.

\section{Conclusions}\label{sec:11}

We have implemented and validated a local deforming patch that makes the imposed compression--expansion history a controlled variable while the local gas and stellar populations evolve. The main results are as follows.

\begin{itemize}\csname @itempenalty\endcsname=10000
\item[] With cooling, gravity and mapped feedback active, one patch agrees with an independently evolved domain containing 16 periodic copies to $7.53\times10^{-14}$ throughout compression and return, with a median 15.6-fold lower runtime.

\item[] Magnetic support and age-dependent FUV feedback can be followed and varied within the same framework while the imposed flow is controlled (Sections~\ref{sec:field_strength} and~\ref{sec:fuv}).

\item[] Relative to the shared uncompressed control over each matching interval, the cycling patch formed 37\% less stellar mass over 20 Myr, 202\% more over 50 Myr and 29\% more over 100 Myr (Table~\ref{tab:cycle_change}). KS and PRFM underestimate the simulated stellar mass by about 35\%, 67\% and 52\% in the 20, 50 and 100 Myr cycling runs and by 44\% (KS) and 53\% (PRFM) in the control over 35--135 Myr (Table~\ref{tab:prediction_errors}). Neither prescription reproduces the amount of stellar mass formed or how it varies among the cycling runs.

\item[] In a separate MHD test the cycling box formed more stellar mass than its fixed-box control when the sink length was held fixed and less when it shrank during contraction (Section~\ref{sec:field_compression}). Before we can confidently apply these simulations to star formation in galactic disks, we need to establish which results persist across a wider range of parameters and treatments of collapse and accretion.

\end{itemize}

\begin{acknowledgments}
The author acknowledges the use of Astra 6, Fable 5.1 and Opus 5.5 for AI assistance in the preparation of this work.
\end{acknowledgments}

\software{Python, NumPy \citep{Harris2020}, Matplotlib \citep{Hunter2007}, \href{https://github.com/jeonggyukim/pyathena}{pyathena} (stellar-population and cooling tables). Appendix~\ref{app:principal_spec} describes the local MHD solver.}

\onecolumngrid

\appendix
\raggedbottom
\makeatletter
\setlength{\@fptop}{0pt}
\setlength{\@fpsep}{18pt}
\setlength{\@fpbot}{0pt plus 1fil}
\makeatother
\renewcommand{\topfraction}{0.92}
\renewcommand{\bottomfraction}{0.85}
\renewcommand{\textfraction}{0.08}
\renewcommand{\floatpagefraction}{0.90}
\setcounter{topnumber}{2}
\setcounter{bottomnumber}{2}
\setcounter{totalnumber}{3}

\section{Verification of the deforming patch: representation, convergence and cost}\label{sec:3}\label{sec:7}

A deforming patch follows a prescribed large-scale flow while the local gas and magnetic fields evolve nonlinearly. Removing that background velocity changes the solver variables, the velocities of periodic images and the frame in which finite-region sources act. None of these changes may alter the physics, because an omitted image jump or source-frame term can mimic a response to compression. A fixed box cannot test terms that act only during deformation.

Deliberately altered reconstruction, time quadrature and boundary terms test whether the comparisons detect the errors they target. These comparisons underpin Section~\ref{sec:validation}, which treats coordinate agreement, numerical convergence and stellar-response sensitivity separately.

\subsection{One-dimensional physical and residual equations}\label{sec:3.1}

The representation test isolates the coordinate transformation with one-dimensional isothermal dynamics, rotation and a supplied trajectory. Two independent states, never reset from each other, evolve on the same linearly deforming domain from the same physical initial conditions, with the same Roe/van Leer/H-correction flux, second-order Runge--Kutta stages, timestep sequence and physical cell widths. A separate solver-matching control (Figure~\ref{fig:1}) shows why comparing different flux solvers or limiter conventions can obscure this test.

In the one-dimensional notation used only here, the coordinate map is $x=X(t)+J(t)\xi$, with moving origin $X$ and dimensionless stretch (volume Jacobian) $J$. The prescribed longitudinal and transverse velocities are $\beta=V+AJ\xi$ and $\chi=W+BJ\xi$, with origin values $V$ and $W$, $A=\dot J/J$ and transverse velocity gradient $B$ ($\beta$ is a velocity, not plasma beta; $A$ is not the patch area and $B$ is not a magnetic field). With residual longitudinal and transverse velocities $u$ and $z$ ($z$ is not the vertical coordinate), the conserved variables are $M=J\rho$, $U=Mu$ and $Z=Mz$, and the represented physical momenta per reference volume are $P_v=U+\beta M$ and $P_w=Z+\chi M$.

Three discrete requirements matter. First, the residual disturbance is reconstructed and limited, and the exact linear background flow is added at each face; limiting total velocity would change the nonlinear reconstruction. Second, background momentum is subtracted with the same face mass flux: for this cell-centered quadrature, with $F_{M,L}$ and $F_{M,R}$ the left and right face fluxes of $M$ for cell $i$, the physical momentum source corrections $(S_v,S_w)_{\rm corr}$ use the product term $C_i$:

\begin{equation}
C_i=\frac{J}{2}(F_{M,L}+F_{M,R})-U_i,\qquad (S_v,S_w)_{\rm corr}=(AC_i,BC_i).
\end{equation}

Third, a time-dependent velocity origin requires the finite background increments to be multiplied by the appropriate stage mass, since endpoint quadrature of \ensuremath{\dot\beta}M is not the same as the discrete second-order Runge--Kutta update. The independent physical source and transport implementation includes these products; they are representation-consistency terms, not added physical forces.

\begin{table}[!htbp]
\caption{Independent physical and residual equations: matched state errors\label{tab:3}}
\centering\footnotesize\setlength{\tabcolsep}{4pt}\renewcommand{\arraystretch}{1.2}
\begin{tabular}{>{\raggedright\arraybackslash}p{\dimexpr 0.62\textwidth-2\tabcolsep\relax}>{\raggedright\arraybackslash}p{\dimexpr 0.38\textwidth-2\tabcolsep\relax}}
\hline
\textbf{Matched test} & \textbf{Largest scaled state difference} \\ \hline
Undistorted wave, 16 / 64 cells & $3.88\times10^{-14}$ / $4.68\times10^{-14}$ \\
Eightfold compression and return, 16 / 64 cells & $1.28\times10^{-13}$ / $1.84\times10^{-13}$ \\
Actual supplied spiral history, 16 / 64 cells & $7.09\times10^{-13}$ / $2.75\times10^{-13}$ \\
Maximum over 14 runs, each 200 Myr & $1.02\times10^{-12}$ \\
\hline
\end{tabular}
\end{table}

The norm is the largest absolute component of $(\Delta M,\Delta U/c_s,\Delta Z/c_s)$ in audit units ($c_s$ the constant isothermal sound speed, $\Delta$ the difference between formulations), not a fractional star-formation error. The largest velocity difference is $7.16\times10^{-12}$ km s$^{-1}$; no checked stage exceeds $10^{-10}$. Changing the limiter first separates the 16-cell spiral-wave solutions at a transport predictor near 0.95 Myr, and changing the time quadrature separates them at 0.05 Myr. The comparison therefore locates implementation differences (matched runs and ablations in Figure~\ref{fig:2}).

\subsection{MHD transport, geometry and source checks}\label{sec:3.3}

The three-dimensional solver uses minmod piecewise-linear reconstruction, HLLD, unsplit second-order Runge--Kutta and face-staggered upwind constrained transport, following \citet{MignoneDelZanna2021}. Invalid intermediate Riemann states fall back to HLLE, failed full transport steps are retried with first-order HLLE, optional timestep subdivision is recorded, and there is no Alfv\'{e}n-speed cap. These earlier tests use no magnetic-energy correction (the later cycling fallback and its explicit budget are in Appendix~\ref{sec:8.4}). Cooling modifies only gas thermal energy. All seven programs of the final MHD verification suite pass.

\begin{table}[!htbp]
\caption{MHD component verification\label{tab:4}}
\centering\footnotesize\setlength{\tabcolsep}{4pt}\renewcommand{\arraystretch}{1.2}
\begin{tabular}{>{\raggedright\arraybackslash}p{\dimexpr 0.58\textwidth-2\tabcolsep\relax}>{\raggedright\arraybackslash}p{\dimexpr 0.42\textwidth-2\tabcolsep\relax}}
\hline
\textbf{Check} & \textbf{Measured result} \\ \hline
Alfv\'{e}n relative rms, 16 \ensuremath{\rightarrow} 32 cells & 0.262 \ensuremath{\rightarrow} 0.0989 (2.65-fold reduction) \\
HLLD induction-flux consistency, 10{,}000 states & $2.77\times10^{-14}$ maximum normalized error \\
Shock density L1, 32 / 64 versus 128 cells & 0.0216 / 0.0110 \\
Uniform fourfold compression-return scaling & $1.87\times10^{-15}$ maximum error \\
Uniform cycle work / initial energy on return & $-1.56\times10^{-15}$ \\
Compressed discrete Poisson relative residual & $2.46\times10^{-13}$ \\
Physical-frame supernova mapping, six branch/flow cases & $5.92\times10^{-16}$ state; $1.10\times10^{-13}$ energy \\
Magnetic flux imported through a closed vertical boundary & Exactly zero \\
\hline
\end{tabular}
\end{table}

The shock test uses \ensuremath{\gamma} = 5/3 and a 128-cell numerical reference, not the canonical \ensuremath{\gamma} = 2 Brio--Wu problem or an exact solution. Source tests cover momentum-conserving (MC), Sedov--Taylor (ST) and ejecta-stage (EJ) feedback during compression and expansion, including a periodic seam, with formation mass and velocity mapping errors of $2.22\times10^{-16}$ and $4.16\times10^{-17}$. A full 30 Myr uncompressed regression reproduces the earlier baseline's final gas and face-field file byte for byte, so the new background-flow source mapping is inactive when $H=0$.

The checks catch specific implementation and scheme differences. In a closed-boundary test, a diode applied only to gas states, with the edge electromotive-force velocity unclamped, imports 128 code units of horizontal magnetic flux; the corrected operator imports zero. Source-frame checks require the physical $H\delta x$ offsets in collapse and feedback. A transport comparison also found that a Rusanov scheme with local Lax--Friedrichs constrained-transport fluxes was too diffusive to reproduce the same collapse history. The archived failing variants document the sensitivity of these checks. The reported compression runs use the corrected boundary and source mappings and the scheme above.

\begin{figure}[!htbp]
\centering
\includegraphics[width=\textwidth,height=0.63\textheight,keepaspectratio]{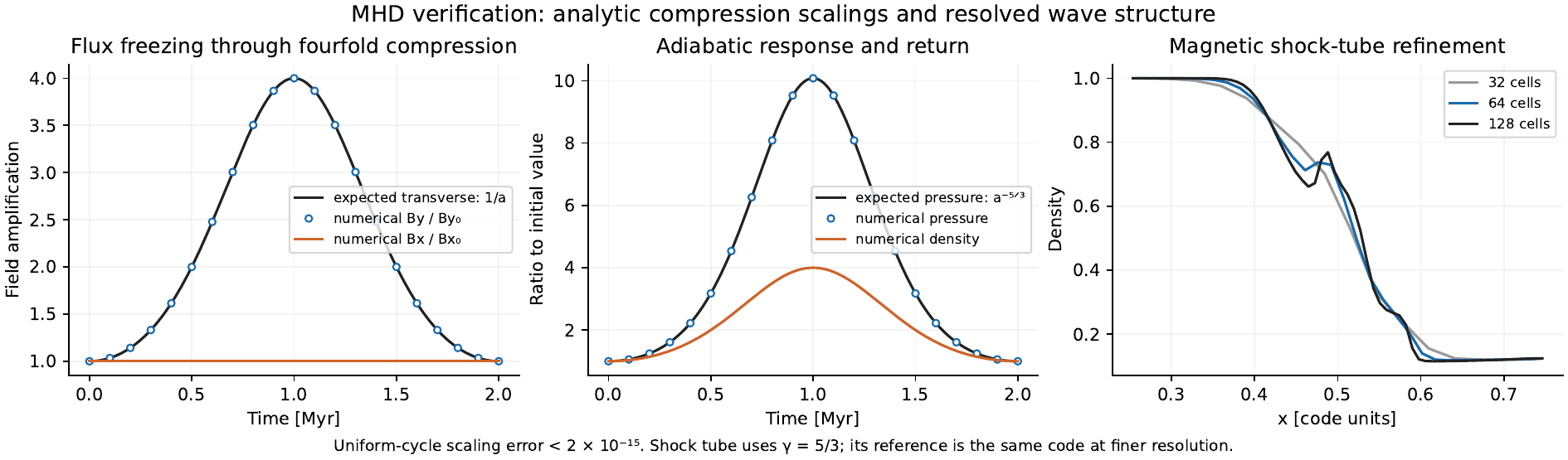}
\caption{\textbf{Flux freezing and MHD shock refinement.} \textbf{Forcing scope: mixed verification suite, including prescribed compression.} Uniform compression and return recover the prescribed flux-freezing and adiabatic scalings. Shock-tube density errors decrease on refinement against a 128-cell numerical reference (not an analytic truth) with \ensuremath{\gamma} = 5/3. Wave, source, gravity and boundary checks are summarized in this appendix. \emph{What this establishes:} The magnetic deformation maps, work bookkeeping and transport have independent numerical checks.\label{fig:9}}
\end{figure}

\subsection{An exact advected pattern through compression and return}\label{sec:7.1}

The smooth problem is a periodic 64 pc cube compressed fourfold and returned over 2 Myr, without cooling, gravity or stellar sources. The initial density is \ensuremath{\rho}\textsubscript{0}[1 + 0.3 sin(k\ensuremath{\xi}) cos(k\ensuremath{\eta})], with mean density $\rho_0$ = 0.03 M\textsubscript{\ensuremath{\odot}} pc$^{-3}$ and wavenumber $k$ = 2\ensuremath{\pi}/(64 pc); initial pressure is uniform at the value for \ensuremath{\rho}\textsubscript{0} and 6000 K. Residual velocities are (3, 2, 0) km s$^{-1}$ and the uniform field is (0.02, 0.05, 0.015) in rationalized code units. With $v_{x0}$ and $v_{y0}$ the initial residual velocity components and $s_x(t)$ the displacement in the reference $x$ coordinate, the analytic density is

\begin{equation}
\rho(\xi,\eta,t)=\frac{\rho_0}{a(t)}\left\{1+0.3\sin[k(\xi-s_x)]\cos[k(\eta-v_{y0}t)]\right\},\quad s_x=v_{x0}\int_0^t a^{-2}(t^\prime)\,dt^\prime.
\end{equation}

Cell averages include the exact $\operatorname{sinc}^2(k\Delta/2)$ factor, where $\operatorname{sinc}u=\sin u/u$. Velocity, pressure and field follow the deformation scalings of Section~\ref{sec:2.1}; no state is reset to the analytic answer. The displacement integral uses 20{,}000-panel Simpson quadrature and its full-cycle value agrees with the independent modified-Bessel expression to $2.22\times10^{-16}$ relative error. The pattern is independent of z, so the density norms sample one horizontal plane. The density L1 error is the summed absolute cell-average error divided by the summed exact density, not by the perturbation amplitude.

\begin{table}[!htbp]
\caption{Exact smooth-MHD spatial refinement\label{tab:9}}
\centering\footnotesize\setlength{\tabcolsep}{4pt}\renewcommand{\arraystretch}{1.2}
\begin{tabular}{>{\raggedright\arraybackslash}p{\dimexpr 0.18\textwidth-2\tabcolsep\relax}>{\raggedright\arraybackslash}p{\dimexpr 0.2\textwidth-2\tabcolsep\relax}>{\raggedright\arraybackslash}p{\dimexpr 0.2\textwidth-2\tabcolsep\relax}>{\raggedright\arraybackslash}p{\dimexpr 0.21\textwidth-2\tabcolsep\relax}>{\raggedright\arraybackslash}p{\dimexpr 0.21\textwidth-2\tabcolsep\relax}}
\hline
\textbf{Cells / axis} & \textbf{Initial cell width [pc]} & \textbf{Peak radial cell width [pc]} & \textbf{Peak L1 [\%]} & \textbf{Return L1 [\%]} \\ \hline
8 & 8 & 2 & 3.07 & 5.34 \\
16 & 4 & 1 & 1.12 & 1.84 \\
32 & 2 & 0.5 & 0.416 & 0.802 \\
64 & 1 & 0.25 & 0.122 & 0.237 \\
\hline
\end{tabular}
\end{table}

All four meshes use the same 1600 timesteps, which satisfy the fine-grid stability condition. The finest adjacent spatial orders, 1.77 at maximum compression and 1.76 after return, are below the formal smooth second-order value for this limited scheme. On the 16\textsuperscript{3} mesh, 400, 800 and 1600 steps give adjacent-state temporal orders 1.99 and 2.00; these differences use a\ensuremath{\rho}/\ensuremath{\rho}\textsubscript{0} and are thus normalized by the mean density at the sampled phase. No first-order retries or timestep subdivisions occur (Figure~\ref{fig:16} shows both refinements).

\subsection{Timestep refinement with a fixed feedback history}\label{sec:7.2}

The 4096-cell source problem of Appendix~\ref{sec:7.5} is replayed with every saved timestep divided by 1, 2, 4 and 8, with identical physical mesh, compression, initial progenitor and supernova time. Cooling, gas and particle gravity, particle drift and mapped feedback evolve independently in every replay, and each run has exactly one ST-stage event, isolating source integration from changes in independently generated stellar events.

\begin{table}[!htbp]
\caption{Timestep refinement with fixed feedback events\label{tab:10}}
\centering\footnotesize\setlength{\tabcolsep}{4pt}\renewcommand{\arraystretch}{1.2}
\begin{tabular}{>{\raggedright\arraybackslash}p{\dimexpr 0.27\textwidth-2\tabcolsep\relax}>{\raggedright\arraybackslash}p{\dimexpr 0.24\textwidth-2\tabcolsep\relax}>{\raggedright\arraybackslash}p{\dimexpr 0.24\textwidth-2\tabcolsep\relax}>{\raggedright\arraybackslash}p{\dimexpr 0.25\textwidth-2\tabcolsep\relax}}
\hline
\textbf{Subdivision factors} & \textbf{Density L1 [\%]} & \textbf{Field L1 [\%]} & \textbf{Max.\ scaled state} \\ \hline
1 / 2 & 0.0698 & 0.0651 & 0.212 \\
2 / 4 & 0.0301 & 0.0279 & 0.155 \\
4 / 8 & 0.0109 & 0.0089 & 0.0711 \\
\hline
\end{tabular}
\end{table}

Density and field L1 differences decrease at every refinement, to 0.0109\% and 0.0089\% for the finest pair. Each L1 norm is normalized by the finer solution's summed absolute values; the field norm samples the three stored face components of each cell. The last column is the local norm $\max|\Delta q|/(1+|q_{\rm coarser}|)$ over gas and sampled field components in code units, where $q$ is a sampled state component (not the shear parameter) and $\Delta q$ the difference between adjacent-timestep solutions. This nonzero local maximum should not be confused with the far smaller volume-integrated differences. No run has positivity retries or subdivisions.

\subsection{Timestep and source-scale sensitivity in the stratified run}\label{sec:7.4}

These controls use the 30 Myr stratified setup (physics in Appendix~\ref{sec:5.1}) with initial gas column $20\,M_\odot\,{\rm pc}^{-2}$, initial thermal-to-magnetic pressure ratio $\beta_0=10$ and 8 pc cells in a $256\times256\times1024$ pc domain, compressed twofold and returned over 10--30 Myr.

\begin{table}[!htbp]
\caption{Timestep and source-scale tests. All variants start with 8 pc cells.\label{tab:13}}
\centering\footnotesize\setlength{\tabcolsep}{4pt}\renewcommand{\arraystretch}{1.2}
\begin{tabular}{>{\raggedright\arraybackslash}p{\dimexpr 0.36\textwidth-2\tabcolsep\relax}>{\raggedright\arraybackslash}p{\dimexpr 0.23\textwidth-2\tabcolsep\relax}>{\raggedright\arraybackslash}p{\dimexpr 0.1\textwidth-2\tabcolsep\relax}>{\raggedright\arraybackslash}p{\dimexpr 0.18\textwidth-2\tabcolsep\relax}>{\raggedright\arraybackslash}p{\dimexpr 0.13\textwidth-2\tabcolsep\relax}}
\hline
\textbf{Cycling-run variant} & \textbf{Stars [M\textsubscript{\ensuremath{\odot}}]} & \textbf{Super\-novae} & \textbf{First-order retries} & \textbf{Timestep splits} \\ \hline
Default (sink length follows the mesh) & 22{,}300 & 29 & 48 & 0 \\
Half Courant factor & 17{,}000 & 30 & 252 & 0 \\
Fixed physical sink length of 8 pc & 43{,}200 & 51 & 2 & 0 \\
64 pc physical supernova cap & 22{,}300 & 29 & 48 & 0 \\
\hline
\end{tabular}
\end{table}

Halving the Courant factor (0.24 to 0.12) changes the final formed mass by $-23.6$\%. At 18 Myr, the last common output before the first supernova in either run, the difference is only +0.133\%, and over 10--30 Myr the normalized time-integrated L1 difference in volume-rms B is 0.156\% (relative to the default history). The fixed-event test above separately demonstrates decreasing integration differences. Because the same seed does not preserve the source history once collapse and accretion change, this is a sensitivity of the coupled calculation, and the evidence does not assign all of it to stochastic sampling.

\begin{figure}[!htbp]
\centering
\includegraphics[width=\textwidth,height=0.63\textheight,keepaspectratio]{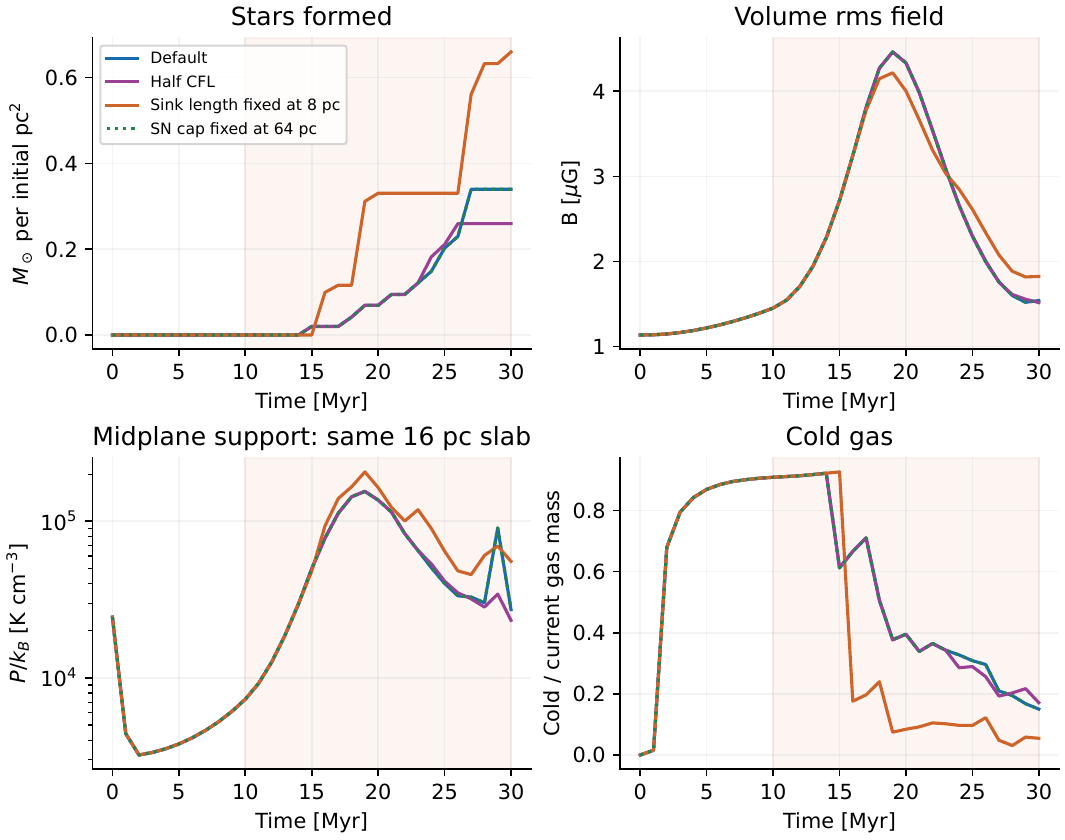}
\caption{\textbf{Timestep and source-scale variants of the cycling run.} All runs start with 8 pc cells and last 30 Myr, contracting from 10 to 20 Myr and expanding from 20 to 30 Myr. The default, halved Courant factor, fixed 8 pc sink length and fixed 64 pc supernova cap share initial conditions and seed. Native midplane bands here are 16 pc thick in every run. The inactive supernova cap's dotted history overlays the byte-identical default. \emph{What this establishes:} Fixing the sink scale and changing the timestep of the coupled calculation alter stellar histories; the fixed-event convergence test separates source integration from changes in the generated stellar population.\label{fig:19}}
\end{figure}

The physical sink-length control fixes the sink length at 8 pc, which sets the collapse threshold. Accretion draws on gas within 12 pc of a sink, with 4 pc softening in its binding check; new sinks are excluded within 24 pc of an existing one, and the binding-energy denominator uses 12 pc. The search is widened to the same physical radius in compressed x. The binding-mass neighborhood stays at 27 cells, so not every spatial component of the closure is frozen. The final formed mass changes by +94.2\%: the stellar response depends strongly on this explicit source parameter (Figure~\ref{fig:19}).

The 64 pc supernova cap never binds in this experiment (every default selected radius lies between 24 and 28 pc), so the capped run's final state, history, stellar births and supernova records match the default byte for byte. The cap therefore does not cause the measured difference here, although this run does not test an actively cap-limited cavity. The default repeat is byte-identical to the archived output at the same initial cell width, so the optional controls preserve the default algorithm.

Across the 30 Myr runs in this test, maximum saved relative mass and energy ledger residuals are $7.26\times10^{-14}$ and $6.98\times10^{-12}$, and maximum normalized \ensuremath{\nabla} \ensuremath{\cdot} B is $1.86\times10^{-14}$. First-order positivity retries are retained and reported; the default-code component suite again passes all seven verification programs. These checks establish numerical accounting and regression consistency.

\subsection{Stratified refinement with equal physical measurement volumes}\label{sec:7.3}

We repeat the same stratified calculation with 16, 8 and 4 pc initial cells, the finest reaching 2 \ensuremath{\times} 4 \ensuremath{\times} 4 pc at peak compression, with the same domain, initial perturbation seed, source laws and physical duration. Adaptive timesteps and stellar events respond freely, so these runs measure the sensitivity of the coupled model.

\begin{table}[!htbp]
\caption{Stratified cycling runs at three resolutions. Field strength is measured at maximum compression (20 Myr) and final stellar mass after return (30 Myr).\label{tab:11}}
\centering\footnotesize\setlength{\tabcolsep}{4pt}\renewcommand{\arraystretch}{1.2}
\begin{tabular}{>{\raggedright\arraybackslash}p{\dimexpr 0.16\textwidth-2\tabcolsep\relax}>{\raggedright\arraybackslash}p{\dimexpr 0.23\textwidth-2\tabcolsep\relax}>{\raggedright\arraybackslash}p{\dimexpr 0.28\textwidth-2\tabcolsep\relax}>{\raggedright\arraybackslash}p{\dimexpr 0.1\textwidth-2\tabcolsep\relax}>{\raggedright\arraybackslash}p{\dimexpr 0.23\textwidth-2\tabcolsep\relax}}
\hline
\textbf{Initial cell width [pc]} & \textbf{B rms at 20 Myr [\ensuremath{\mu}G]} & \textbf{Stars at 30 Myr [M\textsubscript{\ensuremath{\odot}}]} & \textbf{Super\-novae} & \textbf{Gas retained [\%]} \\ \hline
16 & 4.07 & 23{,}700 & 24 & 98.2 \\
8 & 4.33 & 22{,}300 & 29 & 98.3 \\
4 & 4.44 & 52{,}000 & 45 & 96.1 \\
\hline
\end{tabular}
\end{table}

The calculation with 4 pc initial cells has 1{,}050{,}000 cells and 7{,}370 steps (332 first-order retries, 0 subdivisions) and took 3.19 hours of recorded solver time under concurrent project workloads. This timing is not an efficiency ratio for the moving method, because the calculations with 16 and 8 pc initial cells differ in cell count, timestep and feedback history.

At maximum compression (20 Myr), volume-rms field changes by 6.49\% and then 2.60\%, a decreasing refinement difference at a common time set by the prescribed flow (the field itself may peak earlier). The final stellar mass changes by $-6.2$\% and +134\%, so the stellar response does not establish convergence. Figure~\ref{fig:17} shows the complete histories, including cold gas and gas retention, so that agreement is assessed throughout the cycle and not only at peak compression. Over the full 10--30 Myr cycle, normalized time-integrated L1 differences in volume-rms B are 3.57\% and 3.17\% for successive refinements of the initial cell width (16 to 8 pc, then 8 to 4 pc), and cold-mass-fraction differences are 38.4\% and 13.2\%. Both norms use common integer-Myr outputs and the finer history's time integral in the denominator.

\begin{figure}[!htbp]
\centering
\includegraphics[width=\textwidth,height=0.63\textheight,keepaspectratio]{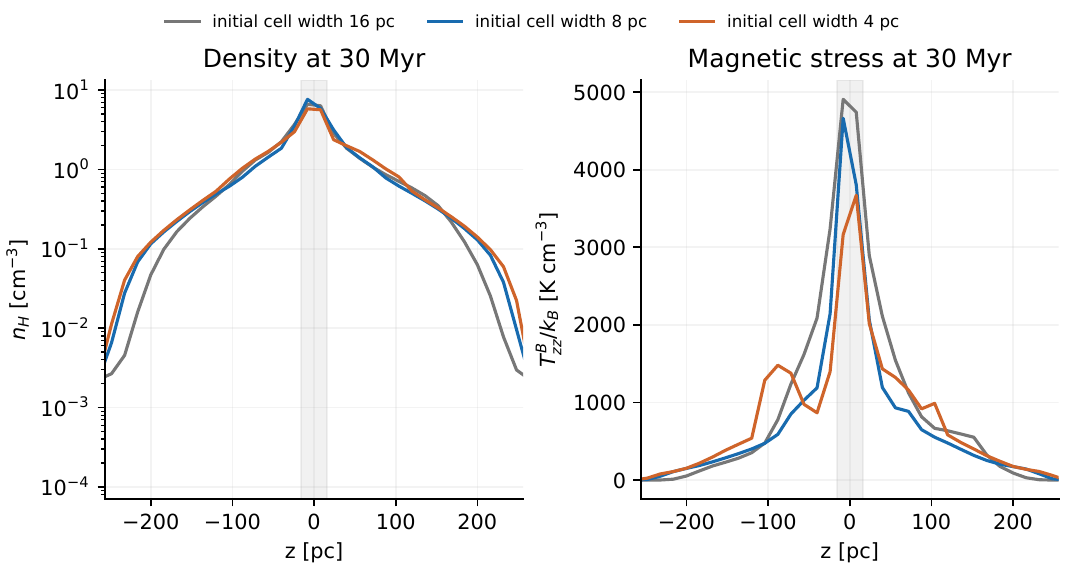}
\caption{\textbf{Compare the same physical region at every resolution.} \textbf{Forcing scope: prescribed compression and return.} Final vertical density and Maxwell-stress profiles use common vertical bins 16 pc thick; the shaded |z| < 16 pc slab is the physical volume used for the table in Appendix~\ref{sec:7.3}. Pressure and stress are computed from native fields before averaging, and magnetic stress is averaged quadratically. \emph{What this establishes:} A fixed physical averaging volume removes the unequal two-cell-band measurement from the refinement comparison.\label{fig:18}}
\end{figure}

The old native midplane history averages two vertical cells, i.e.\ 32, 16 and 8 pc bands at the three resolutions, so it is excluded from the cross-resolution comparison. Instead, the final saved fields are averaged over the same |z| < 16 pc slab at every resolution (Table~\ref{tab:12}) and in common 16 pc vertical bins (Figure~\ref{fig:18}); these are direct volume averages of thermal pressure and Maxwell stress, not stress reconstructed from an averaged magnetic field. In Table~\ref{tab:12}, $T^B_{zz}=(B_x^2+B_y^2-B_z^2)/2$ is the vertical Maxwell stress and $B_{\rm rms}=\langle B^2\rangle_V^{1/2}$, with $\langle\cdot\rangle_V$ a volume average over the stated region; pressure and stress are in K cm$^{-3}$.

\begin{table}[!htbp]
\caption{Measurements in equal physical regions after return ($30$ Myr, $a=1$): the $|z|<16$ pc slab. $P_{\rm th}$ is the thermal pressure.\label{tab:12}}
\centering\footnotesize\setlength{\tabcolsep}{4pt}\renewcommand{\arraystretch}{1.2}
\begin{tabular}{>{\raggedright\arraybackslash}p{\dimexpr 0.18\textwidth-2\tabcolsep\relax}>{\raggedright\arraybackslash}p{\dimexpr 0.2\textwidth-2\tabcolsep\relax}>{\raggedright\arraybackslash}p{\dimexpr 0.22\textwidth-2\tabcolsep\relax}>{\raggedright\arraybackslash}p{\dimexpr 0.22\textwidth-2\tabcolsep\relax}>{\raggedright\arraybackslash}p{\dimexpr 0.18\textwidth-2\tabcolsep\relax}}
\hline
\textbf{Initial cell width [pc]} & \textbf{n\textsubscript{H} [cm$^{-3}$]} & \textbf{P\textsubscript{th}/k\textsubscript{B}} & \textbf{T\textsuperscript{B}\textsubscript{zz}/k\textsubscript{B}} & \textbf{B rms [\ensuremath{\mu}G]} \\ \hline
16 & 6.51 & 23{,}800 & 4{,}820 & 4.44 \\
8 & 6.78 & 20{,}200 & 4{,}230 & 4.53 \\
4 & 5.74 & 77{,}700 & 3{,}420 & 4.81 \\
\hline
\end{tabular}
\end{table}

When the initial cell width is halved from 8 to 4 pc, the final common-slab thermal pressure changes by a factor of 3.84, so pressure, like stellar production, has not converged over this 30 Myr interval. Refinement changes unresolved source scales as well as discretization. The collapse scale changes as cells become smaller and thinner, and the minimum supernova radius falls from 48 to 24 to 12 pc as the initial cell width is refined, so these runs do not isolate the continuum limit of a fixed physical star-formation law. The MC/ST/EJ event counts (16/6/2, 21/7/1 and 26/19/0 for initial cell widths of 16, 8 and 4 pc) show that the histories also sample different remnant-stage mixtures and injection masses under the same hybrid feedback law, whose action responds to resolution and the evolving gas.

\subsection{Matched cost with cooling, gravity and mapped feedback}\label{sec:7.5}

The benchmark compares a 128 pc local cube of 16\textsuperscript{3} cells with a 64 \ensuremath{\times} 64 \ensuremath{\times} 16-cell domain containing 16 horizontal copies. Both have 8 pc initial cells (4 \ensuremath{\times} 8 \ensuremath{\times} 8 pc at maximum twofold compression) and evolve the same 0.2 Myr cycle with the same 121 saved timesteps and scheme, giving 495,616 versus 7,929,856 cell-steps. Cooling, gas self-gravity, particle gravity and drift, and mapped feedback are active. One identical 200 M\textsubscript{\ensuremath{\odot}} progenitor per copy supplies a prescribed supernova at peak compression, with a common 32 pc maximum source radius; new star formation is disabled to keep events identical, and each gas state evolves independently.

The initial mode has mean density 0.003 M\textsubscript{\ensuremath{\odot}} pc$^{-3}$, a 20\% cosine-product perturbation with a 128 pc horizontal period, 6000 K gas, smooth residual velocities of amplitude (2, 1, 0.2) km s$^{-1}$ and a divergence-free nonuniform field; the archived benchmark source gives the complete analytic initial fields. The progenitor lies at (6, 61, 1) pc in each copy, so its feedback kernel can cross the x seam. Vertical boundaries are open and external stellar/halo gravity is off; this repeated mode is separate from the stratified production experiment.

Because the explosion occurs at maximum compression, where $H=0$, it tests feedback on the compressed mesh and across periodic images but not a finite background-flow velocity boost at injection. The MC/ST/EJ source tests during compression and expansion (Appendix~\ref{sec:3.3}) supply that check. The timing still includes the complete source-mapping implementation.

The maximum scaled gas/field difference is $7.53\times10^{-14}$; maximum relative mass and energy ledger residuals are $9.64\times10^{-14}$ and $3.83\times10^{-13}$, with no first-order retries or subdivisions. The norm samples gas and the three face components indexed with each cell; the divergence check also includes the extra normal face at the open upper boundary. Both calculations reproduce the same periodic local problem; Appendix~\ref{sec:3.2} quantifies the disturbance content selected by this periodicity.

\begin{table}[!htbp]
\caption{Matched cost with gravity, cooling and feedback\label{tab:14}}
\centering\footnotesize\setlength{\tabcolsep}{4pt}\renewcommand{\arraystretch}{1.2}
\begin{tabular}{>{\raggedright\arraybackslash}p{\dimexpr 0.12\textwidth-2\tabcolsep\relax}>{\raggedright\arraybackslash}p{\dimexpr 0.28\textwidth-2\tabcolsep\relax}>{\raggedright\arraybackslash}p{\dimexpr 0.2\textwidth-2\tabcolsep\relax}>{\raggedright\arraybackslash}p{\dimexpr 0.22\textwidth-2\tabcolsep\relax}>{\raggedright\arraybackslash}p{\dimexpr 0.18\textwidth-2\tabcolsep\relax}}
\hline
\textbf{Repeat} & \textbf{Execution order} & \textbf{Local [s]} & \textbf{16 copies [s]} & \textbf{Ratio} \\ \hline
1 & small-large & 0.736 & 11.4 & 15.5 \\
2 & large-small & 0.712 & 11.2 & 15.8 \\
3 & small-large & 0.708 & 11.1 & 15.6 \\
\hline
\end{tabular}
\end{table}

The median paired runtime ratio is 15.6 (range 15.5--15.8; Figure~\ref{fig:20}). Three serial pairs alternate execution order on one arm64 macOS host (Apple Clang 21.0.0, -O3, C++17) after all project simulation jobs had finished; external user workloads are not controlled. Timed intervals cover the full step and timestep evaluation, including source and gravity work, and exclude initialization and output equally. This is the cost of equal local physics at equal physical cell widths, timesteps and duration, not a speedup over the calculation of \citet{KimKimOstriker2020}.

\subsection{Physical-image boundary check and negative control}\label{sec:3.4}

\begin{figure}[!htbp]
\centering
\includegraphics[width=\textwidth,height=0.63\textheight,keepaspectratio]{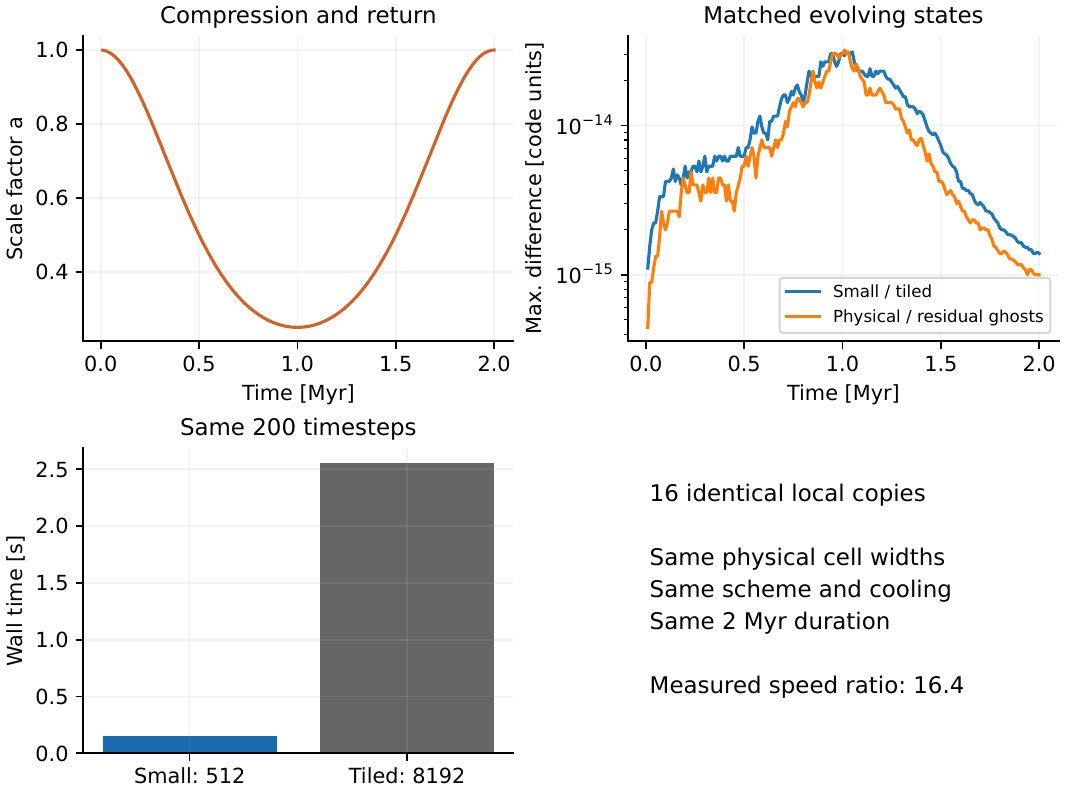}
\caption{\textbf{Matched MHD boundaries and computational cost.} \textbf{Forcing scope: prescribed compression and return.} A 512-cell box and an 8192-cell domain of 16 repeated copies evolve the same nonlinear disturbance through fourfold compression and return, with identical methods, 200 timesteps, physical cell widths, cooling and 2 Myr duration; a separate ghost construction uses physical velocity images. Gravity and stellar sources are off. The states agree to about $3\times10^{-14}$; omitting the physical-image velocity jump gives a 2975-code-unit operator discrepancy. The timings (0.156 s and 2.56 s, a ratio of 16.4) confirm the repeated-volume saving in this simpler problem.\label{fig:10}}
\end{figure}

A separate boundary-equivalence test compares a 512-cell box with an 8192-cell domain of 16 copies of the same nonlinear local disturbance, both evolved with identical timesteps, cell widths and schemes (setup in Figure~\ref{fig:10}). A third path uses explicit physical-velocity image ghosts. Gravity, particle formation and feedback are off; they have separate operator/source checks.

\begin{table}[!htbp]
\caption{Matched-domain boundary equivalence and numerical budgets\label{tab:5}}
\centering\footnotesize\setlength{\tabcolsep}{4pt}\renewcommand{\arraystretch}{1.2}
\begin{tabular}{>{\raggedright\arraybackslash}p{\dimexpr 0.58\textwidth-2\tabcolsep\relax}>{\raggedright\arraybackslash}p{\dimexpr 0.42\textwidth-2\tabcolsep\relax}}
\hline
\textbf{Quantity} & \textbf{Small versus tiled / physical images} \\ \hline
Maximum absolute state difference & $3.11\times10^{-14}$ / $3.20\times10^{-14}$ \\
Maximum normalized \ensuremath{\nabla} \ensuremath{\cdot} B & $2.50\times10^{-15}$ \\
Mass / total-energy ledger residual & $7.11\times10^{-15}$ / $4.32\times10^{-14}$ \\
First-order retries & 0 in either calculation \\
\hline
\end{tabular}
\end{table}

Omitting the physical-image velocity jump gives a maximum right-hand-side discrepancy of 2975 code units. This broken variant is evaluated only as an operator; only correctly transformed states are evolved. Because the three correct routes agree to within the errors in Table~\ref{tab:5}, this discrepancy shows that the boundary test is sensitive to the missing velocity jump.

\subsection{Disturbances represented by the periodic patch}\label{sec:3.2}

\begin{figure}[!htbp]
\centering
\includegraphics[width=\textwidth,height=0.63\textheight,keepaspectratio]{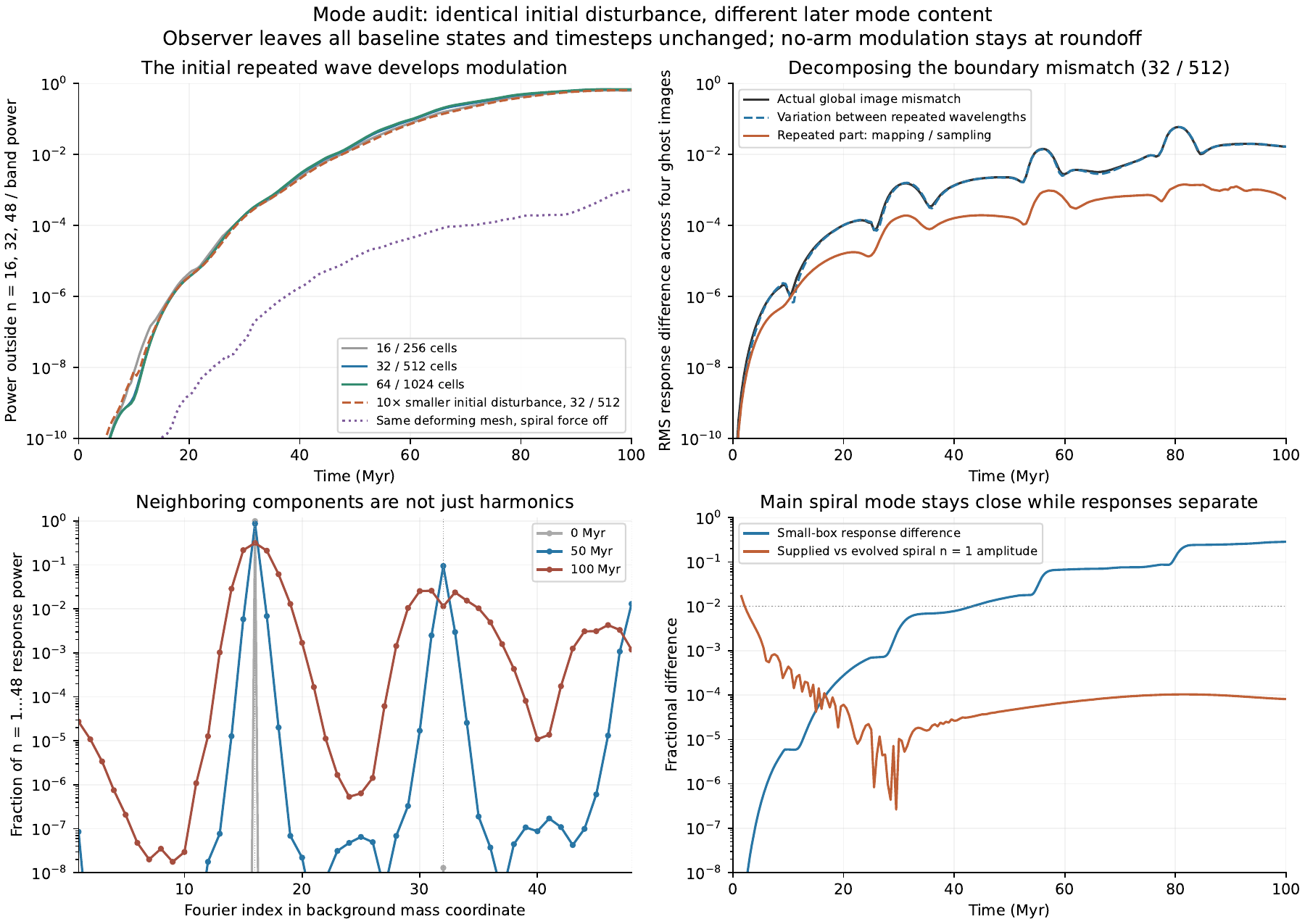}
\caption{\textbf{Disturbance content in the spatial arm-to-arm test.} \textbf{Forcing scope: spatial spiral/shock precursor.} Spectra and image diagnostics of the saved full-domain response, in background mass coordinate. Repeated local waves allow multiples of index 16; neighboring indices 15 and 17 reflect an arm-phase-dependent envelope and are not the galaxy's spiral arm number. The plotted spectral fractions are diagnostic norm fractions, not physical energies or small-box error percentages. \emph{What this establishes:} The initially identical sinusoid evolves into different disturbance content under the spatially varying operator; this global-envelope comparison is distinct from the matched local-model verification.\label{fig:3}}
\end{figure}

The image conditions require the residual disturbance to repeat over the patch, while the supplied background specifies the global flow. In the spatial arm-to-arm diagnostic, Fourier index 16 places one wavelength per patch across the arm interval, and repetition admits multiples of 16. The spatially varying operator couples the initial sinusoid to neighboring indices 15 and 17 and other nonrepeating components. At 100 Myr, the fraction of response power outside multiples of 16 is 63.2\%, 67.1\% and 68.3\% for 256, 512 and 1024 full-domain cells, and 65.0\% for an initial disturbance ten times smaller (Figure~\ref{fig:3}).

In the 32/512 case at 50 Myr, the nonrepeating response accounts for 98.0\% of the squared image mismatch under the signed inner-product decomposition, whereas the dominant supplied spiral density coefficient differs from the original driver by only 0.00458\% in amplitude and $7.72\times10^{-5}$ radians in phase. The evidence therefore points to different evolving disturbance content, not a grossly different supplied spiral. Spatially varying coefficients can couple index 16 to 15 and 17 even in linear response; a term-by-term coupling budget is left for future work.

These measurements quantify the disturbance content that the periodic approximation selects. Representing the arm-phase-dependent envelope requires an additional mapping. Adding the spectral diagnostic leaves the baseline histories byte-identical.
\section{Earlier hydrodynamic controls without imposed compression}\label{sec:4}

\subsection{Physical ingredients}\label{sec:4.1}

The uncompressed hydrodynamic runs use a 256 \ensuremath{\times} 256 \ensuremath{\times} 1024 pc stratified domain with 16 pc cells. We repeated the central-column case ($20\,M_\odot\,{\rm pc}^{-2}$) with 8 pc cells to measure how refinement changes the stellar mass formed. Initial gas surface densities $\Sigma_0$ of 10, 20 and 40 M\textsubscript{\ensuremath{\odot}} pc$^{-2}$ are fixed initial columns, not enforced constant densities. The gas starts at 6000 K in hydrostatic stratification with a smooth 10\% density perturbation and 2 km s$^{-1}$ velocity perturbations, then cools, collapses, forms particles and loses gas through open vertical boundaries, without replenishment.

Gas and particles (deposited by cloud-in-cell) enter a self-gravity solve that uses fast Fourier transforms and a Green function, periodic horizontally and isolated vertically, including the uniform sheet mode. With $G$ the gravitational constant, $\Sigma_{*,0}$ and $H_{*,0}$ the fixed old-stellar surface density and scale height, and $\rho_d$ the halo mass density, the external vertical acceleration $g_z$ is

\begin{equation}
g_z=-2\pi G\Sigma_{*,0}\tanh(z/H_{*,0})-2\pi G\rho_d z,
\end{equation}

with \ensuremath{\Sigma}\textsubscript{*,0} = 35 M\textsubscript{\ensuremath{\odot}} pc$^{-2}$, H\textsubscript{*,0} = 350 pc and \ensuremath{\rho}\textsubscript{d} = 0.01 M\textsubscript{\ensuremath{\odot}} pc$^{-3}$. Cooling uses the Koyama--Inutsuka low-temperature fit and a pinned Athena high-temperature fit. With $\Sigma_{*,<10\,{\rm Myr}}$ the surface density of stars younger than 10 Myr and $n_{\rm H}$ the hydrogen-nucleus number density, the heating per H nucleus is $2\times10^{-26}$[0.1 + \ensuremath{\Sigma}\textsubscript{*,<10 Myr}/(0.025 M\textsubscript{\ensuremath{\odot}} pc$^{-2}$)] erg s$^{-1}$, a young-star FUV proxy without radiation transfer. Temperature and density floors are 10 K and n\textsubscript{H} = $10^{-8}$ cm$^{-3}$. The conversion uses fixed values of 1.1 particles and 1.4 hydrogen masses per H nucleus at all temperatures, omitting ionization-dependent molecular weight.

Cluster particles form in cold gas (temperature $T$ < 300 K) that converges along all three axes, sits at a potential minimum, passes an approximate 27-cell binding test and exceeds the Larson--Penston density threshold $\rho_{\rm LP}$, where $\Delta_{\min}$ is the current minimum physical cell width:

\begin{equation}
\rho_{\rm LP}=\frac{8.86(P/\rho)}{\pi G\Delta_{\min}^2}.
\end{equation}

New particles are separated by at least 3\ensuremath{\Delta}\textsubscript{min}, and cold, bound, infalling excess gas within 1.5\ensuremath{\Delta}\textsubscript{min} can accrete. No PRFM star-formation rate, effective equation of state or fixed efficiency per free-fall time is imposed. Each accreted stellar cohort samples a Poisson number of supernovae with a mean of one per 100 M\textsubscript{\ensuremath{\odot}}, with independent delays uniform from 3 to 40 Myr; each event returns up to 10 M\textsubscript{\ensuremath{\odot}}. The principal runs instead sample supernovae from the age-dependent stellar-population rates in Appendix~\ref{app:principal_sources}.

\subsection{Feedback in evacuated cavities}\label{sec:4.2}

The hybrid feedback prescription tested here uses the shell-formation mass $M_{\rm sf}=1679\,M_\odot\,(n_{\rm H}/{\rm cm^{-3}})^{-0.26}$, following the stage criterion of \citet{KimOstriker2017}. The injection search starts at three cells in the coarsest direction and grows in half-cell steps up to min(128 pc, half the shortest horizontal box length). The enclosed mass $M$ in Table~\ref{tab:6} includes ejecta (this $M$ differs from the reference-volume density of Appendix~\ref{sec:3.1}). An evacuated cavity reaching this cap becomes an ejecta-stage event, so no terminal-momentum model is forced onto vanishing gas mass.

\begin{table}[!htbp]
\caption{Hybrid feedback stages and deposited sources\label{tab:6}}
\centering\footnotesize\setlength{\tabcolsep}{4pt}\renewcommand{\arraystretch}{1.2}
\begin{tabular}{>{\raggedright\arraybackslash}p{\dimexpr 0.25\textwidth-2\tabcolsep\relax}>{\raggedright\arraybackslash}p{\dimexpr 0.75\textwidth-2\tabcolsep\relax}}
\hline
\textbf{Stage} & \textbf{Selection and deposited source} \\ \hline
MC: momentum-conserving & M/M\textsubscript{sf} \ensuremath{\geq} 1 at minimum radius. Terminal momentum, with positive kinetic-energy increment capped at $10^{51}$ erg. \\
ST: Sedov--Taylor & 0.027 \ensuremath{\leq} M/M\textsubscript{sf} < 1. Total $10^{51}$ erg, 72\% thermal and 28\% kinetic. \\
EJ: ejecta & M/M\textsubscript{sf} < 0.027 at maximum radius, or the last smaller sphere before a discrete dense-wall jump. Total $10^{51}$ erg as kinetic ejecta. \\
\hline
\end{tabular}
\end{table}

\begin{figure}[!htbp]
\centering
\includegraphics[width=\textwidth,height=0.63\textheight,keepaspectratio]{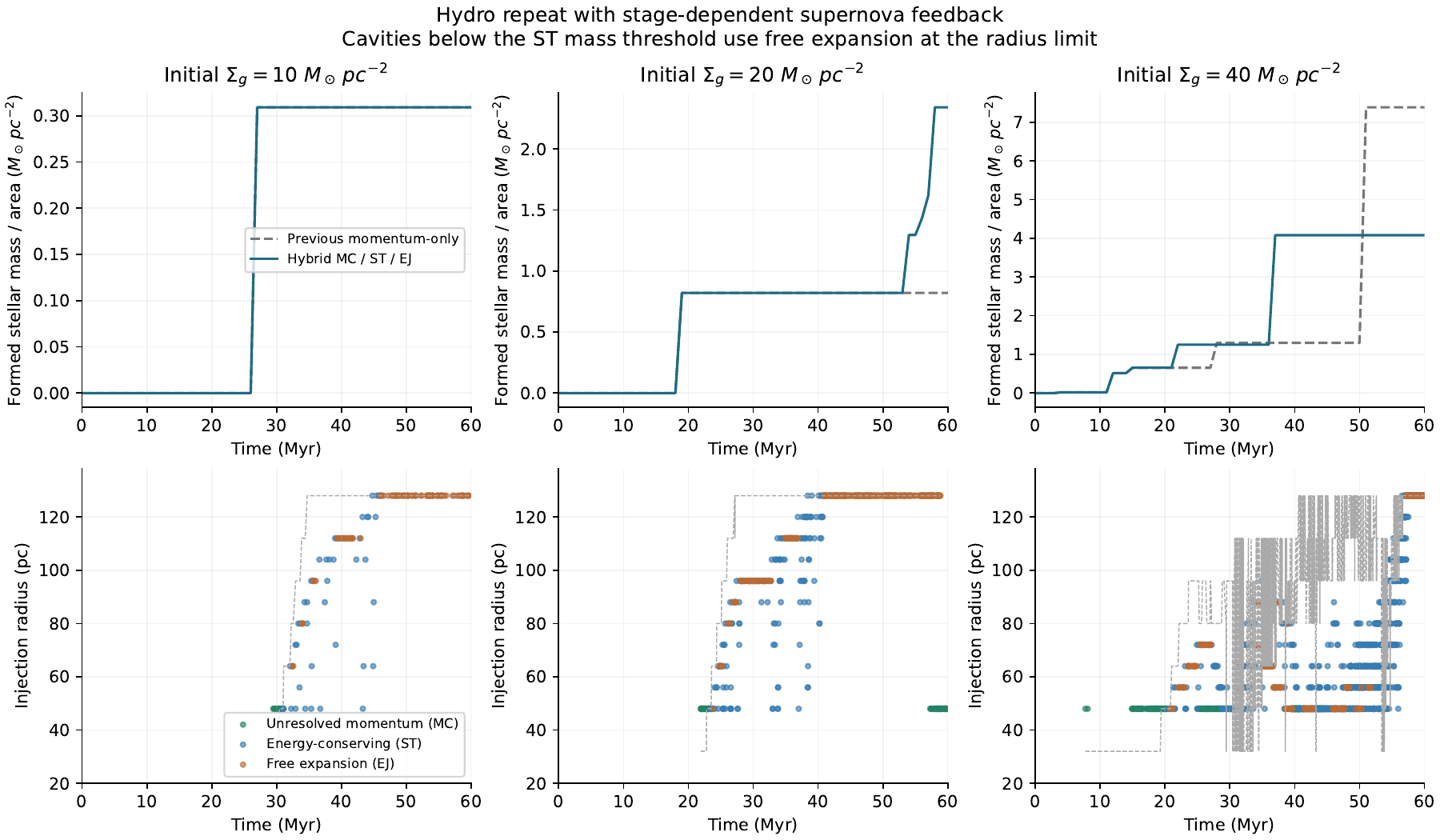}
\caption{\textbf{Revised feedback in evacuated cavities.} \textbf{Forcing scope: no imposed box compression.} Old and revised hydro histories, with injection-radius and remnant-stage diagnostics. The revised MC/ST/EJ prescription preserves pre-existing cavity energy, handles the low-mass ejecta stage and changes the minimum injection radius, so the comparison tests a full source-model revision. \emph{What this establishes:} Evacuated-cavity treatment affects later bursts; it is a measurable local model choice, not a moving-boundary effect.\label{fig:4}}
\end{figure}

ST/EJ homogenization preserves existing mass, momentum and total gas energy, including hot cavity energy, and turns unresolved velocity variance into heat. Removing the mass-weighted mean radial kick enforces zero net impulse in the progenitor frame; the kick amplitude is solved from the discrete kinetic-energy increment. The MC branch uses the \citet{KimOstriker2015} terminal-momentum fit with its density argument clamped to 0.1--100 cm$^{-3}$; kinetic energy lost when a kick brakes inflow becomes heat. These are source-model choices, not a transport fix.

\begin{figure}[!htbp]
\centering
\includegraphics[width=\textwidth,height=0.63\textheight,keepaspectratio]{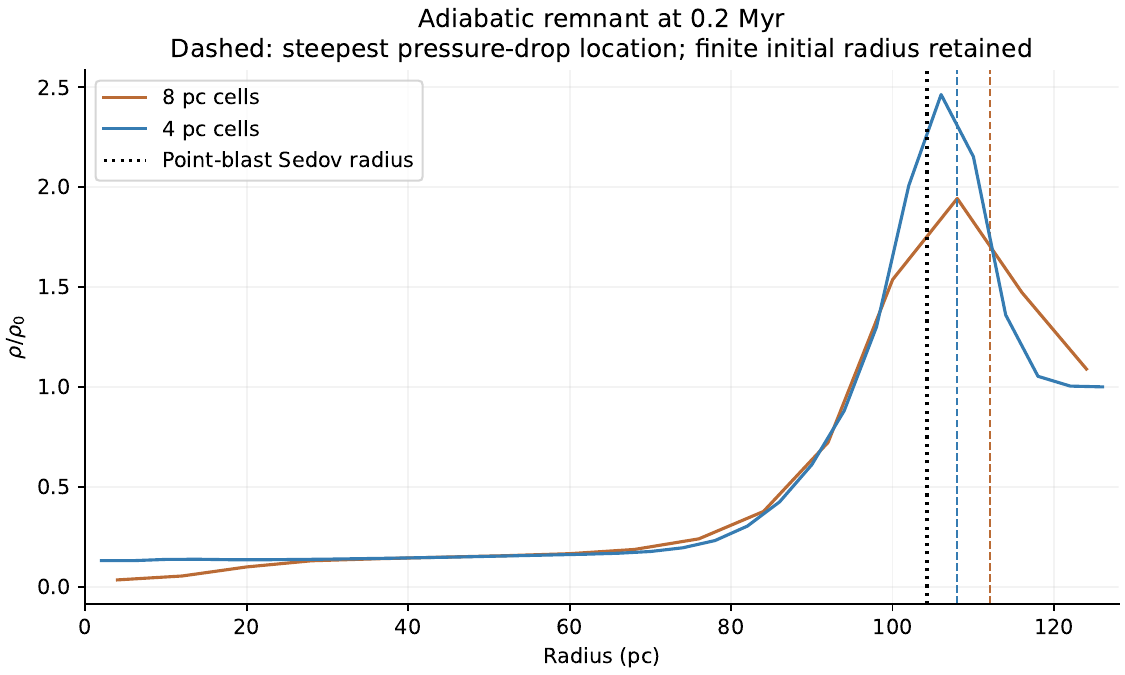}
\caption{\textbf{Adiabatic remnant refinement.} \textbf{Forcing scope: no imposed box compression.} The same saved radial profiles on grids with 8 and 4 pc cells, 0.2 Myr after a finite-radius ST injection. The shock-center pressure drop lies at 112 and 108 pc, versus an analytic point-explosion (Sedov) radius of 104 pc, with no fitted initial age. The earlier diffuse outer-density-contour diagnostic, which failed the 10\% fine-grid radius threshold, is retained with the unchanged original profiles. \emph{What this establishes:} The shock-center radius error falls from 7.4\% to 3.6\%, verifying adiabatic injection/propagation with a finite initial kernel, not radiative-remnant calibration.\label{fig:5}}
\end{figure}

Boost, periodic-seam, repeated-explosion, cavity-wall, inflow-braking and energy-budget tests pass. In the completed hydro repeat, all 2462 ST/EJ events inject $10^{51}$ erg to within $5.45\times10^{-15}$ relative error; the run also has 572 MC events and no event with negative net feedback energy. The no-feedback history is byte-identical to its previous version. In the adiabatic remnant check (Figure~\ref{fig:5}) the shock radius is defined by the steepest pressure drop, and the hot-cooling ordinary differential equation agrees with an independent small-step fourth-order Runge--Kutta integration to $6.53\times10^{-5}$.

\subsection{Completed hydro results}\label{sec:4.3}

\begin{table}[!htbp]
\caption{Completed hydrodynamic applications\label{tab:7}}
\centering\footnotesize\setlength{\tabcolsep}{4pt}\renewcommand{\arraystretch}{1.2}
\begin{tabular}{>{\raggedright\arraybackslash}p{\dimexpr 0.32\textwidth-2\tabcolsep\relax}>{\raggedright\arraybackslash}p{\dimexpr 0.1\textwidth-2\tabcolsep\relax}>{\raggedright\arraybackslash}p{\dimexpr 0.09\textwidth-2\tabcolsep\relax}>{\raggedright\arraybackslash}p{\dimexpr 0.2\textwidth-2\tabcolsep\relax}>{\raggedright\arraybackslash}p{\dimexpr 0.1\textwidth-2\tabcolsep\relax}>{\raggedright\arraybackslash}p{\dimexpr 0.19\textwidth-2\tabcolsep\relax}}
\hline
\textbf{Case} & \textbf{Cell width [pc]} & \textbf{Myr} & \textbf{Stars [M\textsubscript{\ensuremath{\odot}}]} & \textbf{Super\-novae} & \textbf{Time [s]} \\ \hline
\ensuremath{\Sigma}\textsubscript{0} = 10 & 16 & 60 & 20{,}300 & 157 & 257 \\
\ensuremath{\Sigma}\textsubscript{0} = 20 & 16 & 60 & 153{,}000 & 579 & 411 \\
\ensuremath{\Sigma}\textsubscript{0} = 40 & 16 & 60 & 268{,}000 & 1785 & 370 \\
\ensuremath{\Sigma}\textsubscript{0} = 20, fine & 8 & 30 & 59{,}600 & 67 & 900 \\
\ensuremath{\Sigma}\textsubscript{0} = 20, seed 19 & 16 & 60 & 52{,}500 & 446 & 282 \\
\ensuremath{\Sigma}\textsubscript{0} = 20, no feedback & 16 & 60 & 1{,}160{,}000 & 0 & 9.1 \\
\hline
\end{tabular}
\end{table}

Every hydro repeat completed, with zero transport fallbacks and maximum relative mass and energy ledger residuals of $4.93\times10^{-13}$ and $2.93\times10^{-11}$. The revised prescription changes burst histories substantially (Figure~\ref{fig:4}). Over the common first 30 Myr, the central-column calculations with 16 and 8 pc cells form 53{,}800 and 59{,}600 M\textsubscript{\ensuremath{\odot}} (10.8\% apart), and the alternative seed gives a different 60 Myr history (Figure~\ref{fig:7}). At 60 Myr the three baseline runs retain 81\%, 76\% and 50\% of their initial gas. These inexpensive runs test model sensitivity even though their density evolves and their star formation is bursty (Figure~\ref{fig:8} shows the resulting structure).

\section{Earlier compression--expansion cycles and fixed-box controls}\label{sec:5}

\subsection{Initial field, geometry and retained closures}\label{sec:5.1}

With $\beta_0$ the initial thermal-to-magnetic pressure ratio and $\bar P(z)$ the horizontally averaged thermal pressure, the default field is $B_y(z)=[2\bar P(z)/\beta_0]^{1/2}$, with B\textsubscript{x} = B\textsubscript{z} = 0 and \ensuremath{\beta}\textsubscript{0} = 10. We also include a \ensuremath{\beta}\textsubscript{0} = 1 case and exactly zero-field controls. Adding this divergence-free field to the hydro initial profile does not construct a new magnetostatic equilibrium, so vertical readjustment is part of the experiment. These runs use the cooling, gravity and supernova feedback of Appendix~\ref{sec:4}. As in the principal runs, the gas obeys ideal MHD and sink accretion leaves magnetic flux on the mesh (Section~\ref{sec:2.3}). These 30 Myr tests omit rotation and background shear and use no variable-ionization network.

\begin{figure}[!htbp]
\centering
\includegraphics[width=\textwidth,height=0.63\textheight,keepaspectratio]{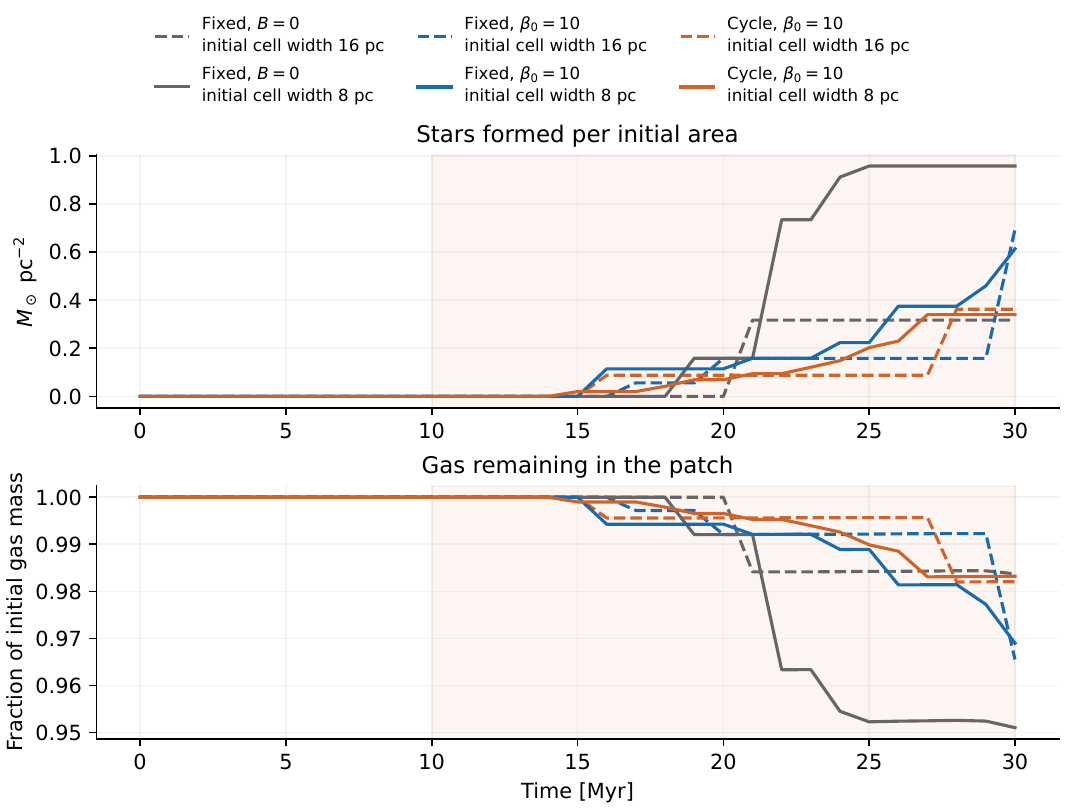}
\caption{\textbf{Stellar response and gas retention.} The fixed and cycling boxes at $\beta_0=10$ and the zero-field fixed box are those of Figure~\ref{fig:11}. Stars are normalized by the initial area for this cumulative plot; gas retention is \ensuremath{\Sigma}\textsubscript{gas}(t)a(t)/\ensuremath{\Sigma}\textsubscript{0}. \emph{What this establishes:} At $\beta_0=10$, halving the initial cell width from 16 to 8 pc lowers the final stellar mass by 11.6\% (fixed box) and 6.2\% (cycling box). The zero-field result is substantially more sensitive. The result for initially 4 pc cells is in Figure~\ref{fig:17}.\label{fig:12}}
\end{figure}

In the 30 Myr stratified runs, the radial width halves and returns to its initial value according to

\begin{equation}
a(t)=\exp\!\left\{-\frac{1}{2}\ln2\left[1-\cos\!\left(2\pi\frac{t-10\,{\rm Myr}}{20\,{\rm Myr}}\right)\right]\right\},\quad 10\leq t/{\rm Myr}\leq30.
\end{equation}

Outside that interval $a=1$; at 20 Myr $a=1/2$. Initial 16 pc cells reach 8 \ensuremath{\times} 16 \ensuremath{\times} 16 pc and initial 8 pc cells 4 \ensuremath{\times} 8 \ensuremath{\times} 8 pc, so resolution improves along the compressed direction. The minimum physical cell width sets the current sink threshold and particle kernel, the binding neighborhood stays at 27 cells, and the maximum physical feedback radius changes with box width.

\subsection{The nine selected 30 Myr runs}\label{sec:5.2}

\begin{table}[!htbp]
\caption{Fixed-box and cycling runs at 30 Myr\label{tab:8}}
\centering\footnotesize\setlength{\tabcolsep}{4pt}\renewcommand{\arraystretch}{1.2}
\begin{tabular}{>{\raggedright\arraybackslash}p{\dimexpr 0.32\textwidth-2\tabcolsep\relax}>{\raggedright\arraybackslash}p{\dimexpr 0.1\textwidth-2\tabcolsep\relax}>{\raggedright\arraybackslash}p{\dimexpr 0.2\textwidth-2\tabcolsep\relax}>{\raggedright\arraybackslash}p{\dimexpr 0.09\textwidth-2\tabcolsep\relax}>{\raggedright\arraybackslash}p{\dimexpr 0.13\textwidth-2\tabcolsep\relax}>{\raggedright\arraybackslash}p{\dimexpr 0.16\textwidth-2\tabcolsep\relax}}
\hline
\textbf{\ensuremath{\Sigma}\textsubscript{0}, field, geometry} & \textbf{Initial cell width [pc]} & \textbf{Stars [M\textsubscript{\ensuremath{\odot}}]} & \textbf{Super\-novae} & \textbf{1st order} & \textbf{Time [s]} \\ \hline
10, \ensuremath{\beta}\textsubscript{0}=10, fixed & 16 & 1{,}500 & 3 & 0 & 8.5 \\
20, \ensuremath{\beta}\textsubscript{0}=10, fixed & 16 & 45{,}500 & 32 & 32 & 22.2 \\
40, \ensuremath{\beta}\textsubscript{0}=10, fixed & 16 & 172{,}000 & 384 & 150 & 361 \\
20, B=0, fixed & 16 & 20{,}800 & 46 & 38 & 69.3 \\
20, \ensuremath{\beta}\textsubscript{0}=1, fixed & 16 & 1{,}240 & 0 & 0 & 11.2 \\
20, \ensuremath{\beta}\textsubscript{0}=10, cycle & 16 & 23{,}700 & 24 & 12 & 37.0 \\
20, \ensuremath{\beta}\textsubscript{0}=10, fixed & 8 & 40{,}200 & 40 & 54 & 1{,}030 \\
20, B=0, fixed & 8 & 62{,}700 & 98 & 246 & 1{,}130 \\
20, \ensuremath{\beta}\textsubscript{0}=10, cycle & 8 & 22{,}300 & 29 & 48 & 387 \\
\hline
\end{tabular}
\end{table}

\begin{figure}[!htbp]
\centering
\includegraphics[width=\textwidth,height=0.63\textheight,keepaspectratio]{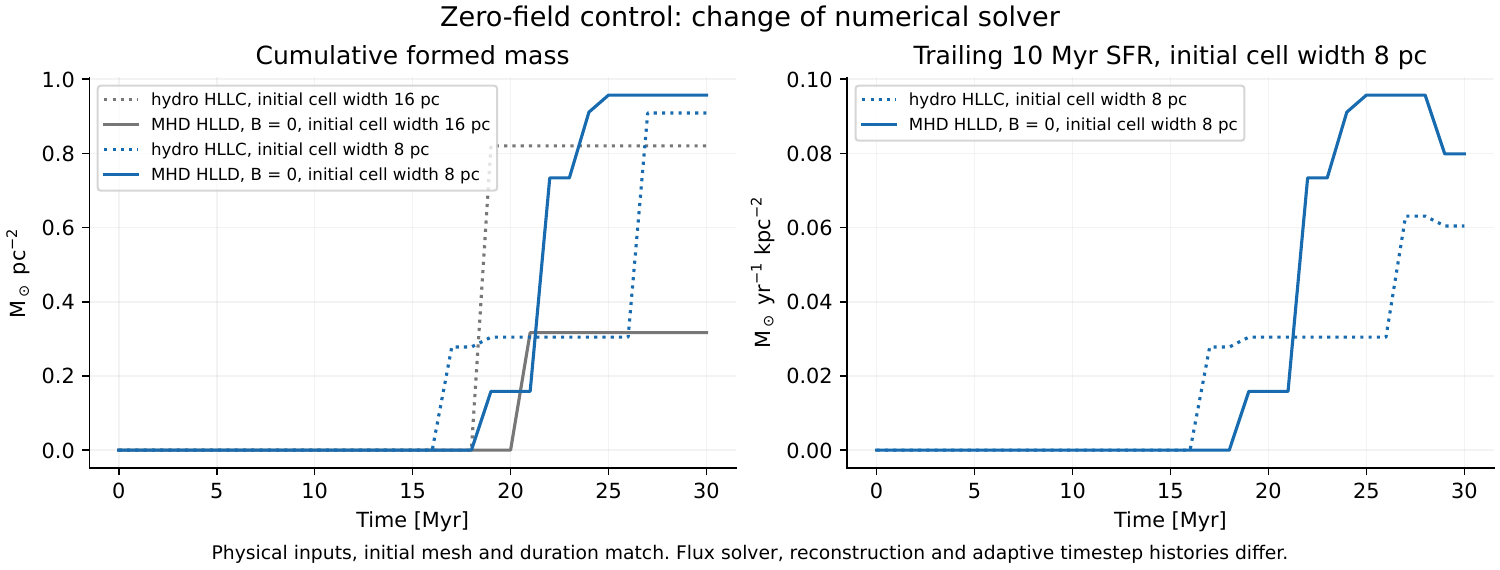}
\caption{\textbf{Separating a solver change from magnetic physics.} \textbf{Forcing scope: no imposed box compression.} Zero-field MHD is compared with the previous hydro solver at the same initial columns, mesh sizes and physical duration. The Riemann solvers (HLLC, Harten--Lax--van Leer with contact, versus HLLD), reconstruction and subsequent adaptive timesteps differ. The right panel compares trailing 10 Myr star-formation rates on the finer grid. \emph{What this establishes:} With 8 pc cells the final cumulative stellar masses agree to within 5.3\%, but burst timing differs and the result with 16 pc cells differs much more; this control prevents attributing a solver difference to magnetic fields.\label{fig:14}}
\end{figure}

All cases use seed 7 and the same initial physical box; the two cycle rows use the corrected physical-frame source mapping. Maximum saved mass and total-energy ledger residuals are $5.77\times10^{-14}$ and $7.18\times10^{-13}$, and maximum normalized magnetic divergence is $4.64\times10^{-14}$. Of 35{,}200 steps, 580 are first-order, with no timestep subdivisions; this fallback is part of the method and is reported, not hidden as an energy correction.

Both resolutions show amplification of the magnetic field and vertical magnetic stress during compression (Figure~\ref{fig:11}). At an initial column of $20\,M_\odot\,{\rm pc}^{-2}$, halving the initial cell width from 16 to 8 pc lowers the $\beta_0=10$ stellar mass (Figure~\ref{fig:12}) but raises the zero-field fixed-box stellar mass by a factor of 3.02. This reverses the ordering of the zero-field and magnetized fixed boxes, so we assign no resolution-independent sign to the magnetic effect on star formation. The initial-column and stronger-field cases show that a range of local conditions is accessible (Figure~\ref{fig:13}).

Switching from the earlier hydro solver to the MHD solver also changes the numerical scheme, as the zero-field control shows (Figure~\ref{fig:14}). This calls for matched solver controls when interpreting physics, not for discarding the local method.

\section{PRFM interfaces and earlier pressure diagnostics}\label{sec:6}

The simulation computes stellar masses and formation histories from its collapse, accretion and feedback rules. We evaluate PRFM afterwards using the saved simulation outputs.

We implement the main-text solar-neighborhood PRFM prescription of \citet{Jeffreson2026} through separate volumetric, integrated and measured-pressure interfaces. With $\Sigma_{\rm SFR}^{\rm PRFM}$ the predicted star-formation rate per unit area, $P_{\rm tot}$ the total vertical supporting pressure, $\Upsilon_{\rm tot}$ the total feedback yield (pressure divided by star-formation rate per unit area, with units of velocity), and $k_B$ Boltzmann's constant, the pressure/yield relation is

\begin{equation}
\Sigma_{\rm SFR}^{\rm PRFM}=P_{\rm tot}/\Upsilon_{\rm tot},\qquad\Upsilon_{\rm tot}=1030\,{\rm km\,s^{-1}}\left(\frac{P_{\rm tot}/k_B}{10^4\,{\rm K\,cm^{-3}}}\right)^{-0.21}.
\end{equation}

For MHD, the measured support includes the midplane thermal pressure $P_{\rm th}$, the vertical turbulent Reynolds stress $P_{\rm turb}$ and the vertical Maxwell stress:

\begin{equation}
P_{\rm tot}=P_{\rm th}+P_{\rm turb}+\frac{B_x^2+B_y^2-B_z^2}{2}.
\end{equation}

Using magnetic energy density instead of this stress would be a different diagnostic. Hydro post-processing also evaluates PRFM locally and after vertical integration, averaging over horizontal bins or matching column areas rather than inserting cloud peaks into a coarse-element law (Figure~\ref{fig:6}). For the 30 Myr MHD runs we evaluate the yield at the measured midplane stress. For the three principal cycling runs and their shared control we instead infer equilibrium pressure from the gas and stellar columns and stellar vertical motions of the simulation (Section~\ref{sec:pair_measurements}). Time averages of predictions are kept separate from predictions at averaged inputs.

\begin{figure}[!htbp]
\centering
\includegraphics[width=\textwidth,height=0.63\textheight,keepaspectratio]{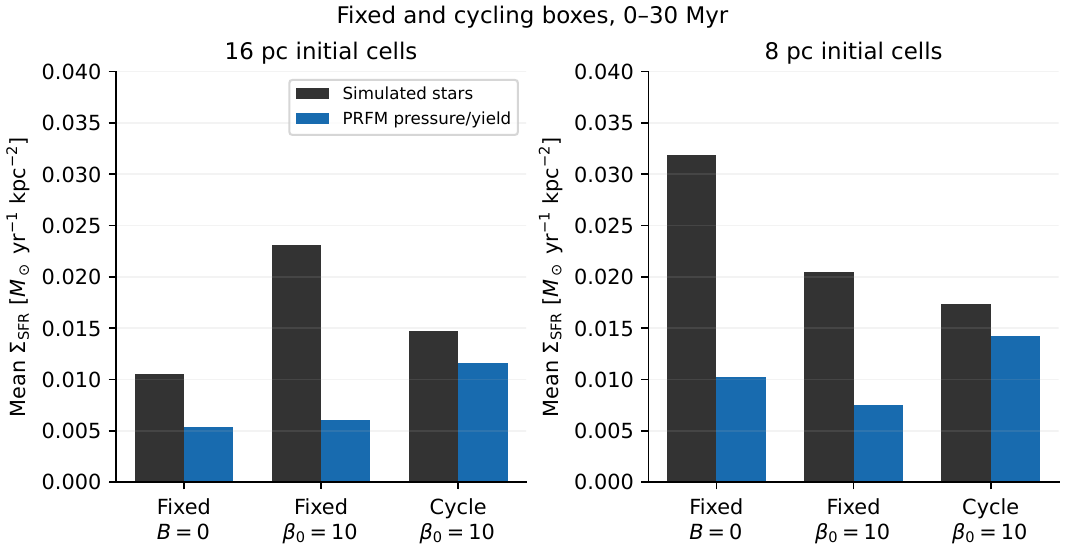}
\caption{\textbf{Mean star-formation rate per unit physical area over 0--30 Myr.} The cycling box is compressed and returned between 10 and 30 Myr. Black bars: mass added at birth and by accretion, each event's mass divided by the physical patch area at that time. Blue bars: time-averaged PRFM rate inferred from measured midplane thermal, Reynolds and Maxwell stresses. The native two-cell slab is 32 pc thick (left) and 16 pc thick (right); Appendix~\ref{sec:7.3} uses a common physical slab for the resolution comparison.\label{fig:15}}
\end{figure}

The PRFM implementation passes 17 checks, with a maximum relative residual of $4.55\times10^{-13}$ over 90 equilibrium cases. The volume branch keeps the published factor of two and low-pressure dispersion floor; the integrated branch by default uses the published dispersion fit without that floor, with a separate sensitivity option. For the auxiliary gas-height calculation used with the volumetric prescription, we solve the vertical-balance equation of \citet{Jeffreson2026} (their Eq.~8). Its cubic form has the squared effective velocity dispersion, $\sigma_{\rm eff}^2$, in the constant term, as in their Eq.~9. Their printed Eq.~27 omits the square. This is our implementation choice, not an audit of their Arepo code.

For a changing horizontal area, with $\Delta t$ the measurement-window duration and $\Delta M_{*,j}$ the mass of star-formation event $j$ (including accretion) at time $t_j$, the measured mean star-formation rate per unit area $\overline{\Sigma}_{\rm SFR}$ divides each event's mass by the physical patch area $A(t_j)$ at that time, sums the events and divides by the window duration:

\begin{equation}
\overline{\Sigma}_{\rm SFR}=\frac{1}{\Delta t}\sum_{j\in\Delta t}\frac{\Delta M_{*,j}}{A(t_j)}.
\end{equation}

Figure~\ref{fig:15} applies this diagnostic over 0--30 Myr; the measured-pressure prescription gives a lower mean surface rate than the simulation in all six runs.
\section{Earlier repeated-passage PRFM tests and numerical audits}\label{sec:8}

This experiment tests PRFM through three successive passages. Repeating the 20 Myr cycles with initially 8 pc cells tests the effect of halving the cell width from 16 pc while holding the physical sink and feedback lengths fixed. The four cycling runs use three uncompressed controls, one for each combination of resolution and perturbation seed. With initially 16 pc cells, the 20 and 40 Myr cycles with seed 7 share one control. The two other cycling runs (initial cell width 16 pc with seed 11, and 8 pc with seed 7) each have a separate control.

\subsection{A controlled history with independent local physics}\label{sec:8.1}

This experiment compares stellar mass formed over three compression--expansion cycles with a matched uncompressed control and with PRFM predictions for each run. The stratified domain starts with a gas column of 20 M\textsubscript{\ensuremath{\odot}} pc$^{-2}$ and \ensuremath{\beta}\textsubscript{0} = 10. A 30 Myr uncompressed preparation precedes three twofold x compression-and-return cycles (scale factor of Appendix~\ref{sec:5.1}, started at 30 Myr, period T = 20 or 40 Myr), with maximum compression at 40, 60 and 80 Myr (fast case) or 50, 90 and 130 Myr (slow case). Controls stay uncompressed over the same intervals.

Gas, magnetic fields, particles, accreted stellar cohorts, scheduled explosions and random-generator state all persist between passages, with no reset at $a=1$. The fine cycling run branches at 30 Myr from the control checkpoint (one cluster, 27 birth/accretion cohorts, 244 pending supernovae), so the pair differs only in the imposed cycles. Both resolutions share the same physical Fourier perturbations and phases; cell-wise random noise is not resampled.

The refinement test in Appendix~\ref{sec:7.3} varies source lengths with the mesh. The repeated-passage runs instead use fixed physical lengths. For sinks, a 16 pc sink length sets the collapse threshold, accretion draws on gas within 24 pc of a sink, new sinks are excluded within 48 pc of an existing one, and binding is evaluated over a 48 pc cube (integrated by cell-volume overlap on either mesh, also under deformation). Supernova injection, separately, uses a 48 pc minimum radius with 8 pc increments and a 64 pc cap. At maximum compression the 16 and 8 pc cells become 8 \ensuremath{\times} 16 \ensuremath{\times} 16 and 4 \ensuremath{\times} 8 \ensuremath{\times} 8 pc. The Courant factor is 0.24, with no altered physical clock or hot-gas timestep shortcut. Source prescriptions thus match more closely under refinement, but the coupled collapse and feedback histories may still differ.

The local source model, though independent of PRFM, is not exact. Collapse and accretion follow the excess-density rule (Appendix~\ref{sec:4}), moving qualifying cold, bound gas directly into a cluster without a separate star-formation efficiency per free-fall time. The uniform heating proxy counts stellar mass added by birth or accretion in the preceding 10 Myr, with an abrupt age cutoff, and sampled supernova delays span 3--40 Myr. These fixed choices can produce sharply concentrated birth episodes and explicit memory times. They remain part of the interpretation of this experiment, and the MHD burst widths are not calibrated stellar-population predictions.

\subsection{Independent PRFM predictions and measurement definitions}\label{sec:8.2}

The primary comparator is the published solar-neighborhood PRFM prescription \citep{Jeffreson2026}, solved for its own equilibrium pressure, thickness, velocity dispersion, feedback yield and star-formation rate from the MHD gas and stellar columns, the young-star vertical velocity moments and the prescribed old-star and halo environment. Given neither measured MHD pressure nor measured star-formation rate, it tests the instantaneous closure on the matched gas and stellar columns and stellar vertical motions from the simulation. A separate autonomous PRFM column evolves its own gas consumption and population-mean mass return (10\% returned uniformly over ages 3--40 Myr), and new stars keep the predicted effective dispersion. This column has no explicit wind, magnetic field or supernova energy injection. Volumetric PRFM on fixed 16 pc horizontal slabs is a supplementary frozen-state diagnostic.

Eligibility uses the mean gas density of the full 1024 pc column and the published threshold of $0.13$ H cm$^{-3}$. Simulated totals count stellar mass added by birth and subsequent accretion after preparation; PRFM totals integrate $A(t)\Sigma_{\rm SFR}(t)$ over the same interval. The initial PRFM dynamical time $t_{\rm dyn}$ (Appendix~\ref{app:prfm_spec}) is 37.5 Myr; over three 20 or 40 Myr cycles, star formation and outflow deplete the gas available to later passages.

\subsection{Stellar mass over three passages}\label{sec:8.3}

\begin{table}[!htbp]
\caption{Stellar mass over three consecutive 20 Myr passages, 30--90 Myr ($M_\odot$). Each row compares one cycling run with the uncompressed run at the same resolution and seed. Both PRFM forms use those same two histories.\label{tab:15}}
\centering\footnotesize\setlength{\tabcolsep}{4pt}\renewcommand{\arraystretch}{1.2}
\begin{tabular}{>{\raggedright\arraybackslash}p{\dimexpr 0.22\textwidth-2\tabcolsep\relax}>{\raggedright\arraybackslash}p{\dimexpr 0.13\textwidth-2\tabcolsep\relax}>{\raggedright\arraybackslash}p{\dimexpr 0.13\textwidth-2\tabcolsep\relax}>{\raggedright\arraybackslash}p{\dimexpr 0.13\textwidth-2\tabcolsep\relax}>{\raggedright\arraybackslash}p{\dimexpr 0.13\textwidth-2\tabcolsep\relax}>{\raggedright\arraybackslash}p{\dimexpr 0.13\textwidth-2\tabcolsep\relax}>{\raggedright\arraybackslash}p{\dimexpr 0.13\textwidth-2\tabcolsep\relax}}
\hline
\textbf{Initial cell width and seed} & \textbf{MHD cycle} & \textbf{MHD control} & \textbf{PRFM cycle} & \textbf{PRFM control} & \textbf{Volumetric PRFM cycle} & \textbf{Volumetric PRFM control} \\ \hline
16 pc, seed 7 & 46{,}300 & 99{,}600 & 44{,}500 & 33{,}400 & 132{,}000 & 92{,}500 \\
16 pc, seed 11 & 101{,}000 & 217{,}000 & 36{,}500 & 29{,}500 & 147{,}000 & 95{,}700 \\
8 pc, seed 7 & 359{,}000 & 270{,}000 & 29{,}800 & 21{,}800 & 97{,}900 & 55{,}100 \\
\hline
\end{tabular}
\end{table}

\begin{figure}[!htbp]
\centering
\includegraphics[width=\textwidth,height=0.63\textheight,keepaspectratio]{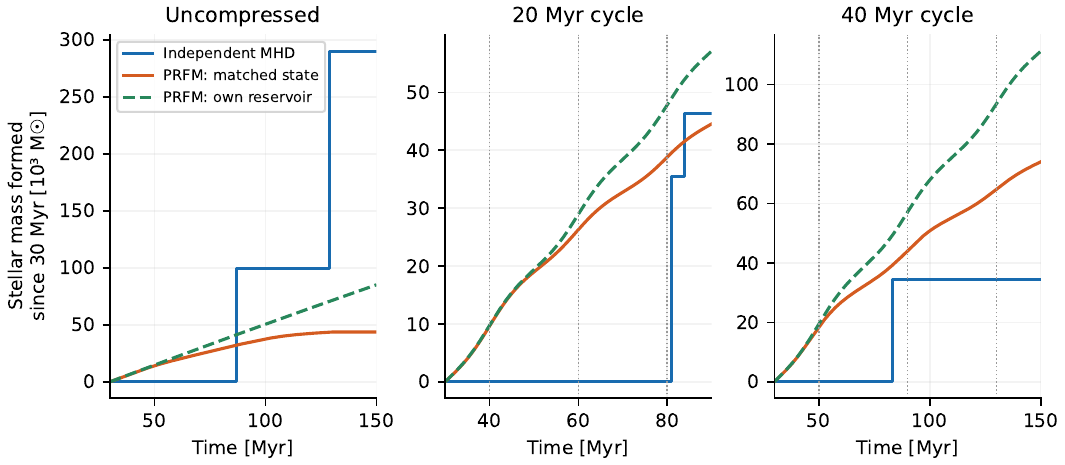}
\caption{\textbf{Stellar production across repeated passages.} Cumulative stellar mass (birth plus subsequent accretion) after the 30 Myr preparation, for the seed-7 uncompressed control with 16 pc initial cells and the 20 and 40 Myr cycles; PRFM uses the matched gas and stellar columns and stellar vertical motions from the simulation, or its own independently evolving reservoir. Dotted lines mark maximum compression. The same control supplies the 30--90 Myr comparison for the 20 Myr cycles and the 30--150 Myr comparison for the 40 Myr cycles. At both durations the cycling run forms less stellar mass than the control over the corresponding interval. Matched-state PRFM predicts 96\% of the simulated cycling mass and 34\% of the control mass over 30--90 Myr, and 214\% and 15\% over 30--150 Myr.\label{fig:21}}
\end{figure}

\begin{figure}[!htbp]
\centering
\includegraphics[width=\textwidth,height=0.63\textheight,keepaspectratio]{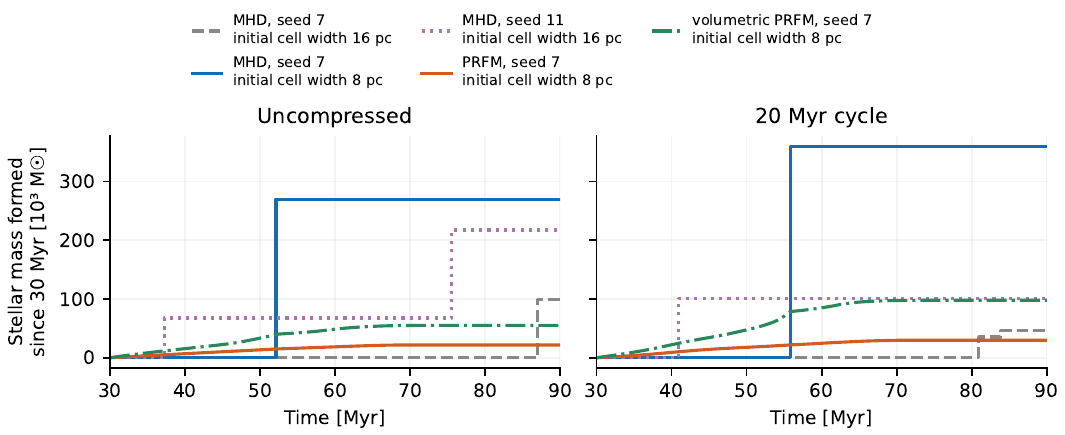}
\caption{\textbf{Cumulative stellar mass in the uncompressed controls and repeated 20 Myr cycles.} Two realizations start with 16 pc cells and one with 8 pc cells. Masses include births and subsequent accretion after 30 Myr. Both PRFM predictions use histories from the finer grid with seed 7.\label{fig:22}}
\end{figure}

Table~\ref{tab:15} lists the simulated and predicted totals; both prescriptions use the published coefficients, and Figures~\ref{fig:21} and~\ref{fig:22} show how they accumulate.

With initially 16 pc cells, each seed's repeated 20 Myr cycles form about 47\% of the stellar mass of its uncompressed control (Table~\ref{tab:15}). Matched-state PRFM predicts 96\% of the simulated cycling mass and 34\% of the control mass for seed 7, and 36\% and 14\% for seed 11. Volumetric PRFM predicts 286\% and 93\% for seed 7, and 145\% and 44\% for seed 11. Both forms therefore err very differently in each cycling run and in its control.

Over three 40 Myr cycles (30--150 Myr), the cycling run and control form $34{,}500\,M_\odot$ and $290{,}000\,M_\odot$. Matched-state PRFM overpredicts the cycling run ($74{,}000\,M_\odot$) and underpredicts the control ($43{,}700\,M_\odot$). Volumetric PRFM gives $331{,}000$ and $130{,}000\,M_\odot$.

\begin{figure}[!htbp]
\centering
\includegraphics[width=\textwidth,height=0.63\textheight,keepaspectratio]{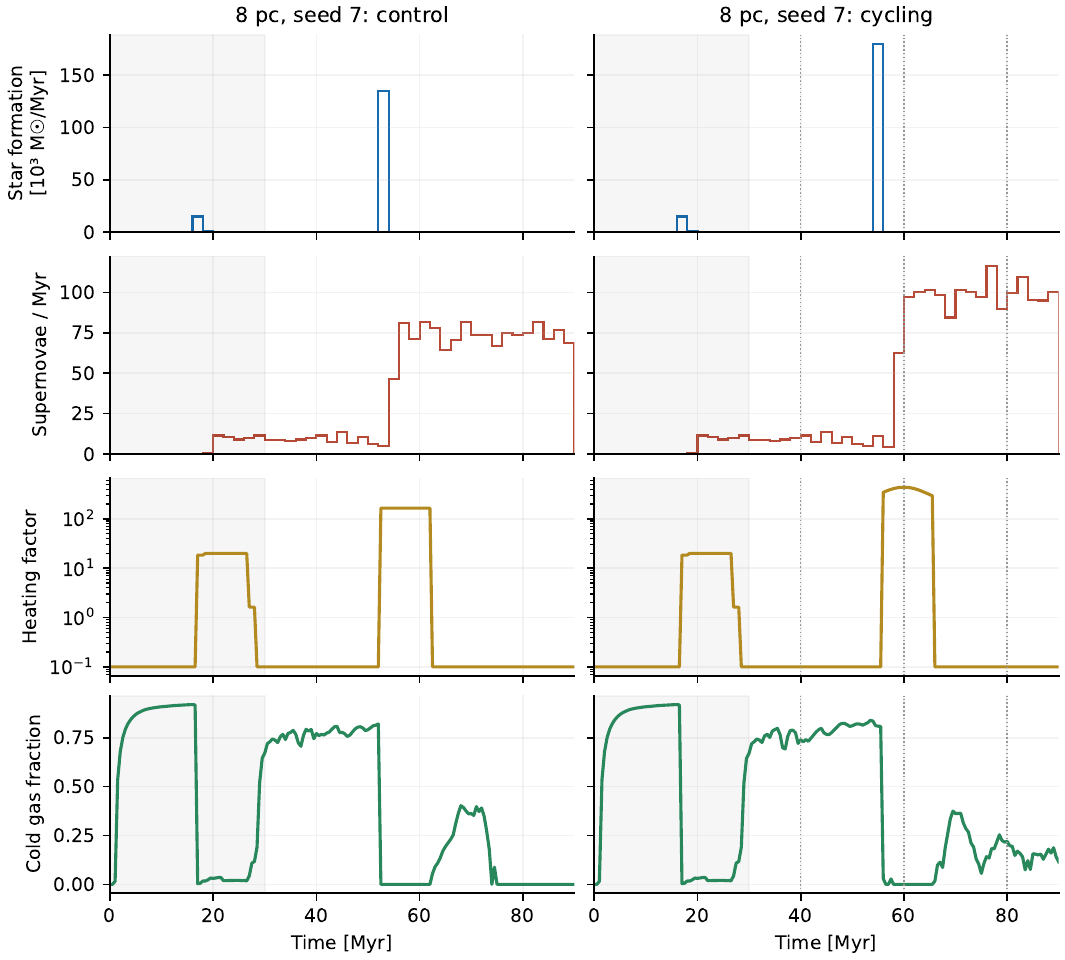}
\caption{\textbf{Feedback retained between passages.} Simulated star-formation rate (birth plus subsequent accretion), supernovae, young-star heating and cold gas for the repeated 20 Myr cycles and their uncompressed control, both with an initial cell width of 8 pc. Stars, cohorts and scheduled explosions survive each return to the original box size.\label{fig:23}}
\end{figure}

With source lengths held fixed, halving the initial cell width to 8 pc reverses the simulated cycle effect from a deficit to a 33\% excess (Table~\ref{tab:17}). On the finer grid, matched-state PRFM predicts about 8\% of the simulated stellar mass in both the cycling run and the control, and volumetric PRFM predicts 27\% of the simulated cycling mass and 20\% of the control mass. The autonomous PRFM column gives 16\% of the simulated cycling mass.

Figure~\ref{fig:23} traces star formation, supernovae, heating and the cold-gas fraction across the passages. At 90 Myr the fine cycling and control runs retain 0.87\% and 0.13\% of the initial gas as gas and have lost 71.9\% and 79.0\% through vertical outflow. Stars account for the remainder in the explicit mass ledger, which includes returned ejecta. Later passages therefore test an evolved finite reservoir, not a replenished, stationary star-forming disk.

\begin{table}[!htbp]
\caption{Stellar-mass sensitivity to spatial refinement ($M_\odot$, birth plus subsequent accretion).\label{tab:17}}
\centering\footnotesize\setlength{\tabcolsep}{4pt}\renewcommand{\arraystretch}{1.2}
\begin{tabular}{>{\raggedright\arraybackslash}p{\dimexpr 0.28\textwidth-2\tabcolsep\relax}>{\raggedright\arraybackslash}p{\dimexpr 0.26\textwidth-2\tabcolsep\relax}>{\raggedright\arraybackslash}p{\dimexpr 0.26\textwidth-2\tabcolsep\relax}>{\raggedright\arraybackslash}p{\dimexpr 0.2\textwidth-2\tabcolsep\relax}}
\hline
\textbf{Refinement case} & \textbf{Mass, initially 16 pc cells} & \textbf{Mass, initially 8 pc cells} & \textbf{Change} \\ \hline
control & 99{,}600 & 270{,}000 & +171\% \\
cycle & 46{,}300 & 359{,}000 & +676\% \\
\hline
\end{tabular}
\end{table}

Two resolutions over the same 30--90 Myr interval cannot determine a convergence order for a stochastic, nonlinear stellar population; exact numerical equivalence and smooth convergence were tested separately in Appendix~\ref{sec:3}.

Each cycling/control pair shares its preparation, but the two resolutions enter the cycles in different states. By 30 Myr, the preparation had formed 45{,}500 M\textsubscript{\ensuremath{\odot}} on the grid with initially 16 pc cells and 32{,}500 M\textsubscript{\ensuremath{\odot}} on the grid with initially 8 pc cells ($-28.4$\%). By then, the runs with initial cell widths of 16 and 8 pc had produced 32 and 106 supernovae, respectively, despite the smaller fine-grid mass. Their cold-gas mass fractions were 0.000 and 0.671. Under the same event-delay prescription, the independently formed populations are thus at different stages of their feedback history. The comparison therefore refines the complete preparation-plus-response problem; it is not a replay of one state on two meshes.

Old stars supply a fixed potential but have no measured velocities; the baseline PRFM input gives them 12.9 km s$^{-1}$ from the isolated stellar sheet. A sensitivity calculation including the initial gas and halo in the mass-weighted vertical Jeans relation gives 17.9 km s$^{-1}$, held fixed. This lowers the predicted masses on the 16 pc grid by 10--12\%. The halo follows the inner Navarro--Frenk--White approximation of \citet[][Section 2.2]{Jeffreson2026}: $\Omega_d^2=2\pi G\rho_d$, with $\Omega_d$ the effective halo frequency and $\rho_d$ the local halo density.

\begin{figure}[!htbp]
\centering
\includegraphics[width=\textwidth,height=0.63\textheight,keepaspectratio]{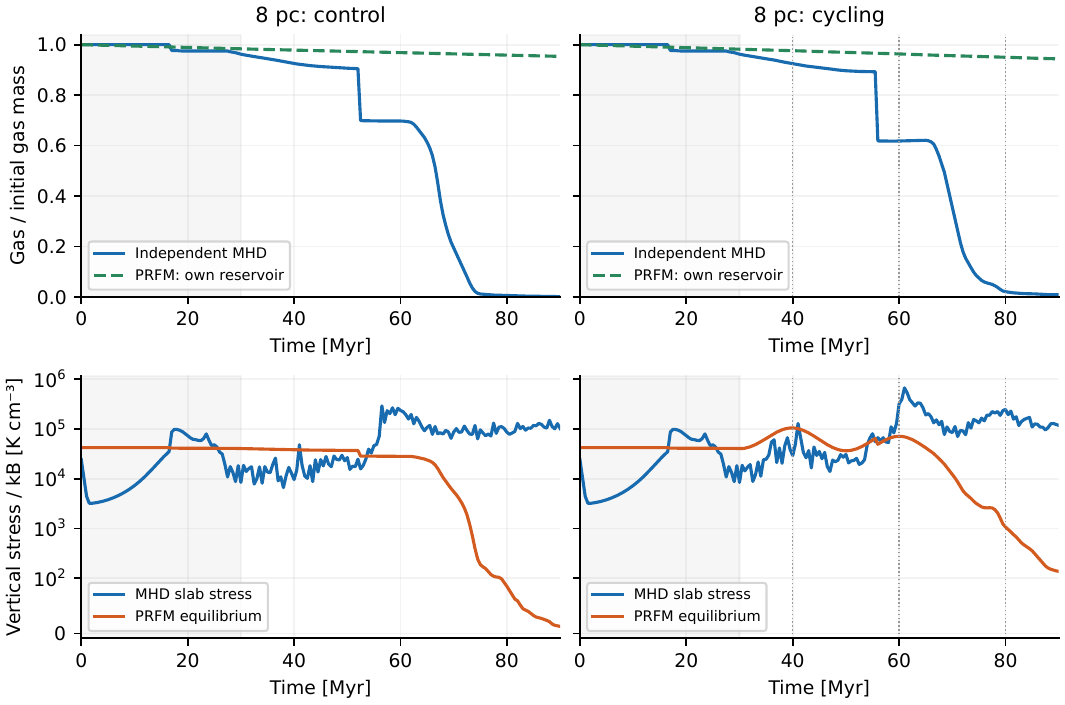}
\caption{\textbf{Reservoir evolution and vertical stress.} Retained gas and central |z| < 16 pc stress (thermal, turbulent and vertical Maxwell terms) in the repeated 20 Myr cycles and their fixed-box control, both with an initial cell width of 8 pc. The figure shows the measured stress alongside the equilibrium pressure inferred by PRFM from the simulated gas and stellar columns and stellar vertical motions.\label{fig:24}}
\end{figure}

\subsection{Energy correction, accuracy checks and computational accounting}\label{sec:8.4}

The additional-seed cycle with 16 pc initial cells exposed a reproducible failure at 64.72 Myr in a low-thermal-energy cell at the lower open vertical boundary, not at the moving periodic seam. On the same frozen input, the finite-volume induction update kept the thermal energy positive, but the constrained-transport field made it negative. If ordinary second- and first-order attempts fail, an optional first-order retry now adds $(B_{\rm CT}^2-B_{\rm FV}^2)/2$ to the total energy density, where $B_{\rm CT}$ and $B_{\rm FV}$ are the cell-centered constrained-transport and finite-volume field magnitudes. The correction is applied only in cells with nonpositive thermal energy and only if the finite-volume thermal state is positive. The correction imposes no pressure floor, and the Runge--Kutta-weighted correction energy is logged separately. The magnetic-energy consistency idea follows \citet{MignoneDelZanna2021}; the selective fallback policy is specific to this implementation.

The repaired case matches the saved histories through 64.5 Myr and completes with 10 cell-stage corrections totaling $1.30\times10^{-9}$ of its injected supernova energy. Frozen-failure, repeated-domain, inactive-branch and restart tests pass, and earlier cases without an accepted subdivision never reach the new branch. The fine cycle needs 63 corrections (96.3 code energy units, $5.70\times10^{-10}$ of its injected supernova energy), the fine control none, and neither fine run adds mass or energy through floors. Per-case correction counts and ledger residuals are archived.

Across completed cases the maximum normalized magnetic divergence is $3.61\times10^{-5}$, and the maximum absolute relative mass and energy ledger residuals are $5.55\times10^{-14}$ and $1.51\times10^{-8}$; the fine control thus crosses the original $10^{-8}$ energy-accounting target. Frozen cooling audits at 30 and 35 Myr isolate accumulation roundoff: compensated bookkeeping changes the recorded sum but no gas-state byte. A separate, explicit $10^{-6}$ long-run failure guard is applied; production cooling and thermodynamic evolution were not changed to improve the ledger.

The fine control also misses the original $10^{-10}$ instantaneous-field-normalized divergence target, first in the saved 80 Myr output. Checkpoints at 70, 75 and 80 Myr share the maximum absolute divergence $1.74\times10^{-13}$ \ensuremath{\mu}G pc$^{-1}$, so the rising ratio reflects the declining field in its denominator. At the final checkpoint the absolute error is still $1.74\times10^{-13}$ \ensuremath{\mu}G pc$^{-1}$, or $8.55\times10^{-14}$ when normalized by the shared 30 Myr field scale.

The fine cycle crosses the same relative divergence target in its 81.5 Myr saved output. Flux freezing maps divergence as 1/a, with a the volume Jacobian of the deformation, so the appropriate geometric comparison of the absolute residual is a \ensuremath{\times} max|\ensuremath{\nabla}\ensuremath{\cdot}B|. Its frozen 75 and 80 Myr states both give $5.875\times10^{-14}$ in code units (four significant figures; the unrounded values differ fractionally by less than $2\times10^{-12}$). At 90 Myr, back at $a=1$, the absolute divergence is $1.71\times10^{-13}$ \ensuremath{\mu}G pc$^{-1}$. These supplementary comparisons do not replace the original criteria: all three missed targets keep their pass/fail flags, and no magnetic cleaning or field reset was introduced.

The long fine calculations use four workers each for independent transport planes, preserving accumulation order and random sampling. Serial and threaded gas/field outputs agree byte for byte on frozen benchmarks, and history prefixes match through 34 Myr (control) and 35 Myr (cycle). After an interruption, the fine cycle restarted from 35 Myr, and its replay through 35.5 Myr matches all five archived histories byte for byte. The method's cost is measured by the matched serial benchmark of Appendix~\ref{sec:7.5}, not by these concurrent runs.

\section{Diagnostics with prescribed compression and expansion}\label{app:cycle_diagnostics}

These diagnostics are distinct from the uncompressed cloud-assembly tests. The energy and feedback histories compare the 20 Myr passage with the first 20 Myr of the shared control after branching. The FUV intervention changes only the continued radiation from one early cluster during the expansion branch.

Table~\ref{tab:expansion_gradient} measures the instantaneous effect of the outward gradient in saved cycling states. In physical coordinates $\partial v_x/\partial x=H+\partial w_x/\partial x$, where $w_x$ is the residual radial velocity. The radial convergence test keeps each state's density, residual velocity field, gravitational potential and magnetic field and varies only $H$.

In the 20 Myr passage, omitting the gradient raises the converging fraction at all three snapshots. At 84.5 Myr in the 50 Myr passage, the search domain holds only $0.50\,M_\odot$ of cold gas, none of it converging in all three directions in any trial. At 134.5 Myr in the 100 Myr passage, omitting the gradient adds $162\,M_\odot$ of converging cold mass. In all five snapshots, no cell above the sink density threshold is a local potential minimum, so every trial gives zero instantaneous sink candidates.

\begin{figure}[!htbp]
\centering
\includegraphics[width=\textwidth,height=0.63\textheight,keepaspectratio]{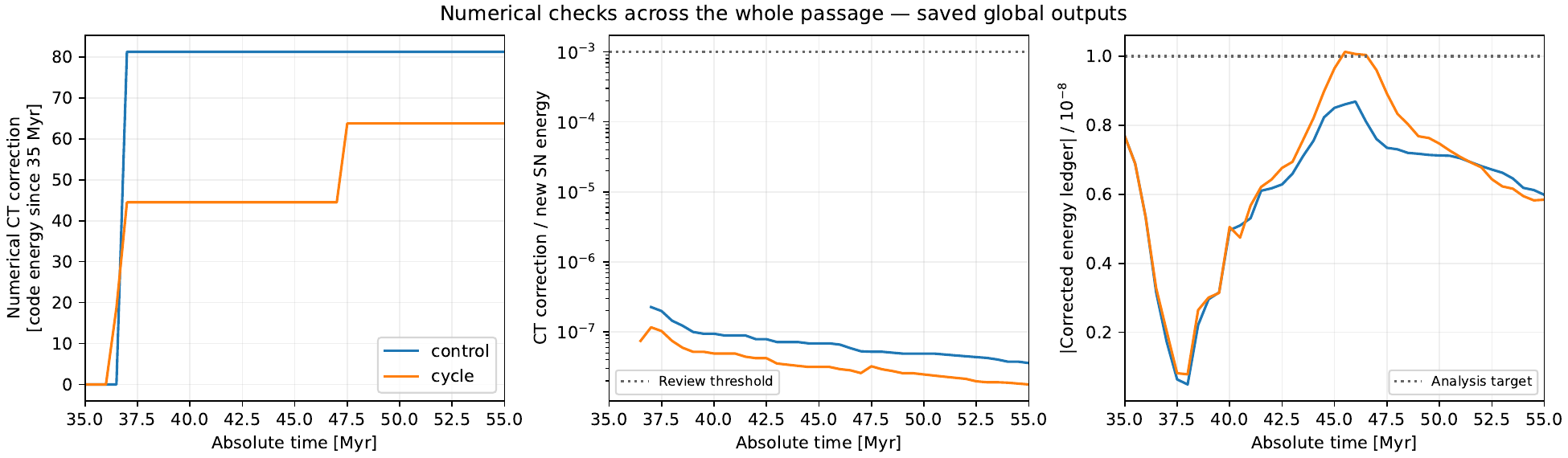}
\caption{\textbf{Energy history of the matched first passage.} The box contracts and expands over 35--55 Myr; the control keeps its original width. The ledger residual and the recorded constrained-transport energy correction are separate diagnostics. The cycle's ledger residual reaches $1.01\times10^{-8}$, above the tighter analysis target but below the integration guard.\label{fig:38}}
\end{figure}

\begin{figure}[!htbp]
\centering
\includegraphics[width=\textwidth,height=0.63\textheight,keepaspectratio]{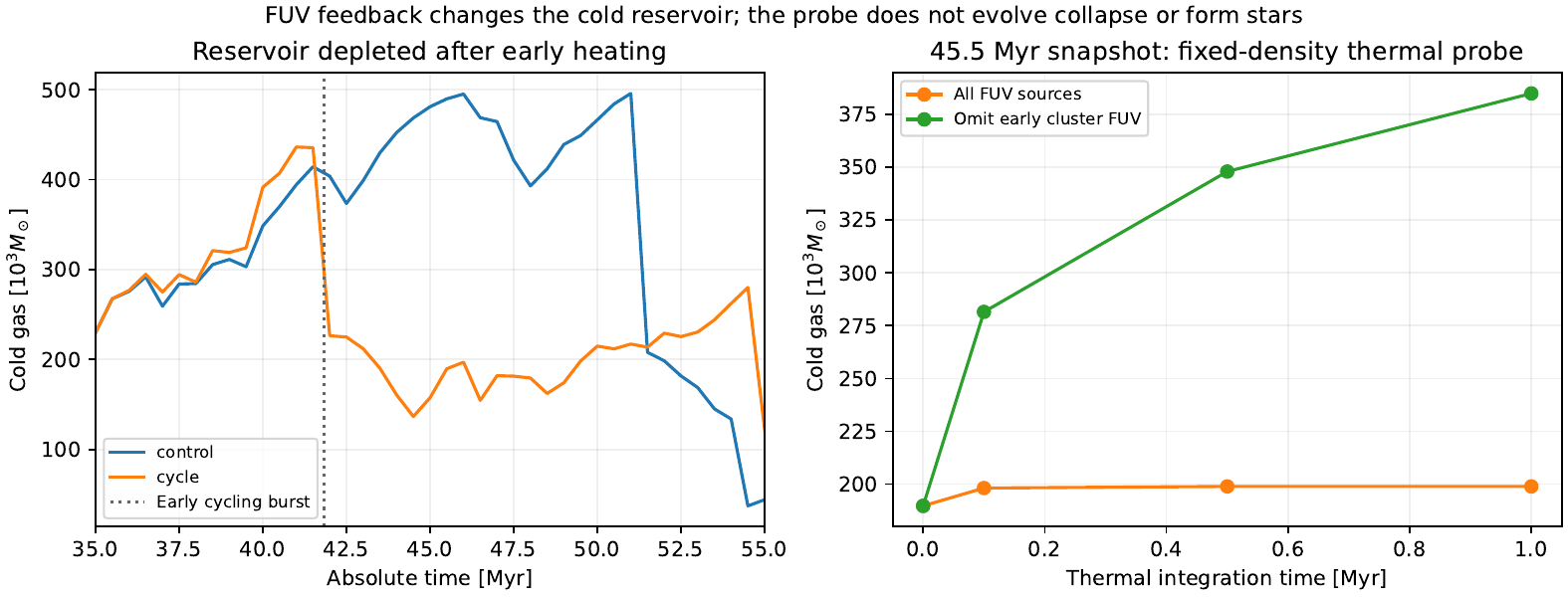}
\caption{\textbf{Cold gas during the prescribed cycle.} Cold gas and feedback in the 20 Myr cycling run and the uncompressed control around the early burst and expansion of the 35--55 Myr passage. These histories motivate a separate test of continued FUV from the early cluster.\label{fig:40}}
\end{figure}

Section~\ref{sec:9} presents the 20 Myr cycling run, the control and the matched FUV intervention, and Figure~\ref{fig:39} (Section~\ref{sec:fuv}) shows the feedback history. Figures~\ref{fig:38} and~\ref{fig:40} add the saved energy-correction and cold-gas diagnostics.

\section{Data provenance}\label{app:provenance}

Table~\ref{tab:20} identifies the records underlying the numerical tests and comparisons. Readers interested in the simulation data or analysis code may contact the author.

\begin{columntableblock}
\caption{Simulation and verification records and the sections reporting their results.\label{tab:20}}
\centering\footnotesize\setlength{\tabcolsep}{4pt}\renewcommand{\arraystretch}{1.2}
\begin{tabular}{>{\raggedright\arraybackslash}p{\dimexpr 0.42\textwidth-2\tabcolsep\relax}>{\raggedright\arraybackslash}p{\dimexpr 0.58\textwidth-2\tabcolsep\relax}}
\hline
\textbf{Evidence} & \textbf{Record and discussion} \\ \hline
Reduced equation audit & Physical/residual comparison suite; Appendix~\ref{sec:3.1} \\
Spatial disturbance-mode audit & Mode comparison records; Appendix~\ref{sec:3.2} \\
Revised hydro suite & Uncompressed hydrodynamic run records; Appendix~\ref{sec:4} \\
Adiabatic remnant / hot cooling & Remnant and cooling verification; Appendix~\ref{sec:4.2} \\
Final MHD component verification & Transport, geometry and source checks; Appendix~\ref{sec:3.3} \\
Uniform flux-freezing cycle & Analytic compression test; Appendix~\ref{sec:3.3} \\
Selected uncompressed MHD runs & Fixed-box histories and measured-support PRFM diagnostic; Appendices~\ref{sec:5} and~\ref{sec:6} (Figure~\ref{fig:15}) \\
Selected MHD cycling runs & Histories with corrected source mapping; Appendix~\ref{sec:5} \\
Field-strength and cycling runs & Paired outcomes, seeds, birth events and reproduction checks; Section~\ref{sec:field_strength} \\
Static source regression & Source-mapping regression; Appendix~\ref{sec:3.3} \\
PRFM formula checks & Formula and equilibrium checks; Appendix~\ref{sec:6} \\
MHD refinement and cost through contraction and return & Refinement and matched-cost records; Appendix~\ref{sec:7} \\
Default verification repeat & Repeated verification suite; Appendix~\ref{sec:7.4} \\
Repeated mode / PRFM / input audits & Repeated-passage records; Appendix~\ref{sec:8} \\
Uncompressed control and 20 Myr passage & First-passage comparison report; Section~\ref{sec:9.2} \\
Passage duration and convergence & Full-passage masses and fixed-state convergence tests in Tables~\ref{tab:period_response} and~\ref{tab:expansion_gradient} \\
Matched FUV intervention & Paired intervention completion report; Section~\ref{sec:fuv} \\
\hline
\end{tabular}
\end{columntableblock}

Protocols, initial-state records, software versions and completion reports cover the 30 Myr stratified runs, the repeated passages, the three principal cycling runs and their shared 35--135 Myr control, and the 45.5--55 Myr FUV intervention. Frozen run sources take precedence over later working files, and Appendix~\ref{app:principal_spec} gives the principal equations and source choices. The FUV reference run reproduces the original final physical and random-number-generator fields, apart from documented C++ padding. Interpreting or restarting it requires its source stream and intervention sidecar files.

Of the 31 figures, ten are in the main text (including the field-geometry schematic) and 21 in the appendices. Publication figures have rounded labels, physical descriptions instead of internal names, a compact appendix layout and a three-panel cost redraw. The feedback figure for the 20 Myr cycling run and the control (Figure~\ref{fig:39}) adds the saved vertical Maxwell stress, using the same phase and midplane selection as its thermal and Reynolds stresses. Plotted values are unchanged apart from the declared rebinning of FUV rates from the frozen events, which leaves event times and stellar masses unchanged. Display copies, regeneration scripts and receipts, including the FUV bin edges, document these changes.

The source/cost benchmark uses the corrected v2 fixture. The superseded fixture omitted an open-boundary normal-field face and failed the divergence test despite agreement between domains; it is not the reported validation. The archived stratified cycling runs use the corrected source mapping. The run records distinguish earlier versions of the cycling runs and the magnetic-diode test. State norms keep their declared units and are not combined into one accuracy measure.
\section{Uncompressed hydrodynamic results}\label{supp:hydro}

Figures~\ref{fig:6}--\ref{fig:8} show the fixed-box hydro runs in Appendix~\ref{sec:4}.

\begin{appendixfigure}
\centering
\includegraphics[width=\textwidth,height=0.53\textheight,keepaspectratio]{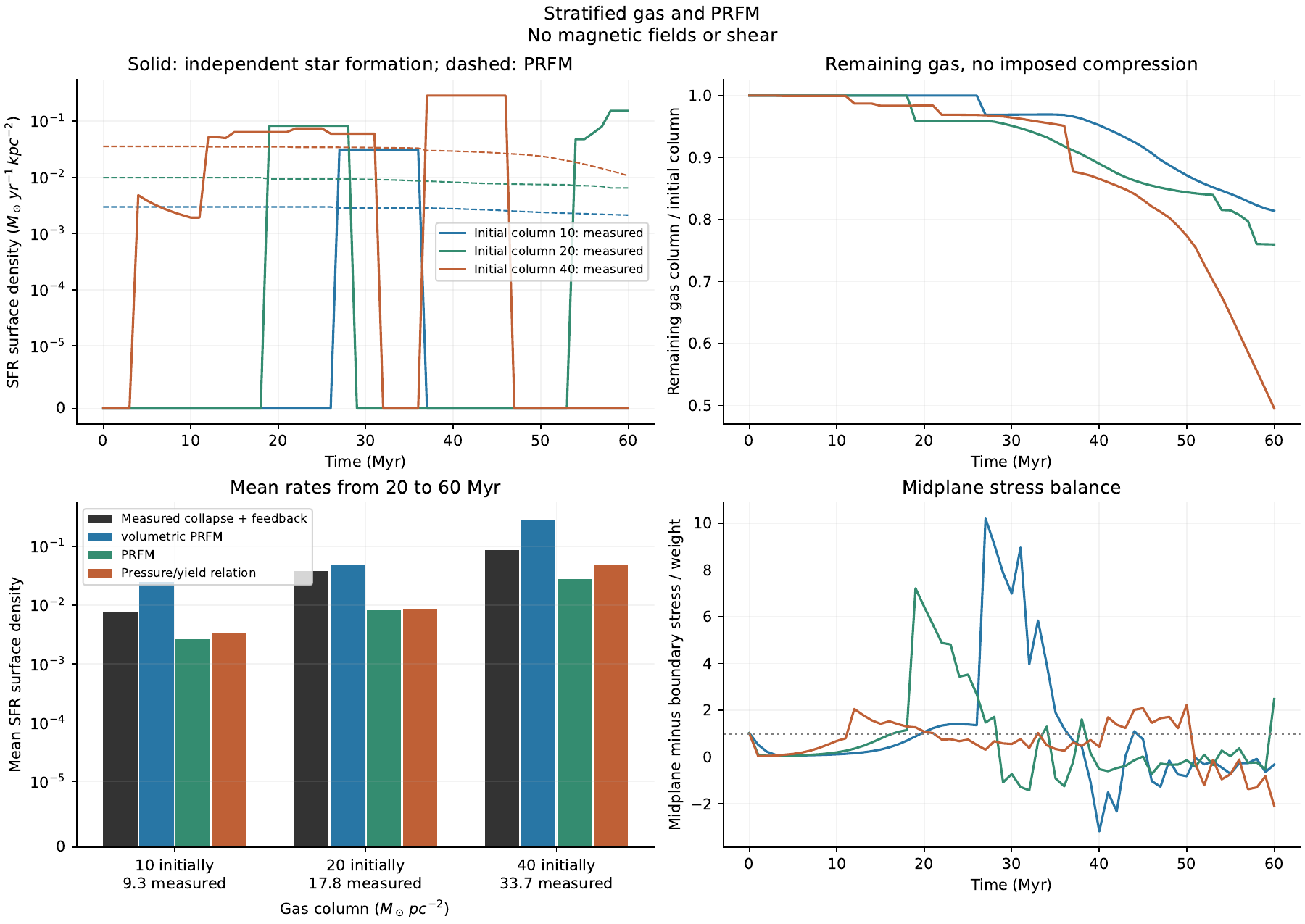}
\caption{\textbf{Hydrodynamic simulations and PRFM predictions.} Top: star formation and remaining gas at three initial columns. Lower left: measured 20--60 Myr mean rates against volumetric PRFM, integrated PRFM and measured pressure divided by feedback yield. Lower right: vertical stress against gas weight.\label{fig:6}}
\end{appendixfigure}

\begin{appendixfigure}
\centering
\includegraphics[width=\textwidth,height=0.63\textheight,keepaspectratio]{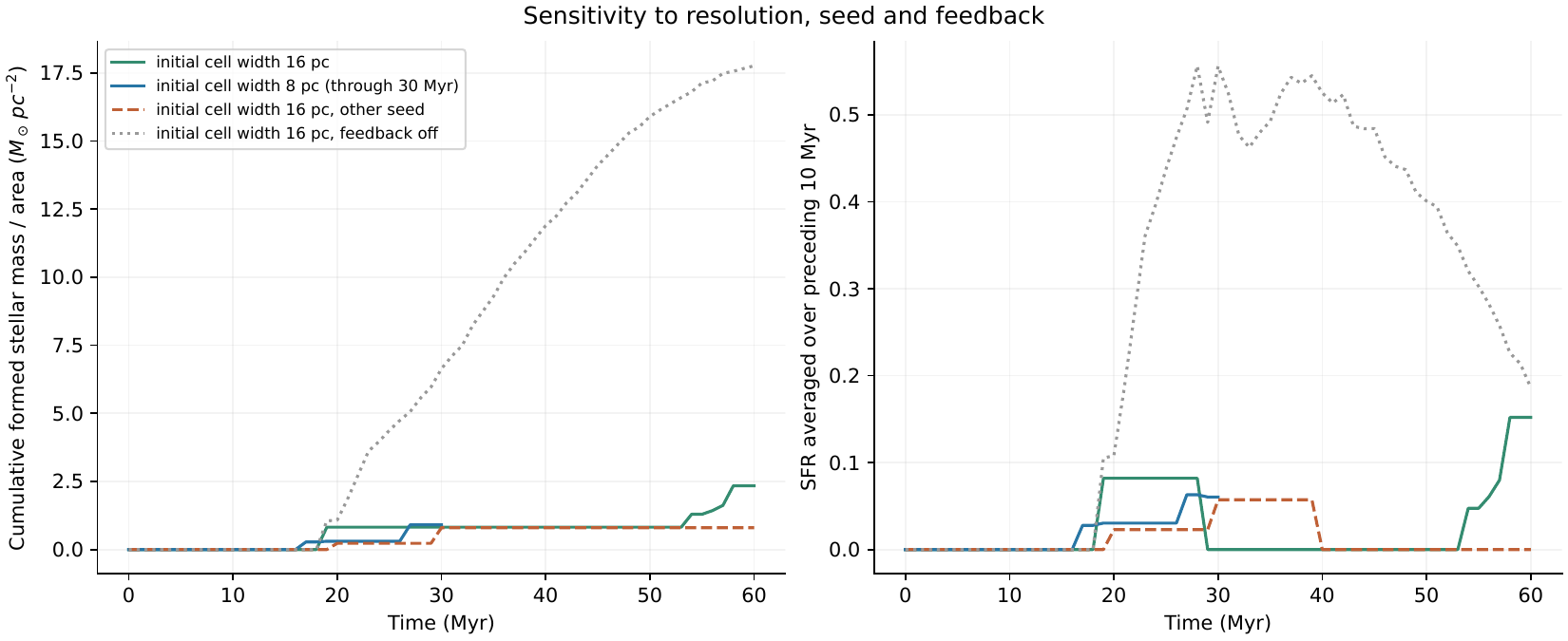}
\caption{\textbf{Resolution, seed and feedback in the hydrodynamic runs.} At the central gas column, the mass formed by 30 Myr differs by 10.8\% when the cell width is halved from 16 to 8 pc. The figure also shows a second perturbation seed with 16 pc initial cells, followed to 60 Myr, and a feedback-off control.\label{fig:7}}
\end{appendixfigure}

\begin{appendixfigure}
\centering
\includegraphics[width=\textwidth,height=0.63\textheight,keepaspectratio]{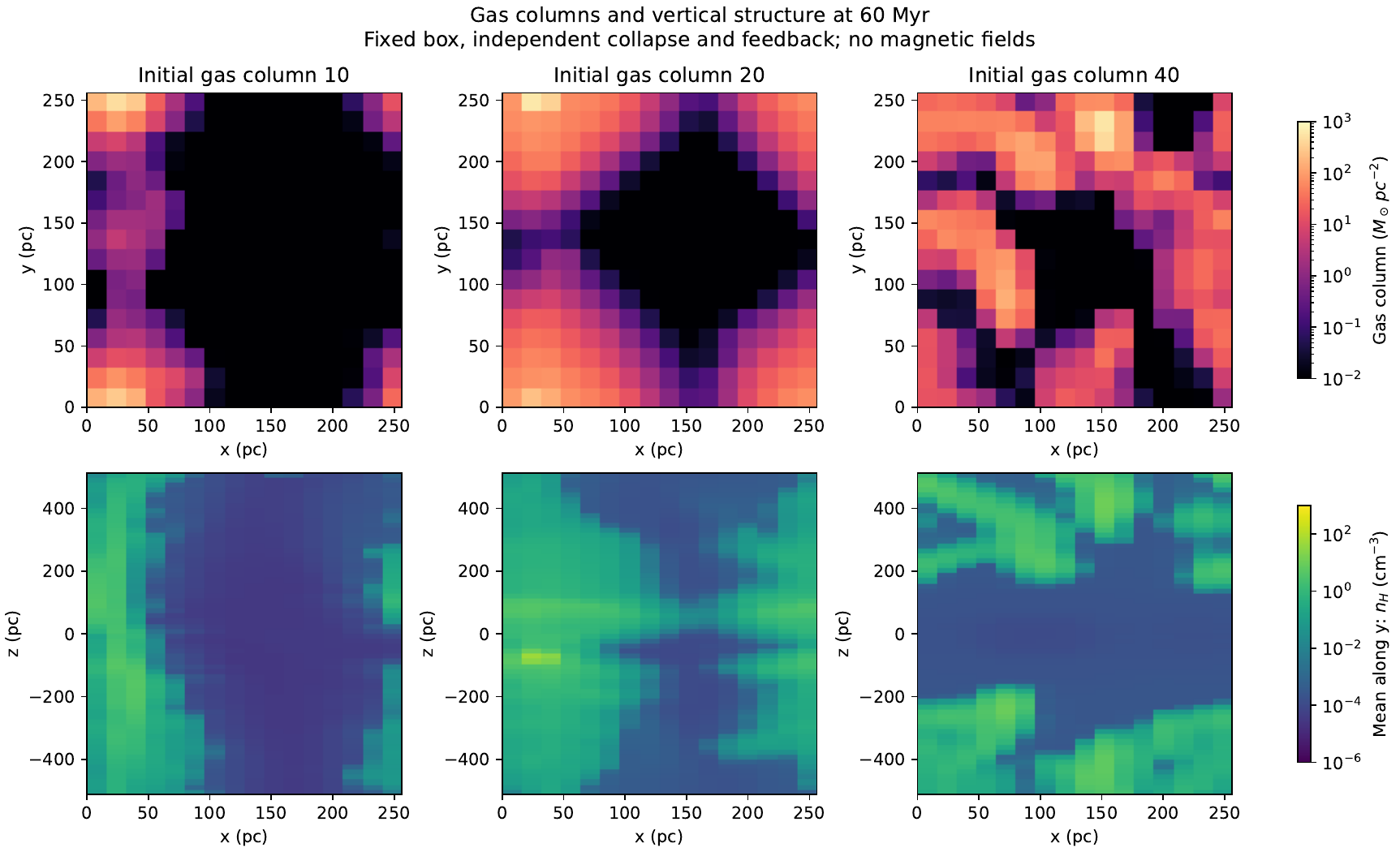}
\caption{\textbf{Gas distribution in the uncompressed hydrodynamic runs at 60 Myr.} Upper panels: gas column density; lower panels: density averaged along $y$. Each fixed box evolves with cooling, self-gravity and stellar feedback.\label{fig:8}}
\end{appendixfigure}

\FloatBarrier
\section{Uncompressed initial-column and field-strength experiments}\label{supp:field}

\begin{appendixfigure}
\centering
\includegraphics[width=0.92\textwidth,height=0.63\textheight,keepaspectratio]{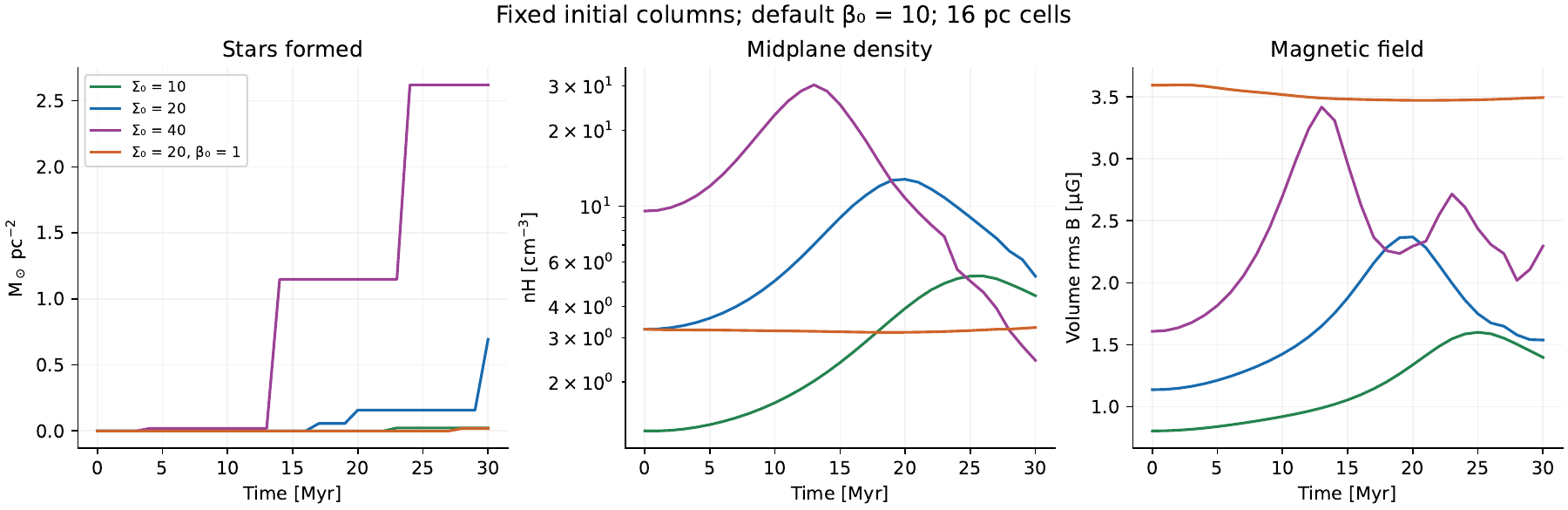}
\caption{\textbf{Magnetized fixed-box runs.} The fixed boxes with 16 pc cells start from gas columns of 10, 20 and $40\,M_\odot\,{\rm pc}^{-2}$ with $\beta_0=10$; the $20\,M_\odot\,{\rm pc}^{-2}$ case also runs with $\beta_0=1$. Each evolves for 30 Myr (Appendix~\ref{sec:5}).\label{fig:13}}
\end{appendixfigure}

\FloatBarrier
\section{Solver-matching control for the spiral-shock precursor}\label{supp:solver_control}

This isothermal, rotating, one-dimensional shock calculation shows a prerequisite for a coordinate comparison: both representations must use the same hydrodynamic method. It is a solver control for the independent physical/residual verification of Appendix~\ref{sec:3.1}.

\begin{appendixfigure}
\centering
\includegraphics[width=\textwidth,height=0.38\textheight,keepaspectratio]{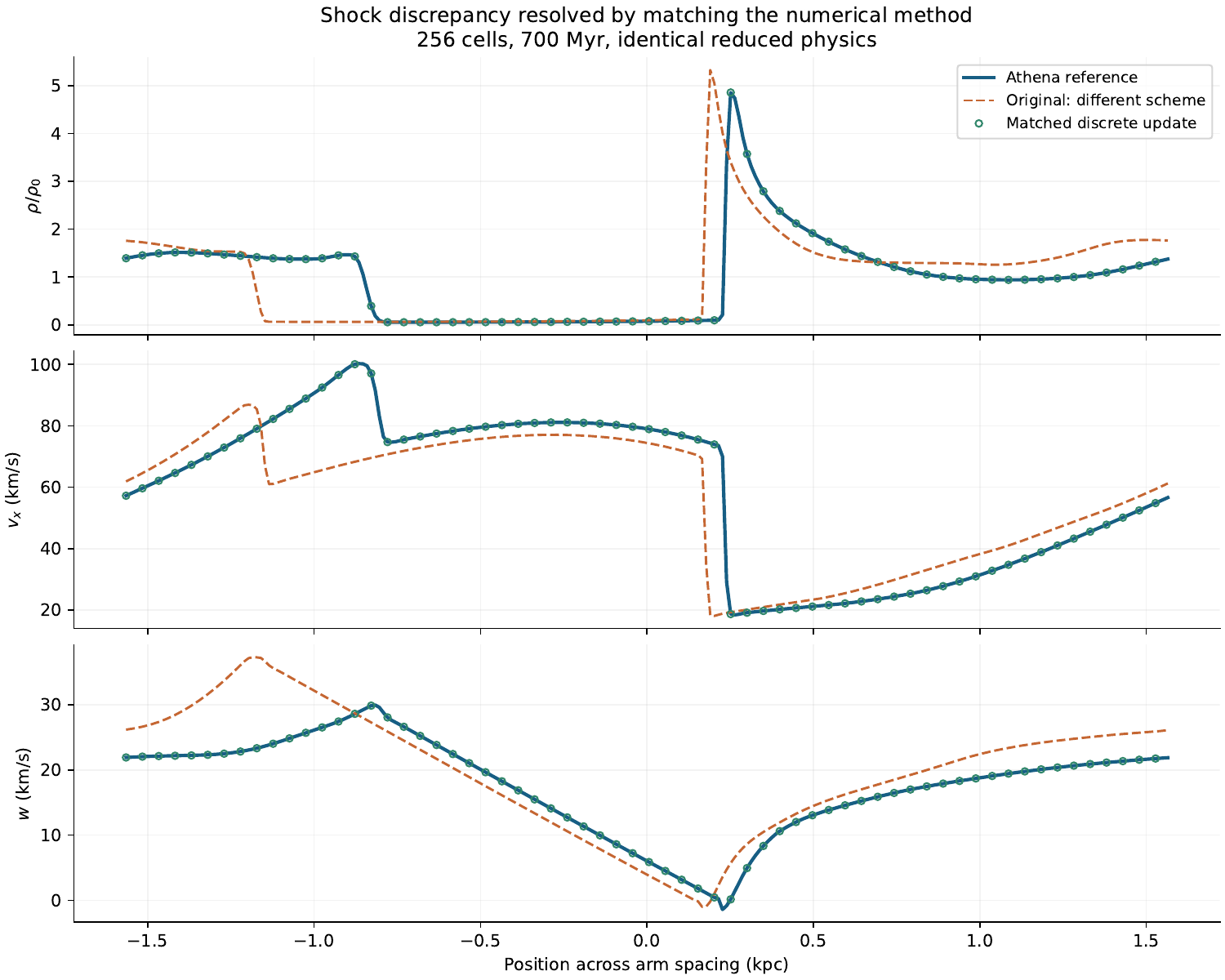}
\caption{\textbf{Solver-matching control for a one-dimensional spiral shock.} The isothermal comparison uses the same reference hydrodynamic discretization. Matching the flux solver and limiter removes the shock discrepancy that comes from comparing different numerical methods. The independent physical/residual test of Appendix~\ref{sec:3.1} holds those choices fixed, so that it tests the coordinate transformation alone.\label{fig:1}}
\end{appendixfigure}

\FloatBarrier
\section{Principal shearing calculation and integrated PRFM specification}\label{app:principal_spec}

The three principal cycling runs and their shared uncompressed control use the same shearing-box solver and source prescriptions. We apply the same conditional PRFM calculation to their saved histories. Appendices~\ref{sec:5} and~\ref{sec:8} specify the separate 30 Myr and repeated-passage experiments.

Table~\ref{tab:period_response} gives the underlying stellar masses and cluster birth counts for Tables~\ref{tab:prediction_errors} and~\ref{tab:cycle_change}.

\begin{table}[!htbp]
\caption{Stellar mass formed over each complete passage ($M_\odot$). All three control rows come from one uncompressed run. They give cumulative totals over 35--55, 35--85 and 35--135 Myr. The birth count gives the number of new clusters. Stellar mass includes subsequent accretion. The difference rows give the simulated cycle minus control, calculated before rounding; each prediction is compared only with the simulated mass of its own run (Table~\ref{tab:prediction_errors}).\label{tab:period_response}}
\centering\footnotesize\setlength{\tabcolsep}{2.8pt}
\begin{tabular}{@{}rlrrrr@{}}
\hline
$T$ (Myr) & Run & Births & Simulated & KS & PRFM \\
\hline
20 & Shared control & 2 & $17{,}888$ & $6{,}366$ & $5{,}608$ \\
 & Cycle & 2 & $11{,}336$ & $7{,}313$ & $7{,}404$ \\
 & Difference & & $-6{,}552$ & & \\
\hline
50 & Shared control & 2 & $17{,}888$ & $15{,}615$ & $13{,}010$ \\
 & Cycle & 7 & $54{,}089$ & $17{,}768$ & $17{,}554$ \\
 & Difference & & $+36{,}201$ & & \\
\hline
100 & Shared control & 5 & $54{,}194$ & $30{,}365$ & $25{,}275$ \\
 & Cycle & 5 & $70{,}092$ & $34{,}071$ & $33{,}119$ \\
 & Difference & & $+15{,}899$ & & \\
\hline
\end{tabular}
\end{table}

\subsection{Combined background, images and sources}\label{app:principal_map}

Let $s(t)$ be the dimensionless lattice shear and $y_0$ the lattice-origin offset. Between lattice relabelings, $\dot y_0=0$ and
\begin{equation}
 x=a\xi,\quad y=\eta+s\xi+y_0,\quad z=\zeta,\qquad
 \dot s=-q\Omega a,\qquad \mathbf v_0=(Hx,-q\Omega x,0).
 \label{eq:combined_map}
\end{equation}
The stored residual velocity $\mathbf u=\mathbf v-\mathbf v_0$ subtracts shear as well as the compression removed in $\mathbf w$ (Section~\ref{sec:2.1}). Define $\mathcal D=\partial_t|_{\xi,\eta,\zeta}+\mathbf u\cdot\nabla_x$. Physical derivatives are $\partial_x=a^{-1}(\partial_\xi-s\partial_\eta)$, $\partial_y=\partial_\eta$ and $\partial_z=\partial_\zeta$. With $\mathbf F=-\rho^{-1}\nabla(P+B^2/2)+\rho^{-1}(\mathbf B\cdot\nabla)\mathbf B-\nabla\Phi$, the equations before radiative and stellar source updates are
\begin{align}
 \mathcal D\rho&=-\rho(\nabla\cdot\mathbf u+H),\\
 \mathcal D u_x&=F_x-Hu_x+2\Omega u_y,\quad
 \mathcal D u_y=F_y-(2-q)\Omega u_x,\quad \mathcal D u_z=F_z,\\
 \mathcal D\mathbf B&=(\mathbf B\cdot\nabla)\mathbf u-\mathbf B\nabla\cdot\mathbf u
       +H(B_x\hat{\mathbf x}-\mathbf B)-q\Omega B_x\hat{\mathbf y}.
 \label{eq:combined_induction}
\end{align}
Here $\Phi$ includes gas/particle self-gravity and the imposed vertical potential; radial tidal and Coriolis terms are combined explicitly. Inserting $\mathbf v_0$ into the physical rotating-frame equations gives the additional imposed acceleration
\begin{equation}
 \mathbf f_{\rm imposed}=(\dot H+H^2)x\hat{\mathbf x}+(2-q)\Omega Hx\hat{\mathbf y}.
 \label{eq:combined_forcing}
\end{equation}
Gas and particles both receive this background; for $\Omega=0$ it reduces to the nonrotating imposed acceleration.

For a reference radial extent $L_\xi$, a radial image is displaced by $(aL_\xi,sL_\xi,0)$ and differs in physical velocity by $(HL_x,-q\Omega L_x,0)$, where $L_x=aL_\xi$, so the azimuthal image offset $\Delta y=sL_\xi$ obeys $d\Delta y/dt=-q\Omega L_x(t)$ between remaps. Residual states are periodic on this lattice (matching at fixed physical $y$ would give different boundaries). Azimuthal images differ by $(0,L_y,0)$ with $L_y=256$ pc. Particle wrapping, nearest-image separation and all finite source regions use the same lattice.

Stored reference-mesh face fields are $(b_\xi,b_\eta,b_\zeta)=(B_x,aB_y-sB_x,aB_z)$. A uniform field reconstructs as
\begin{equation}
 B_x=b_\xi,\qquad B_y=(b_\eta+s b_\xi)/a,\qquad B_z=b_\zeta/a,
 \label{eq:combined_flux}
\end{equation}
which supplies $\dot B_y=-HB_y-q\Omega B_x$. Symmetric epicyclic velocity maps bracket the compression map, and Simpson quadrature advances $\dot s=-q\Omega a$. When $|s|>1/2$, set $k=-\operatorname{round}(s)$, $s\leftarrow s+k$ and $y_0\leftarrow y_0-k\Delta/2$. This integer relabeling leaves physical cell centers invariant: cell-centered states permute, face fluxes change through a shared edge electromotive force, and particles wrap. The physical background work on residual gas-plus-magnetic energy is
\begin{equation}
 \dot W_{\rm bg}=\int\!\left\{-H\left[\rho u_x^2+P+\frac{B_y^2+B_z^2-B_x^2}{2}\right]
             +q\Omega(\rho u_xu_y-B_xB_y)\right\}dV.
 \label{eq:combined_work}
\end{equation}
The energy budget accumulates the full change from each background-flow update, including rotation and shear, and records the remap energy change separately.

For a nearest-image source--cell displacement $\delta\mathbf x$, transfers, mergers and supernova injection use the common-frame boost $\mathbf v_0(\delta\mathbf x)=(H\delta x,-q\Omega\delta x,0)$. The source-energy identity of Section~\ref{sec:2.3} generalizes by replacing $v_{0,i}\Delta m_{x,i}^{\rm phys}$ with $\mathbf v_{0,i}\cdot\Delta\mathbf m_i^{\rm phys}$ and $v_{0,i}^2$ with $|\mathbf v_{0,i}|^2$; the velocity-frame conversion leaves magnetic fields unchanged.

Self-gravity uses horizontal Fourier modes on the sheared lattice and an isolated vertical Green function for the discrete Poisson operator. Horizontal derivatives and the Laplacian include the same $a,s$ metric; particle deposition and force interpolation share triangular-shaped-cloud weights. The external potential is
\begin{equation}
 \Phi_{\rm ext}(z)=2\pi G\Sigma_{\rm old}\bigl(\sqrt{z^2+h_{\rm old}^2}-h_{\rm old}\bigr)
             +2\pi G\rho_{\rm dm}R_0^2\ln[1+(z/R_0)^2],
 \label{eq:principal_potential}
\end{equation}
with $\Sigma_{\rm old}=42\,M_\odot\,{\rm pc}^{-2}$, $h_{\rm old}=245$ pc, $R_0=8000$ pc and the halo density of Section~\ref{sec:pair_setup}. Open vertical gas and electromotive-force boundaries follow Section~\ref{sec:2.2}; particles crossing a vertical boundary are removed and their mass recorded.

The component tests for the solver used in the principal runs cover mapped fluxes, uniform fields on skewed meshes, combined compression/flux freezing, integer shear remapping, transport after remapping, particle gravity and epicyclic orbits. The mapped-flux error is $1.11\times10^{-15}$, the largest normalized divergence in the remap/transport check is $1.95\times10^{-15}$, and particle-orbit errors fall by factors of four under timestep halving. These checks complement the pure-compression tests.

\subsection{Source tables and finite physical regions}\label{app:principal_sources}

Table~\ref{tab:principal_sources} gives the spatial choices. Potential searches and control regions use mapped index neighbors; exclusion and merging use physical center-to-center separations, and supernova injection uses a physical radius. Each distance is measured to the nearest image. Accretion/merging stops at the first supernova or at an age of 40 Myr; exclusion of new sinks applies to non-runaway clusters younger than 40 Myr. The 30 Myr stratified runs require a separate binding-energy test for sink creation (Appendices~\ref{sec:4.1} and~\ref{sec:5.1}). In the principal runs, the density, potential and convergence criteria in Table~\ref{tab:principal_sources} determine where a new sink can form, subject to the exclusion distance. Binding energy remains a diagnostic and a criterion in the reduced cloud model of Part~II.

Stellar-population and cooling tables come from the \href{https://github.com/jeonggyukim/pyathena}{\texttt{pyathena}} package, whose tabulated FUV luminosities and supernova rates we use instead of the separate Padova approximation. We recorded the input arrays and software version used in the runs.

The cooling arrays follow standard TIGRESS: corrected \citet{KoyamaInutsuka2002} neutral-gas cooling, plus solar-metallicity collisional-ionization-equilibrium cooling and molecular weights from \citet{SutherlandDopita1993}, as described by \citet{KimOstriker2017}. They have 801 samples indexed by $T_1=(P/\rho)m_{\rm H}/k_B$, with physical temperature from the tabulated molecular weight. Interpolation is linear, with exact integration within each linear isochoric interval. Photoelectric heating vanishes in hot ionized gas according to the heating table. The gas uses $\rho/n_{\rm H}=1.4271\,m_{\rm H}$; the PRFM diagnostic uses $1.4\,m_{\rm H}$.

The adopted TIGRESS stellar-population functions \citep{KimOstriker2017} derive from STARBURST99 \citep{Leitherer1999}. Their arrays have 201 samples at 0.2 Myr spacing over 0--40 Myr; we interpolate linearly, and emission and rates are zero at ages of 40 Myr or more. Their integrated supernova rate is $0.0105\,M_\odot^{-1}$ (95.6 $M_\odot$ per supernova), close to the rounded TIGRESS value of 95.5 $M_\odot$ for a fully sampled Kroupa initial mass function. The upstream table release omits the original synthesis input deck, so our calculation is specified by the exact arrays and normalization rather than by an assumed stellar-track configuration.

Let $L_{\rm FUV}=\sum_jm_j\Psi(\tau_j)$ sum the tabulated FUV luminosity per mass $\Psi$ over non-runaway clusters of current mass $m_j$ and mass-weighted age $\tau_j$. The spatially uniform heating multiplier is
\begin{equation}
 J=0.0024+\frac{L_{\rm FUV}/A}{6.8\,L_\odot\,{\rm pc}^{-2}}
       \frac{1-e^{-\tau_d/2}+(\tau_d/2)E_1(\tau_d/2)}{\tau_d},\quad
 \tau_d=\kappa_d\Sigma_g,\quad\kappa_d=1000\,{\rm cm^2\,g^{-1}}.
 \label{eq:principal_fuv}
\end{equation}
Here $A$ is the physical horizontal area, $\Sigma_g$ the full gas column (cgs for $\tau_d$) and $E_1(x)=\int_x^\infty e^{-u}du/u$. This is slab attenuation, not cell-by-cell ray tracing.

Supernova candidates are Poisson draws from the age-integrated table scaled by current cluster mass. A fraction $2/3$ are binary candidates; secondary ages are drawn from the normalized supernova-age distribution, and candidates whose secondaries precede the primary are discarded. An accepted binary has an in-situ primary supernova and an isotropically kicked secondary with speed density proportional to $\exp[-v/(50\,{\rm km\,s^{-1}})]$ between 20 and 200 km s$^{-1}$. The checked runaway-to-in-situ event ratio is near one half. A runaway carries no mass, gravity or additional FUV until its scheduled explosion, which returns exactly $10\,M_\odot$ through the parent's mass accounting. Other supernovae return up to $10\,M_\odot$ and use $10^{51}$ erg or terminal momentum. For ambient $n_{\rm H}$ in cm$^{-3}$, shell-formation mass is $1540 n_{\rm H}^{-0.33}M_\odot$ and terminal momentum $2.8\times10^5 n_{\rm H}^{-0.17}M_\odot\,{\rm km\,s^{-1}}$. Enclosed-mass ratios of 1 and 0.027 delimit momentum, resolved-energy and ejecta stages, subject to the radius cap in Table~\ref{tab:principal_sources}.

\subsection{Integrated PRFM solve and eligibility}\label{app:prfm_spec}

The integrated PRFM implementation follows Equations (16) and (28)--(32) of \citet{Jeffreson2026}, written here in our notation. Let $\Sigma_g$ and $\Sigma_\star$ be gas and total stellar columns, $\sigma_\star$ the mass-weighted stellar vertical dispersion, and $P_{\rm eq}$ the inferred total equilibrium pressure. Define the ratio of the stellar and gas scale heights, $r_H=H_\star/H_g=\sigma_\star/\sigma_{\rm eff}$, the effective column $S$ and the dimensionless halo factor $D_h$:
\begin{align}
 \sigma_{\rm eff}&=12\,{\rm km\,s^{-1}}(P_{\rm eq}/P_0)^{0.22},\qquad
 S=\Sigma_g+\frac{2\Sigma_\star}{1+r_H},\\
 D_h&=\left[1+\frac{(16\pi/3)G\rho_{\rm dm}\sigma_{\rm eff}^2}{(\pi G S)^2}\right]^{1/2},\\
 H_g&=\frac{2\sigma_{\rm eff}^2}{\pi G S(1+D_h)},\qquad
 P_{\rm eq}=\frac{\pi G\Sigma_g S}{4}(1+D_h)=\frac{\Sigma_g\sigma_{\rm eff}^2}{2H_g}.
 \label{eq:prfm_integrated}
\end{align}
Here $P_0/k_B=10^4$ K cm$^{-3}$. Pressure and dispersion are solved self-consistently by bracketed bisection in $\ln P_{\rm eq}$, with no low-pressure dispersion floor. With the pressure-dependent yield of Section~\ref{sec:pair_measurements},
\begin{equation}
 t_{\rm dyn}=\frac{2H_g}{\sigma_{\rm eff}},\quad
 t_{\rm dep}=\frac{\Upsilon_{\rm tot}}{\sigma_{\rm eff}}t_{\rm dyn},\quad
 \frac{\Sigma_g}{t_{\rm dep}}=\frac{P_{\rm eq}}{\Upsilon_{\rm tot}}.
 \label{eq:prfm_depletion}
\end{equation}
The integrated renormalization is unity. An auxiliary equilibrium-height calculation, solving the unsquared balance equivalent to the corrected cubic, can supply a height ratio for the volumetric evaluation but is not the pressure solve used here. The volumetric density--pressure law, the renormalization of two and the 12 km s$^{-1}$ floor do not enter this integrated comparison.

The fixed old population has $\Sigma_{\rm old}=42\,M_\odot\,{\rm pc}^{-2}$ and $\rho_{\rm old}(z)=\Sigma_{\rm old}[1+(z/h_{\rm old})^2]^{-3/2}/(2h_{\rm old})$. Its assigned dispersion is the Jeans moment of the 35 Myr starting state
\begin{equation}
 \sigma_{\rm old}^2=\frac{2}{\Sigma_{\rm old}}\int_0^\infty \rho_{\rm old}(z)\,z\,g_+(z)\,dz,
 \label{eq:old_jeans}
\end{equation}
where $g_+$ is the restoring-acceleration magnitude from old stars, the actual simulation halo, and symmetrized gas and live-star profiles at 35 Myr. Gas/live-star terms use $4\pi G\int_0^z\rho(z')dz'$ from the profile layers, each 16 pc thick, with constant enclosed column beyond the box; old-star and halo terms remain analytic to infinity. This gives $\sigma_{\rm old}=19.1$ km s$^{-1}$, with old-star, halo, gas and live-star variance contributions of $218$, $66.1$, $77.7$ and $1.35$ (km s$^{-1}$)$^2$.

At later samples, $\Sigma_l$, $\bar v_l$ and $\sigma_l^2$ are the measured live-particle column, vertical mean velocity and variance, including surviving inherited particles. With $\Sigma_\star=\Sigma_{\rm old}+\Sigma_l$ and zero old-population mean velocity,
\begin{equation}
 \sigma_\star^2=\frac{\Sigma_{\rm old}\sigma_{\rm old}^2+\Sigma_l(\sigma_l^2+\bar v_l^2)}{\Sigma_\star}
             -\left(\frac{\Sigma_l\bar v_l}{\Sigma_\star}\right)^2.
 \label{eq:stellar_mixture}
\end{equation}
Gas and live-star columns span the entire physical patch and vertical box. The old-population values $\Sigma_{\rm old}$ and $\sigma_{\rm old}$ stay fixed and are not fitted to the subsequent star-formation rate.

Each layer $i$ has horizontally averaged density $\bar\rho_i$, thickness $\Delta z=16$ pc, number density $n_i=\bar\rho_i/(1.4m_{\rm H})$ and mass $M_i=A\bar\rho_i\Delta z$; the eligibility threshold is $n_c=0.13$ cm$^{-3}$. After solving the full-column equilibrium, set
\begin{equation}
 f_{\rm elig}=\frac{\sum_i M_i\mathbf 1(n_i>n_c)}{\sum_i M_i},\qquad
 \dot M_\star^{\rm PRFM}=\frac{\sum_i M_i\mathbf 1(n_i>n_c)}{t_{\rm dep}}
                       =f_{\rm elig}A\frac{P_{\rm eq}}{\Upsilon_{\rm tot}}.
 \label{eq:prfm_eligibility}
\end{equation}
Here $\mathbf 1$ is the indicator function. Eligible gas receives the common depletion time; there is no whole-box density gate or eligible-only gravity. The full-column sensitivity sets $f_{\rm elig}=1$ with the same equilibrium solve. A separate halo sensitivity doubles the input halo density while keeping the primary eligibility, because PRFM's published inner Navarro--Frenk--White coefficient is $\Omega_d^2=2\pi G\rho_{\rm dm}$, with $\Omega_d$ the effective halo frequency, whereas the simulation's near-midplane halo acceleration is $-4\pi G\rho_{\rm dm}z$. Over 35--55 Myr this raises the predicted masses in the shared control and the 20 Myr cycling run from $5{,}608$ and $7{,}404\,M_\odot$ to $5{,}950$ and $7{,}820\,M_\odot$. The 30 Myr stratified runs instead use $-2\pi G\rho_d z$, with $\rho_d$ the halo-density parameter (Appendix~\ref{sec:4.1}), which is a different restoring force at the same numerical density. These are independent sensitivities, not a combined alternative fit. Predicted mass uses trapezoidal integration over the saved samples, including the changing area.

\bibliographystyle{aasjournalv7.1-local}
\bibliography{references}
\end{document}